\documentclass[11pt]{article}
\usepackage[mathscr]{eucal}
\usepackage{epsfig,amsfonts}
\usepackage{amsmath}
\usepackage{amsthm,amssymb}
\usepackage{bm}
\usepackage{graphicx}
\usepackage{relsize}
\usepackage{hhline}
\usepackage{comment}
\usepackage{cite}
\usepackage{diagbox}
\usepackage{psfrag}
\usepackage{mathrsfs} 
\usepackage{hyperref}
\usepackage{bm}
\usepackage{array}
\usepackage{tikz}
\usepackage{enumerate}
\usepackage[singlelinecheck=false]{caption}
\usepackage{textcomp}
\usepackage{slashed}
\usepackage{hyperref}
\usepackage{mathrsfs} 

\usepackage{lettrine}

\usepackage{multirow}
\DeclareMathAlphabet{\mathpzc}{OT1}{pzc}{m}{it}
\DeclareMathAlphabet\mathbfcal{OMS}{cmsy}{b}{n} 
\makeatletter
\@addtoreset{equation}{section}
\makeatother

\def\bea{\begin{eqnarray}}
\def\eea{\end{eqnarray}}
\def\be{\begin{equation}}
\def\ee{\end{equation}}

\def\be{\begin{equation}}
\def\ee{\end{equation}}
\def\bdm{\begin{displaymath}}
\def\edm{\end{displaymath}}
\def\bea{\begin{eqnarray}}
\def\eea{\end{eqnarray}}

\def\ri{{\rm i}}

\def\XXint#1#2#3{{\setbox0=\hbox{$#1{#2#3}{\int}$}
    \vcenter{\hbox{$#2#3$}}\kern-.5\wd0}}

\newcommand{\rd}{\mbox{d}}
\newcommand{\re}{\mbox{e}}

\DeclareMathAlphabet{\mathpzc}{OT1}{pzc}{m}{it}
\DeclareMathOperator*{\slim}{slim}

\begin{document}

\begin{titlepage}
$\phantom{I}$
\vspace{2.8cm}

\begin{center}
\begin{LARGE}

{\bf  Scaling limit of  $Q$-functions for the ${\cal Z}_r$ invariant inhomogeneous 
XXZ spin-$\frac{1}{2}$ chain near  free fermion point}

\end{LARGE}

\vspace{1.3cm}
\begin{large}

{\bf 

Gleb A. Kotousov$^{1}$, Sergei L. Lukyanov$^{2}$ and Daria A. Shabetnik$^{2}$}

\end{large}

\vspace{1.cm}
${}^{1}$Institut f$\ddot{{\rm u}}$r Theoretische Physik, 
Leibniz Universit$\ddot{{\rm a}}$t Hannover\\
Appelstra\ss e 2, 30167 Hannover, Germany\\
\vspace{.4cm}
${}^{2}$NHETC, Department of Physics and Astronomy\\
     Rutgers University\\
     Piscataway, NJ 08855-0849, USA\\
\vspace{1.0cm}

\end{center}

\begin{center}

\parbox{13cm}{%
\centerline{\bf Abstract} \vspace{.8cm}
At the beginning of the 70's,  Baxter introduced a multiparametric generalization of the six-vertex model.
This integrable system has been found to exhibit a remarkable variety of critical behaviors. 
The work is part of a series of papers devoted to their systematic study. We focus on the
case when the lattice model possesses an additional ${\cal Z}_r$ symmetry and consider the
critical behavior near the so-called free fermion point. Among other things, discussed is the algebra
of extended conformal symmetry underlying the universal behavior. The main result of the paper
is the class of  differential equations  that describe the scaling limit of the solutions to the Bethe
Ansatz equations. 
This is an instance of the correspondence between Ordinary Differential Equations and
Integrable Quantum Field Theory (ODE/IQFT correspondence).
}
\end{center}

\vfill

\end{titlepage}

\setcounter{page}{2}

\tableofcontents

\section{Introduction \label{sec1}}
The concept of the $Q$-operator first appeared in the
work   of Baxter \cite{Baxter} devoted to the solution of the eight-vertex model.
He recognized that the Bethe Ansatz equations --- a common feature in the analysis
 of quantum integrable systems --- 
can be understood as equations determining 
the zeroes of the eigenvalues of the 
$Q$-operator.  The latter unambiguously specify the states of the model
and are often referred to as the $Q$-functions.
Although emerging in the study of exactly solvable statistical systems,
the $Q$-operators have since become central objects of investigation in Integrable Quantum Field Theory
(IQFT), particularly, in the description of integrable structures in Conformal Field Theory (CFT)  \cite{Bazhanov:1996dr,Bazhanov:1998dq,Gromov:2010km,Jeong:2018qpc,Costello:2021zcl}. The most direct connection 
between the conformal $Q$-functions and their lattice counterparts
arises through the scaling limit of the critical statistical system, where the universal scaling behavior is governed by the underlying CFT.
\medskip

Through numerous examples, it has been established that the CFT
$Q$-functions can be identified with the spectral determinants of certain classes of Ordinary Differential Equations (ODEs). 
These observations were later generalized within the broader framework of integrable quantum field theory, leading to what is now known as the ODE/IQFT correspondence \cite{Voros:1994,Dorey:1998pt,Bazhanov:1998wj,Bazhanov:2003ni}.
\medskip

The basic example of conformal $Q$-operators was introduced in refs.\cite{Bazhanov:1996dr,Bazhanov:1998dq}
 in the context of the quantum KdV theory \cite{Bazhanov:1994ft}.
It was proposed later in the work \cite{Bazhanov:2003ni} that  their eigenvalues coincide with the spectral determinants 
of a class of ODEs
referred to as the Schr\"{o}dinger equation with  monster potentials:
\bea\label{sa903hjbdf32}
\bigg(- {\frac{{\rd}^2}{{\rd} x^{2}}} +
\frac{\ell(\ell+1)}{x^2}+x^{2\alpha}-2\, \frac{{\rd}^2}{{\rd} x^2}\sum_{j=1}^{{\tt L}}\log\big(x^{2\alpha+2}
-z_j\big)\bigg)\, \Psi=E\Psi\ .
\eea
Here $\alpha$ and $\ell$ are given parameters, while
the set of complex numbers $\{z_j\}_{j=1}^{{\tt L}}$ satisfy the
algebraic system
\bea\label{jssysys}
\sum_{m=1\atop m\not= j}^{{\tt L}}
\frac{z_j\,\big(z_j^2+(3+\alpha)(1+2\alpha)\,z_j z_m+\alpha (1+2\alpha)\,z_m^2\big)}{(z_j-z_m)^3}-
\frac{\alpha z_j}{4(1+\alpha)}+\frac{(2\ell+1)^2-4\alpha^2}{16(\alpha+1)}=0\, .
\eea
It turns out 
that for given ${\tt L}$, the number of solutions to the above system coincides with 
the dimension
of the level ${\tt L}$ subspace of the Fock space for the chiral boson, i.e., 
the number of integer partitions of ${\tt L}$ \cite{Bazhanov:2003ni,Masoero}. 
This is consistent with the fact that the
conformal $Q$-functions can be obtained through the scaling limit of
the eigenvalues of the $Q$-operator of the critical XXZ spin-$\frac{1}{2}$ chain, where the underlying CFT
is the free massless Bose field (for details see, e.g., \cite{Kotousov:2019ygw}).

\medskip

The relation to the spin chain gives 
an important insight into the structure of the solutions of  the algebraic system \eqref{jssysys},
which can be described as follows.  
The lattice model depends on  the so-called anisotropy parameter, 
sometimes denoted as $q$, and the critical regime occurs when $q$ is a unimodular number.
For $q=\ri \equiv\sqrt{-1}$ the Hamiltonian of the Heisenberg  XXZ spin-$\tfrac{1}{2}$  chain becomes that of the XX model which, 
by means of the Jordan--Wigner transformation,
can be mapped into a quadratic form of lattice fermion operators and then
diagonalized straightforwardly via the Fourier transform.
As a result, the chiral space of states
appearing in the scaling limit is classified in the same way as for a   free chiral   Dirac  field.
Outside the point $q=\ri$ the Jordan--Wigner transformation maps the Heisenberg  XXZ spin-$\tfrac{1}{2}$  chain
to a system of interacting fermions. Nevertheless, in the spirit of  Landau's Fermi-liquid theory, 
the free fermion classification remains valid for the full critical regime, where $|q|=1$.
Taking into account that  $q$  is related to the parameter $\alpha$ from  the ODE
 \eqref{sa903hjbdf32} as $q=\re^{\frac{\ri\pi}{\alpha+1}}$,
it follows that the solutions of \eqref{jssysys}  for any value of $\alpha>0$
can be labeled in the same way as particle-hole excitations over the free fermion vacuum (see 
ref.\cite{Bazhanov:2003ni}).
\medskip

The  Heisenberg XXZ spin-$\frac{1}{2}$ chain admits a multiparametric integrable deformation,  which is related
to the inhomogeneous six-vertex model introduced by Baxter in ref.\cite{Baxter:1971}. 
Fixing all of the parameters except for $q$ to certain values, 
the system then possesses an additional global ${\cal Z}_r$
symmetry
(for more details see, e.g., the work \cite{Bazhanov:2020new}). 
We'll refer to the lattice model as the ${\cal Z}_r$ invariant inhomogeneous XXZ
spin-$\frac{1}{2}$ chain.
In particular, for $r=1$ it coincides with the  Heisenberg  XXZ spin-$\frac{1}{2}$ chain.
For any positive integer $r$ the model depends only on the anisotropy parameter $q$ and
turns out to be critical for  
\be\label{akjsd892hja}
q=\re^{\ri\gamma}\,,\qquad\qquad \gamma\in(0,\pi)\, .
\ee
However,  the  interval for $\gamma$ is split into $r$  segments of equal width $\frac{\pi}{r}$
in each of which the universal behavior is described differently. Following the pioneering paper 
\cite{Jacobsen:2005xz}, the inhomogeneous XXZ spin-$\frac{1}{2}$ chain with $r=2$
 has been studied extensively in the literature \cite{Ikhlef:2008zz,IJS2,Ikhlef:2011ay,Candu:2013fva,Frahm:2013cma,Bazhanov:2019xvy,Bazhanov:2019xvyA,Robertson:2020imc,Frahm:2021ohj,Frahm:2023ery,Kluemper}.
For generic $r$ the domains  $\pi(1-\frac{1}{r})<\gamma<\pi$  and  $0<\gamma<\frac{\pi}{r}$ 
were explored in the works  \cite{Kotousov:2021vih} and \cite{Kotousov:2023zps}, respectively.
\medskip

In this paper, we take  $r$ odd  and focus on the regime with
\be\label{aksj9812jh23}
\gamma\in\big(\tfrac{\pi}{2}(1-\tfrac{1}{r}),\tfrac{\pi}{2}(1+\tfrac{1}{r})\big)\, \qquad\qquad\qquad
(r\ {\rm odd})\ .
\ee
Its special property is that it contains the value $\gamma=\frac{\pi}{2}$ for which the
Hamiltonian of the ${\cal Z}_r$ invariant inhomogeneous XXZ
spin-$\frac{1}{2}$ chain   becomes particularly simple:
\be\label{kas8932jbncxas}
\mathbb{H}=(-1)^{\frac{r+1}{2}}r\sum_{i}\Bigg(\sigma^+_i\bigg(\prod_{j=1}^{r-1}\sigma^z_{i+j}\bigg)\sigma^-_{i+r}+
\sigma^-_i\bigg(\prod_{j=1}^{r-1}\sigma^z_{i+j}\bigg)\sigma^+_{i+r}\Bigg)\qquad\qquad\qquad (\gamma=\tfrac{\pi}{2})\ ,
\ee
where $\sigma^\pm_i=\frac{1}{2}(\sigma^x_i\pm\ri \sigma^y_i)$.
We refer to it as the ``free fermion point'' since $\mathbb{H}$ can be mapped to a collection of 
$r$ non-interacting lattice fermions. The main output of our work is a proposed class of ODEs, which
describe the scaling limit of the $Q$-functions of
the ${\cal Z}_r$ invariant inhomogeneous XXZ spin-$\frac{1}{2}$ chain in the regime \eqref{aksj9812jh23}.
Similar to the Schr\"{o}dinger equation with monster potentials \eqref{sa903hjbdf32}, the  ODEs
can be classified in the same way as the space of states of 
an $r$-component free chiral   Dirac  field. However,  as $r=1$ the proposed ODEs turn out to give
a different, but equivalent, description to \eqref{sa903hjbdf32},\,\eqref{jssysys}
 of the $Q$-functions for the quantum KdV theory.
\medskip

The paper is organized as follows. Section \ref{sec3} contains the main result --- the new class of
ordinary differential equations that describe the scaling limit of the
spin chain  near the free fermion point.
The basic properties of these ODEs are also discussed.
Section \ref{sec2} is devoted to a detailed analysis of the scaling limit of the lattice system at the free fermion point.
In section \ref{sec4}, we describe the space of states occurring in the scaling limit
as well as the algebra of extended conformal symmetry for the full critical regime \eqref{aksj9812jh23}.
The relation between the parameters of the ODEs and those of the lattice system is clarified in section \ref{sec5}.
Numerical checks of our main conjectures concerning the ODE/IQFT correspondence
are outlined in section \ref{sec6}. In the  section that follows we describe some formal mathematical properties
of the algebraic system, analogous to \eqref{jssysys},
 that determines the positions of the apparent singularities and residues
in the proposed differential equations. The special case $r=1$ and the relation to the
Schr\"{o}dinger operators with monster potentials is discussed in section \ref{sec8}.
The paper concludes with an outline of some remaining open problems.
Technical details related to our study are presented in the Appendices.

\section{Class of ODEs describing the scaling \label{sec3}}
This work is a continuation of the  study of the scaling behavior of the
 inhomogeneous XXZ spin-$\frac{1}{2}$ chain reported in refs.\cite{Bazhanov:2020new, Bazhanov:2019xvy,Bazhanov:2019xvyA,Kotousov:2021vih,Kotousov:2023zps,Gehrmann:2024tue}. 
As such, we recall some basic facts and notations
that were used in those  papers.  
The spin chain is defined on a lattice of $N$ sites, where $N$ is divisible by $r$,
subject to 
 quasi-periodic boundary conditions,
\be\label{BC1b}
\sigma^x_{N+m}\pm\ri\,\sigma^y_{N+m}=\re^{\pm2\pi\ri {\tt k}}
\,\big(\,\sigma^x_{m}\pm\ri\,\sigma^y_{m}\,\big)\,,
\qquad \qquad \sigma_{N+m}^z=\sigma^z_m \, .
\ee
For any
 values of the parameters the system possesses ${\rm U}(1)$ invariance and the corresponding charge coincides with $S^z$ ---
the eigenvalue of the  $z$-projection of the total spin operator:
\be\label{oi90iu3jnbsdsd}
\mathbb{S}^z=\frac{1}{2}\sum_{i=1}^N\sigma^z_i\ .
\ee
Focusing on the case for which the model possesses ${\cal Z}_r$ symmetry,
the Bethe Ansatz equations can be written as
\be\label{bae}
\bigg(
\frac{1+q^{+r} \, (\zeta_j)^r}
{1+q^{-r}\,(\zeta_j)^r }
\bigg)^{\frac{N}{r}}
=-\re^{2\pi\ri {\tt k}}\,q^{2S^z}\,
\prod_{i=1}^{\frac{N}{2}-S^z}\,
\frac{\zeta_i-q^{+2}\,\zeta_j }
{\zeta_i-q^{-2}\,\zeta_j }\  ,
\ee
while the  energy in terms of the Bethe roots $\{\zeta_j\}$ is given by
\be\label{kjas8923hjnbxc}
{\cal E}=2r\sum_{j=1}^{\frac{N}{2}-S^z}\frac{\ri\, (q^r-q^{-r})}{(\zeta_j)^r+(\zeta_j)^{-r}+q^r+q^{-r}}\ .
\ee
It was already mentioned in the Introduction that for $q$ a unimodular number, $q=\re^{\ri\gamma}$ and $\gamma\in(0,\pi)$,
 the spin chain is critical. In each of the $r$ segments
\be\label{ksja892jhds}
\frac{\pi}{r}\,A<\gamma<\frac{\pi}{r}\,(A+1)\qquad\qquad {\rm with}\qquad\qquad A=0,1,2\ldots, r-1
\ee
the field theoretical description of the universal behavior turns out to be different. 
Our current  analysis concerns the case 
\be\label{askjd891jh3ds}
A=\frac{r-1}{2}\qquad\qquad {\rm with}\qquad\qquad  r=1,3,5,\ldots\ .
\ee
We will parameterize   $\gamma$ in terms of $\delta$ as
\bea\label{kas9012jjh23nbsd}
\gamma=\frac{\pi}{2}\,(1-\delta)\ ,\ \ \  \qquad\qquad \ \ |\delta|<\frac{1}{r}\ .
\eea 
\medskip

The differential equation which describes the scaling limit of the $Q$-functions corresponding to the vacuum
state in the sector with given $S^z$ is straightforward to obtain when this quantum number is an integer 
that is divisible by $r$. To do so one may make use of a formal property of the Bethe Ansatz equations.
Since $\frac{N}{2}-S^z$ is divisible by $r$ in this case  it is possible to construct a class of solutions of the form
\be\label{kjsa892j3}
\zeta_m^{(a)}=\re^{\frac{\ri\pi}{r}(2a-2-A)}\big(\tilde{\zeta}_m\big)^{\frac{1}{r}}\,,
\ee
where $a=1,2,\ldots, r$ and $m=1,2,\ldots,\frac{1}{r}\,(\frac{N}{2}-S^z)$. Indeed the substitution of
$\{\zeta_m^{(a)}\}$ back into  \eqref{bae} shows
that the  system is satisfied provided that 
$\tilde{\zeta}_m$  obey the similar system  but with $r=1$ and 
$q$, $N$, $S^z$ replaced by $\tilde{q}=(-1)^Aq^r$, $\tilde{N}=\frac{N}{r}$ and $\tilde{S}^z=\frac{S^z}{r}$,
respectively. These are nothing but the  Bethe Ansatz equations for the  Heisenberg XXZ spin-$\frac{1}{2}$
chain. Furthermore, it turns out that if  $\{\tilde{\zeta}_m\}$  are the solution corresponding to the vacuum of the
XXZ chain, then   $\{\zeta_m^{(a)}\}$ would be the set of vacuum Bethe roots 
for the ${\cal Z}_r$ invariant
spin chain.
Thus one  immediately obtains the differential equation describing the scaling limit
of the vacuum  $Q$-functions for the  ${\cal Z}_r$ invariant  inhomogeneous spin chain
from the results for the Heisenberg XXZ spin-$\frac{1}{2}$ chain.
It is given by
\bea\label{mnasghuiqw}
\bigg(- {\frac{{\rd}^2}{{\rd} x^{2}}} +
\frac{\ell(\ell+1)}{x^2}+x^{2\alpha}\bigg)\, \Psi=E\Psi
\eea
with 
\be\label{sa0923jdsjhA}
\alpha=\frac{1+r\delta}{1-r\delta}\ ,\qquad \qquad\qquad \ell+\frac{1}{2}=\frac{2}{1-r\delta}\ {\tt k} +\frac{1}{r}\ S^z
\ee
and 
$\delta$ is related to the anisotropy parameter $\gamma$ as  in \eqref{kas9012jjh23nbsd}.
Note that the differential equation \eqref{mnasghuiqw} coincides with \eqref{sa903hjbdf32}
specialized to ${\tt L}=0$.
It would be convenient for us to perform the change of variables
\be\label{sa0923jdsjhB}
v=\frac{2}{1-r\delta}\,\log(x)+\log\bigg(\frac{1-r\delta}{2}\bigg)\,,\qquad\qquad\qquad \psi=x^{-\frac{1}{2}}\,\Psi
\ee
and bring the Schr\"{o}dinger equation \eqref{mnasghuiqw} to the form
\be\label{as9021jjbnsd}
\big(- \partial_v^2 +{\tt K}^2+\re^{2v}+\epsilon^r\,\re^v\big)\psi=0\,.
\ee
Here and below we use the shortcut notation for the $v$\,-\,dependent function, 
\be\label{asjk812jdsb}
\epsilon\equiv\epsilon(v)=\mu\,\re^{-\delta v}
\ee
with
\be\label{askj892h2aaa3gh8721}
\mu=\bigg(\frac{1-r\delta}{2}\bigg)^{\frac{1}{r}+\delta}\ (-E)^{\frac{1}{r}}\ .
\ee
Also, ${\tt K}={\tt k}+\tfrac{1-r\delta}{2r}\,S^z$.
\bigskip


For a generic eigenstate of the $Q$-operators,  we propose that
the corresponding eigenvalue
is described by a differential equation of the form
\be\label{askjdjh21A}
\big(-\partial_v^2+U(v)\big)\psi=0\,,
\ee
where
\be\label{askjdjh21B}
U(v)={\tt K}^2+\re^{2v}+\epsilon^r\,\re^v+\sum_{i=1}^{{\tt J}}\bigg(
\frac{3\delta^2\varpi_i^2\epsilon^2}{(\epsilon^2-\varpi_i^2)^2}+\frac{2\delta}{\epsilon^2-\varpi_i^2}\,\big(c_i\epsilon^2+
\varpi_i\epsilon\,\re^v\big)\bigg)\ .
\ee
Again, it should be kept in mind that $\epsilon$ is a function of $v$ given by \eqref{asjk812jdsb}.  
The parameter  ${\tt K}$ is expressed in terms of $S^z$ and ${\tt k}$ 
but the relation turns out to be more complicated than for the vacuum case and will be discussed later
in sec.\,\ref{sec5}.
Compared to the  ODE \eqref{as9021jjbnsd}, the differential equation  \eqref{askjdjh21A},\,\eqref{askjdjh21B}
contains a sum,
which introduces  singularities in the complex $v$-plane at the points where $\epsilon(v)=\pm\varpi_i$.
The set $\{\varpi_i\}_{i=1}^{{\tt J}}$, as well as the coefficients $\{c_i\}_{i=1}^{{\tt J}}$, 
are determined through 
 the system of algebraic equations
\bea\label{kasj892jhsa}
2c_j&=&\varpi_j^r-1+2\delta\sum_{i\ne j}\frac{\varpi_i\varpi_j}{\varpi_j^2-\varpi_i^2}\\[0.2cm]
c_j^2&=& {\tt K}^2+\delta c_j-\frac{\delta^2}{4}+\sum_{i\ne j}\bigg(
\frac{3\delta^2\varpi_i^2\varpi_j^2}{(\varpi_i^2-\varpi_j^2)^2}+\frac{2\delta\,c_i\varpi_j^2}{\varpi_j^2-\varpi_i^2}\bigg)
\qquad\qquad (j=1,2,\ldots,{\tt J})\, .\nonumber
\eea
\medskip

The scaling limit of the $Q$-functions is expressed in terms of the spectral determinants of the ODE. The latter can be defined as follows.
Let's assume, for now, that $\delta<0$.
Then as $v\to-\infty$ the function $\epsilon$ \eqref{asjk812jdsb} tends to zero so that $U(v)\to {\tt K}^2$.  This allows one to introduce  two Jost
solutions  of the ODE \eqref{askjdjh21A},\,\eqref{askjdjh21B} through the asymptotic condition
\be\label{sakj2jsdnnbsa}
\psi_{\pm}\to\re^{\pm{\tt K}v}\qquad\qquad {\rm as}\qquad\qquad v\to-\infty\, \qquad\qquad (\delta<0)
\ee
provided that ${\tt K}$  is pure imaginary, $\Re e({\tt K})=0$. If we treat $\varpi_i$ and $c_i$
entering into $U(v)$ \eqref{askjdjh21B} as independent parameters,  it is clear that
 $\psi_{+}$, as a function of complex  ${\tt K}$, is analytic in the right-half plane $\Re e({\tt K})\ge 0$, while
$\psi_{-}$ is analytic for $\Re  e({\tt K})\le 0$. In fact,  in this case
both  $\psi_{+}$ and $\psi_-$ can be unambiguously analytically continued to
 the whole complex ${\tt K}$-plane and thus defined turn out to be meromorphic functions with the only singularities being
a certain set of simple poles. However, one should keep in mind that $\varpi_i$ and $c_i$ are determined through
the  system \eqref{kasj892jhsa}, where ${\tt K}$ enters as a parameter. Thus they,
and hence also $\psi_\pm$, are multi-valued functions of ${\tt K}$. For our purposes, it is sufficient 
to assume that   the branch of  $\psi_\pm$   may be chosen in such a way that as
functions of ${\tt K}$ they are analytic at least in a certain strip containing the imaginary ${\tt K}$-axis. 
This assumption can be verified for the cases ${\tt J}=1,2$, when the system 
\eqref{kasj892jhsa} admits a relatively simple, closed-form solution (for details see sec.\,\ref{sec71}).

\medskip
In addition to $\psi_\pm$, we need to introduce another solution $\chi(v)$
 that decays as $v\to+\infty$. It is  specified unambiguously for $\delta\ne -\frac{k-1}{rk}$ with $k=1,2,3,\ldots$ through the  asymptotic formula:
\be\label{askj21983hjbnsd}
\chi(v)\asymp\exp\Big(-\tfrac{v}{2}+\re^v\,{}_2 F_1\big(-\tfrac{1}{2},-\tfrac{1}{1+r\delta},1-\tfrac{1}{1+r\delta}\,\big|\,-\epsilon^r(v)\,\re^{-v}\,\big)+o(1)\Big)\qquad {\rm as}\qquad v\to+\infty\ ,
\ee
where ${}_2 F_1$ stands for the conventional Gauss hypergeometric function.
This asymptotic is straightforwardly obtained  from the WKB approximation.
The spectral determinants $D_\pm(\mu)$, subject to the normalization condition
\be\label{kjas891hjnsdm}
D_\pm(0)=1\,,
\ee
can be defined as
\be\label{skaj8721y32jh}
D_\pm(\mu)=
\frac{W_\pm(\mu)}{W_\pm(0)}\qquad\quad {\rm with}\qquad\quad W_\pm(\mu)=W[\chi,\psi_{\pm}]\ \qquad\qquad
(\delta<0)\,,
\ee
where $W[f,g]=f\,\partial_v g-g\,\partial_v f$ stands for the Wronskian.  

\medskip

It turns out that the spectral determinants obey the so-called quantum Wronskian relation
\be\label{asd9h12jhbsa}
\re^{+\ri\pi {\tt K}}\,D_+(q\mu)\,D_-(q^{-1}\mu)
-\re^{-\ri\pi {\tt K}}\,D_-(q\mu)\,D_+(q^{-1}\mu)=2\ri\sin(\pi {\tt K})\qquad\qquad\qquad (\delta<0)
\ee
with $q=\re^{\frac{\ri\pi}{2}(1-\delta)}$.
The proof is especially simple in the case of the vacuum ODE \eqref{as9021jjbnsd}. Then the potential is non-singular
for any value of complex $v$ and the derivation of \eqref{asd9h12jhbsa} follows the original arguments of ref.\cite{Bazhanov:1998wj},
 based on the formal
invariance of the differential equation w.r.t. the transformation
\be\label{ask932j21}
v\mapsto v+\ri\pi\,,\qquad\qquad \mu\mapsto q^{-2}\mu\, .
\ee
For the general ODE \eqref{askjdjh21A},\,\eqref{askjdjh21B} the situation is more delicate. 
At first glance, to meaningfully define the action of the symmetry transformation \eqref{ask932j21} 
on the solutions $\psi_\pm$ and $\chi$,  one should require  
 $U(v)$ to  be free of singularities in  the complex $v$ plane.
 As a matter of fact,  one can allow singularities provided they are apparent, i.e., such that
 the ratio of any two solutions of the ODE is single-valued in their vicinity. A quick inspection of the potential \eqref{askjdjh21B} 
shows that it has poles at 
\be\label{jkdsdas982jh}
v_{j,m}=-\frac{1}{\delta}\,\log\Big(\frac{\varpi_j}{\mu}\Big)+\ri\,\frac{\pi m}{\delta}\qquad {\rm with}\qquad j=1,2,\ldots,{\tt J}\,;\ \ m\in\mathbb{Z}\ .
\ee
In the vicinity of   $v_{j,m}$, the function $U(v)$ \eqref{askjdjh21B} admits the Laurent  expansion of the form
\bea\label{kjas9823jhsd}
U(v)=\frac{3}{4}\ \frac{1}{(v-v_{j,m})^2}+\frac{R_{j,m}}{v-v_{j,m}}+U_{j,m}+O(1)\ .
\eea
That the singularities at $v_{j,m}$ are apparent imposes the constraints
\bea
R_{j,m}^2=U_{j,m}\ ,
\eea 
which, in turn,  lead to eqs.\,\eqref{kasj892jhsa}.
\medskip

The following comment is in order here.  The full set of conditions that the differential equation of the form  $\big(-\partial_v^2+U(v)\big)\psi=0$ has
an apparent singularity at the point $v=v_*$ is well know (see, e.g., ref.\cite{Gao}). Among them is that
$U(v)$ must admit a Laurent expansion at  $v=v_*$ and
\be\label{askd98h12gh3gfgsa}
U(v)=\frac{j_*(j_*+1)}{(v-v_*)^2}+O\big((v-v_*)^{-1}\big)
\ee
with integer or half-integer  $j_*=\frac{1}{2},1,\frac{3}{2},\ldots\ $. In the case of $U(v)$  \eqref{askjdjh21B}, for all the apparent singularities 
with  $v_*=v_{j,m}$ formula \eqref{kjas9823jhsd} means that
 $j_*=\frac{1}{2}$. At the same time, for the Schr\"{o}dinger operator with
monster potential \eqref{sa903hjbdf32},  $j_*=1$.  Nevertheless  it is expected that
the differential  equations \eqref{askjdjh21A}-\eqref{kasj892jhsa} with $r=1$ 
and the ones with monster potential \eqref{sa903hjbdf32},\,\eqref{jssysys} provide equivalent but alternative descriptions of the scaling limit of the 
$Q$-functions for the Heisenberg XXZ spin-$\frac{1}{2}$ chain (see sec.\,\ref{sec8} below).
\medskip

In the above  it was assumed that $-\frac{1}{r}<\delta<0$. Let's briefly explain the modifications required for 
$0<\delta<\frac{1}{r}$.
In this case as $v\to-\infty$ the function $\epsilon=\epsilon(v)$  \eqref{asjk812jdsb} exponentially grows and 
\be
\lim_{v\to-\infty} U(v)={\tt K}^2+2\delta\sum_{i=1}^{{\tt J}} c_i\ \qquad\qquad (\delta>0)\,.
\ee 
As such, it is convenient to introduce the parameter ${\tt M}$ through the relation
\be\label{askj902jhbdssa}
\sum_{i=1}^{{\tt J}} c_i={\tt M}\,{\tt K}+\frac{{\tt M}^2}{2}\,\delta\,.
\ee
Then, in the definition of the Jost solutions $\psi_\pm$ \eqref{sakj2jsdnnbsa}, one should replace ${\tt K}$ 
by ${\tt K}+{\tt M}\,\delta$. At the same time, the asymptotic condition \eqref{askj21983hjbnsd} specifying $\chi(v)$ is still valid. In fact,
for $\delta>0$ it simplifies to
\be
\chi(v)\asymp\exp\Big(-\tfrac{v}{2}-\re^v+o(1)\Big)\qquad {\rm as}\qquad v\to+\infty\qquad\qquad (\delta>0)\ .
\ee
Turning to the definition of 
 the spectral determinants, the normalization condition \eqref{kjas891hjnsdm} requires one to modify \eqref{skaj8721y32jh} as
\be\label{skaj8721y32jhAAA}
D_\pm(\mu)=
\frac{W_\pm(\mu)}{W_\pm(0)}\qquad\quad {\rm with}\qquad\quad W_\pm(\mu)=\mu^{\mp{\tt M}}\,W[\chi,\psi_{\pm}]\ \qquad\qquad
(\delta>0)
\ee
(here the denominator is understood as the limit of $W_\pm(\mu)$ as $\mu\to0$).
A repeat of the derivation of the quantum Wronskian relation, yields
\be\label{asd9h12jhbsaAA}
\re^{+\ri\pi ({\tt K}+{\tt M})}\,D_+(q\mu)\,D_-(q^{-1}\mu)
-\re^{-\ri\pi ({\tt K}+{\tt M})}\,D_-(q\mu)\,D_+(q^{-1}\mu)=2\ri\sin\big(\pi\, ({\tt K}+{\tt M})\big)\qquad (\delta>0)\, .
\ee

\medskip
To make a link to the lattice system, recall that the Hamiltonian of the  inhomogeneous XXZ spin-$\tfrac{1}{2}$  chain
belongs to a commuting family, a prominent member of which is the Baxter $Q$-operator \cite{Baxter}. In fact, there are
 two $Q$-operators,
which we denote by $\mathbb{A}_\pm(\zeta)$ to emphasize that they are normalized as 
\be
\mathbb{A}_\pm(0)=1
\ee
and to keep with the notation of the work \cite{Bazhanov:2020new}.
The details of the construction of $\mathbb{A}_\pm$
as well as their main properties can be found in sec.\,3 of that paper. Of these, the two most important ones are the commutativity condition
\bea\label{tqcomm}
\big[\mathbb{A}_\pm(\zeta_1),\,\mathbb{A}_\pm(\zeta_2)\big]=\big[\mathbb{A}_\pm(\zeta_1),\,\mathbb{A}_\mp(\zeta_2)\big]=0
\eea
and  the  quantum Wronskian relation. The latter, for the case of the ${\cal Z}_r$ invariant inhomogeneous XXZ spin chain, takes the form
\be\label{qwron}
q^{+2\mathbb P}\ 
\mathbb{A}_+\big(q^{+1}\zeta\big)\,\mathbb{A}_-\big(q^{-1}\zeta \big)-
q^{-2\mathbb P}\ 
  \mathbb{A}_-\big(q^{+1}\zeta\big)\,\mathbb{A}_+\big(q^{-1}\zeta
  \big)=\big(q^{+2\mathbb P}-q^{-2\mathbb P}\,\big)\, (1+\zeta^r)^{\frac{N}{r}}
 \ee
with 
\be
q^{2{\mathbb P}}=\re^{\ri\pi{\tt k}}\, q^{\mathbb{S}^z}\ .
\ee
Note that \eqref{qwron}, together with the fact that the 
eigenvalues of $\mathbb{A}_+(\zeta)$ are polynomials in $\zeta$ of order $\frac{N}{2}-S^z$ immediately implies that its zeroes $\zeta_j=\zeta_j^{(+)}$ 
obey the Bethe Ansatz equations \eqref{bae}. Similarly, the eigenvalues of $\mathbb{A}_-(\zeta)$ are polynomials in $\zeta$ of
order $\frac{N}{2}+S^z$ whose zeroes $\zeta_j^{(-)}$ obey the set of coupled equations which can be obtained from  \eqref{bae} 
through the substitution $(S^z,{\tt k})\mapsto (-S^z,-{\tt k})$.
\medskip

Let  $A_\pm(\zeta)$ be the eigenvalues of $\mathbb{A}_\pm(\zeta)$  corresponding to a  state of the lattice system, i.e., the $Q$-functions:
\be\label{kjas982hjbs}
A_\pm(\zeta)=\prod_{j=1}^{\frac{N}{2}\mp S^z}\bigg(1-\frac{\zeta}{\zeta^{(\pm)}_j}\bigg)\ .
\ee
We expect that the scaling relation between them and the spectral determinants as defined in \eqref{skaj8721y32jh},\,\eqref{skaj8721y32jhAAA}
is given by
\be\label{jkas8912jhas}
\slim_{N\to\infty}G^{(N/r)}\Big(-\big( \tfrac{r}{1-r \delta}\big)^{1+r\delta}\mu^r\,\big|\,\tfrac{1}{2}\, (1 - r\delta)\Big)\,
A_\pm\bigg(\Big(\tfrac{r^2N_0}{(1-r\delta)N}\Big)^{\frac{1}{r}+\delta}\,\re^{\frac{\ri\pi(r+1)}{2r}}\,\mu\bigg)=D_\pm(\mu)
\ee
with
\be\label{akjs893jhbas}
G^{(N)}(E\,|\,g)=\exp\Bigg(\
{\displaystyle \sum_{m=1}^{\big[\frac{1}{2(1-g)}\big]}}\
\dfrac{(-1)^{m}\,N}{2m\cos(\pi m g)}\  \bigg(\frac{N_0}{N}\bigg)^{2m(1-g)}\  E^{m}\Bigg)
\ee
and the symbol $[\ldots]$ stands for the integer part.
Also ``slim'', short for scaling limit, is used to emphasize that the limit should
 be considered only for a certain class of states --- the low energy states.
The scaling relation is literally applicable provided that
 $\delta\ne -\tfrac{k-1}{rk}$ with $k=1,2,3,\ldots\ $. Otherwise,  
the function  $G^{(N)}(E\,|\,g)$  is replaced by
\be\label{kas923jnnbc}
G^{(N)}(E\,|\,g)\big|_{g=1-\frac{1}{2k}}=\exp\Bigg(
\dfrac{N_0 E^{k}}{k\pi}\,\log(N B_{k})+
{\displaystyle\sum_{m=1}^{k-1}}\
\dfrac{N}{2m\cos(\frac{\pi m}{2k})}\  \bigg(\frac{N_0}{N}\bigg)^{\frac{m}{k}}\  E^m\Bigg) \,,
\ee
where the constants $B_k$ can be chosen at will.
It is worth mentioning that the precise form of the relation, including the explicit expression for the state independent
 function $G^{(N)}(E\,|\,g)$ and the value of the
numerical  constant,
\be
N_0=\frac{\sqrt{\pi}\,\Gamma(\frac{3}{2}-\frac{r\delta}{1+r\delta})}{r\,\Gamma(2-\frac{r\delta}{1+r\delta})}\ ,
\ee
follows from  a consideration of the ground state and were obtained from the previously established relation for the
Heisenberg XXZ spin-$\tfrac{1}{2}$  chain see, e.g., ref.\cite{Bazhanov:2019xvyA}.\footnote{%
Formula \eqref{jkas8912jhas} can  be re-written in a simpler way by introducing proper notations. 
However, we prefer to use the same ones  as in eqs.\,(5.47) and (5.48)   from ref.\cite{Bazhanov:2019xvyA}.} 
\medskip

A mathematically rigorous proof of  eq.\,\eqref{jkas8912jhas} is an open problem. Nevertheless,
an argument in support of it follows from the quantum Wronskian relation. It should be mentioned that in the case of the lattice system,
the operator valued relation \eqref{qwron} takes the same form for any value of $q$ and generic ${\tt k}$. At the same time, the functional relation
obeyed by the spectral determinants seems different for $\delta<0$ \eqref{asd9h12jhbsa} and  $\delta>0$ \eqref{asd9h12jhbsaAA}. In order 
for the left and right hand sides of \eqref{jkas8912jhas} to obey the same quantum Wronskian relation for any $|\delta|<\frac{1}{r}$ one requires that
\be\label{kjas89jhbnsd}
\re^{2\pi\ri {\tt K}}=\pm\re^{2\pi \ri {\tt k}}\, q^{2{S}^z}
\ee
as well as
\be
\re^{2\pi\ri{\tt M}}=1\, .
\ee
The last condition implies that ${\tt M}$ must be an integer. Our numerical analysis of the  algebraic system \eqref{kasj892jhsa}
suggests that
 indeed, for all solutions $\{\varpi_i,c_i\}_{i=1}^{{\tt J}}$ the sum of $c_i$ is such that \eqref{askj902jhbdssa} holds true with
some integer ${\tt M}$. In the next sections we also give further analytical and numerical evidence in support of the scaling relation.
\medskip

Formula \eqref{jkas8912jhas} can be equivalently written as a relation for the Bethe roots $\zeta^{(\pm)}_j$ themselves rather than the products \eqref{kjas982hjbs}. 
It implies that in the vicinity of $\zeta=0$, the Bethe roots develop a scaling behavior such that the following limits exist:\footnote{%
In writing the relation \eqref{mas9823cas} we assume a certain ordering of the Bethe roots, say,
$$
\big|\zeta_1^{(\pm)}\big|\le\big|\zeta_2^{(\pm)}\big|\le\ldots\le\big|\zeta_{\frac{N}{2}-S^z}^{(\pm)}\big|\ .
$$
}
\be\label{mas9823cas}
\slim_{N\to\infty\atop j-{\rm fixed}}\,
\bigg(\frac{(1-r\delta)N}{r^2N_0}\bigg)^{\frac{1}{r}+\delta}\,\re^{-\frac{\ri\pi(r+1)}{2r}}\ \zeta^{(\pm)}_j=\mu_j^{(\pm)}\, .
\ee
Moreover, the limiting values  coincide with the zeroes of the spectral determinants, i.e., 
\be
D_\pm\big(\mu_j^{(\pm)}\big)=0\ .
\ee
In view of the invariance of the Bethe Ansatz equations \eqref{bae} w.r.t. the transformation $\zeta_j\to\zeta^{-1}_j$ and ${\tt k}\to-{\tt k}$  
one may expect that the similar limits to \eqref{mas9823cas} exist for the roots accumulating at $\zeta=\infty$.
 In this case, the counterpart to \eqref{jkas8912jhas}  would involve the 
polynomials in $\zeta^{-1}$:
\be\label{kajs892hj9012}
\bar{A}_\pm(\zeta)\equiv
\prod_{m=1}^{\frac{N}{2}\mp S^z}\bigg(1-\frac{\zeta_m^{(\pm)}}{\zeta}\bigg)\, .
\ee
These are  eigenvalues of 
\be\label{kjsajh1289}
\bar{\mathbb{A}}_\pm(\zeta)=\zeta^{-\frac{N}{2}\pm S^z}\,\mathbb{A}_\pm(\zeta)\,\big(
\mathbb{A}_\pm^{(\infty)}\big)^{-1}\,,
\ee
where the operator $\mathbb{A}_\pm^{(\infty)}$ belongs to the commuting family and its eigenvalue
for a Bethe state
is given by a product over the corresponding Bethe roots:
\be\label{8932io90218932}
A_\pm^{(\infty)}=\prod_{m=1}^{\frac{N}{2}\mp S^z}\,\big(-\zeta_m^{(\pm)}\big)^{-1}\, .
\ee
Then the scaling of the Bethe roots in the vicinity of $\zeta=\infty$ can be expressed as
\be\label{jkas8912jhasB}
\slim_{N\to\infty}G^{(N/r)}\Big(-\big( \tfrac{r}{1-r \delta}\big)^{1+r\delta}\bar{\mu}^r\,|\,\tfrac{1}{2}\, (1 - r\delta)\Big)\,
\bar{A}_\pm\bigg(\Big(\tfrac{(1-r\delta)N}{r^2N_0}\Big)^{\frac{1}{r}+\delta}\,
\re^{-\frac{\ri\pi(r+1)}{2r}}\,\bar{\mu}^{-1}\bigg)=\bar{D}_\pm(\bar{\mu})\ .
\ee
Here $\bar{D}_\pm(\bar{\mu})$ are the spectral determinants of an ODE similar to \eqref{askjdjh21A},\,\eqref{askjdjh21B}
but with ${\tt K}$ replaced by $\bar{\tt K}$, $\mu$ substituted for $\bar{\mu}$, and the set $\{(\varpi_i,c_i)\}_{i=1}^{{ {\tt J}}}$ swapped for
$\{(\bar{\varpi}_i, \bar{c}_i)\}_{i=1}^{\bar {\tt J}}$. The latter obey an algebraic system that is obtained from \eqref{kasj892jhsa}
through the same formal substitutions. Similar to \eqref{kjas89jhbnsd} the parameter $\bar{\tt K}$ entering into the differential equation, 
the twist parameter ${\tt k}$  and the ${\rm U}(1)$ charge $S^z$ must be related as
\be
\re^{2\pi \ri \bar{{\tt K}}}=\pm\re^{-2\pi \ri {\tt k}}\, q^{2{S}^z}\, .
\ee
\medskip

A final comment concerns the works  \cite{Kotousov:2021vih}  and \cite{Kotousov:2023zps}.
In the latter paper, it is discussed the ${\cal Z}_r$ invariant inhomogeneous XXZ spin-$\frac{1}{2}$ chain 
in the critical regime with $\gamma\in(0,\frac{\pi}{r})$. The differential equations are proposed which
describe the scaling limit of the $Q$-functions, taken similar to as in 
 eqs.\,\eqref{jkas8912jhas} and \eqref{jkas8912jhasB}. 
Among the results of the work  \cite{Kotousov:2021vih}
is a conjecture for the class of differential equations describing the scaling limit of the $Q$-functions
for the  inhomogeneous XXZ spin-$\frac{1}{2}$ chain in a certain domain of
parameters with $\gamma\in\big(\pi(1-\frac{1}{r}),\pi\big)$  for which the ${\cal Z}_r$ symmetry is, in general, explicitly broken.
It is interesting to note that though for  $r=1$ the spin chain becomes the usual Heisenberg XXZ spin-$\frac{1}{2}$ chain,
the corresponding differential equations from \cite{Kotousov:2021vih} are essentially different to both the
Schr\"{o}dinger operator with monster potential \eqref{sa903hjbdf32},\,\eqref{jssysys}
as well as the ones considered in this paper \eqref{askjdjh21A},\,\eqref{askjdjh21B} for $r=1$. 
This special case of the ODEs from the paper \cite{Kotousov:2021vih} 
was  also explored in ref.\cite{MasoeroRaimondo}.





\section{The free fermion point  $(\delta=0)$\label{sec2}}
\subsection{Scaling limit of $A_+(\zeta)$}
In the previous section, the spectral determinants were introduced for all values of $|\delta|<\frac{1}{r}$ apart from the points
$\delta= -\tfrac{k-1}{rk}$  with $k=1,2,3,\ldots\ $.
Of these, the case of   $k\ge 2$ is rather straightforward and requires one
to  more carefully specify the asymptotic \eqref{askj21983hjbnsd}, which
is used in the definition of the subdominant solution $\chi(v)$.\footnote{%
The  quantum Wronskian relation needs to be modified  for $\delta= -\tfrac{k-1}{rk}$ and $k=1,2,3,\ldots\ $.}
In contrast to this, for $k=1$, i.e., $\delta=0$  the analysis is more subtle.  The issue can be traced already at the level of the
differential equations, since a formal substitution of $\delta=0$ into formulae
\eqref{askjdjh21A}-\eqref{kasj892jhsa} leads to the vacuum ODE  \eqref{as9021jjbnsd}.
In this section, rather than performing a detailed study of the limit $\delta\to 0$ of the spectral determinants,
we'll consider the scaling limit of the lattice $Q$-functions instead. It is much simpler to do since  for 
$\delta=0$ the Bethe Ansatz equations decouple and admit an explicit solution. 
\medskip

For our purposes, it is sufficient to focus on the $Q$-function $A_+(\zeta)$. Furthermore, we assume
\be
 S^z\ge 0\ ,
\ee
so that the  number of  zeroes   of $A_+(\zeta)$, which is given by $\frac{N}{2}-S^z$, is less than or equal to 
$\frac{N}{2}$.\footnote{%
The lattice model possesses ${\cal C}{\cal P}$ symmetry, the generator of which intertwines the sectors with
$S^z$ and $-S^z$. Moreover, under the ${\cal C}{\cal P}$ transformation, 
$
{ \hat {\cal C}}{\hat  {\cal P}}\,{\mathbb A}_+(\zeta)\,{ \hat {\cal C}}{\hat  {\cal P}} 
= \bar{{\mathbb A}}_-
\big(\zeta^{-1}\big)\ 
$, where $\bar{{\mathbb A}}_-$ is defined in \eqref{kjsajh1289}. As a result,
in the analysis of the scaling behavior of the spin chain,  it is sufficient to restrict to the case $S^z\ge0$.
The results for $S^z<0$ can  be recovered from those with $S^z>0$ after the identification of the  
${\cal C}{\cal P}$ transformation in the underlying CFT. Such a procedure is discussed in detail
for the ${\cal Z}_2$ invariant spin chain with  $\gamma\in(0,\frac{\pi}{2})$ in sec.\,17.2 of ref.\cite{Bazhanov:2019xvyA}.} 
The Bethe Ansatz equations  \eqref{bae}  for the zeroes $\zeta_j$ at $q=\ri$ boil down to
\be\label{ksa921jd}
\bigg(
\frac{1+ (\ri\,\zeta_j)^r}
{1-(\ri\,\zeta_j)^r }
\bigg)^{\frac{N}{r}}
=\re^{\ri\pi (2 {\tt k}+S^z-1)}\ .
\ee
Hence
\be\label{oi89uy21hdsbn}
(\zeta_j)^r=(-1)^{\frac{r-1}{2}}\,\tan\big(\tfrac{\pi r}{N}\,({\cal I}_j+{\tt k}+\tfrac{S^z-1}{2})\big)\,,
\ee
where  ${\cal I}_j$ are  integers.
Different choices of the set $\{{\cal I}_j\}$ correspond to the different Bethe states of the lattice system. We
refer to the vacuum  in the sector with fixed value of $S^z$ as 
the state which minimizes the  energy given by \eqref{kjas8923hjnbxc} with $q=\ri$. It turns out that
the vacuum distribution of the Bethe roots depends  essentially on the integer
\be\label{AAAask8923nbn43}
{\tt s}=\frac{N}{2}-S^z\ \ \  ({\rm mod}\ r)\, .
\ee
If
the number of roots
$\frac{N}{2}-S^z$ is divisible by $r$, i.e., ${\tt s}=0$,
the vacuum turns out to be non-degenerate.
The corresponding set $\{\zeta_j\}$ breaks up into $r$ groups of equal size
$\{\zeta_m^{(a)}\}$ labelled by the superscript $a=1,2,\ldots,r$  
such that
\be\label{kdnxcdsjhewio}
\arg\big(\zeta_m^{(a)}\big)=\frac{\pi}{r}\,\Big(2a-2-\frac{r-1}{2}\Big)\ .
\ee
In addition, the  integers ${\cal I}_m^{(a)}$  appearing in \eqref{oi89uy21hdsbn} with 
$\zeta_j=\zeta_m^{(a)}$ turn out to be independent of $a$ and 
take the
values
\be\label{oias92jdhhsas}
{\cal I}_m^{(a)}= -\bigg[\frac{(r-1)\,S^z}{2r}+\frac{1}{2}\bigg]+m\,,\qquad\qquad
\qquad m=1,2,\ldots,\frac{N-2S^z}{2r}\qquad\qquad ({\tt s}=0)\,.
\ee
Recall that the brackets $[\ldots]$ stands for the integer part.
It should be emphasized that in writing \eqref{oias92jdhhsas} 
we assume a certain choice for the fundamental domain of the twist parameter  ${\tt k}$ \eqref{BC1b}.
Namely,
if $(r-1)S^z$ is even then $-\frac{1}{2}<{\tt k}\le \frac{1}{2}$, while for
$(r-1)S^z$ odd one should take $0<{\tt k}\le 1$.
\medskip

The description  of  the integers ${\cal I}_m^{(a)}$ is more cumbersome 
when ${\tt s}=1,2,\ldots,r-1$. In this case 
 ${\tt s}$ of the groups 
$\{\zeta_m^{(a)}\}$ have an excess of one additional root compared to the remaining ones  and 
 the vacuum is $\frac{r!}{s!(r-s)!}$ times degenerate.  
Assuming that, say, the
groups with $a=1,2,\ldots,r- {\tt s}$ have one less root, the corresponding   ${\cal I}_m^{(a)}$   are given by 
\be\label{ask9812hj32}
 {\cal I}_m^{(a)}=
\begin{cases}
-\big[\frac{(r-1)\,S^z-{\tt s}}{2r}+\frac{1}{2}\big]+m 
&  {\rm for} \ \ \ \ m=1,2,\ldots,\frac{N-2\,(S^z+{\tt s})}{2r}\,,\ \  \ \ \ \ \ \ \  a=1,\ldots,r-{\tt s}\\[0.3cm]
-\big[\frac{(r-1)\,S^z-{\tt s}}{2r}+\frac{1}{2}\big]
+\sigma_{\tt k}-1+m& {\rm for}\ \ \ \ m=1,2,\ldots,\frac{N-2\,(S^z+{\tt s})}{2r}+1\,,\ \   \ \ a=r-{\tt s}+1,\ldots,r
\end{cases}
\ee
Here  we use the notation $\sigma_{\tt k}\in\{0,1\}$, which 
depends on the twist parameter ${\tt k}$. 
For the definition of this integer, one needs to distinguish between:
\bea
&&{\rm case}\ {\rm (i)}\ :\ \ \ \ \ (r-1)S^z- {\tt s}\ \ {\rm even} \nonumber  \\[-0.25cm]
&&\qquad\qquad\qquad\qquad\qquad\qquad\qquad\quad\, .\\[-0.25cm]
&&{\rm case}\ {\rm (ii)}\ :\ \ \ \  \, (r-1)S^z-{\tt s}\ \ {\rm odd}\nonumber
\eea
Again, choosing the fundamental domain of ${\tt k}$ differently for each of the two cases, we define $\sigma_{\tt k}$ as
\bea\label{kjas892jsdbn}
&&\sigma_{\tt k}=
\begin{cases}
[1-{\tt k}]\ \ &  {\rm with}\ \ -\tfrac{1}{2}<{\tt k}\le \tfrac{1}{2} \ \ \ {\rm for \ case\ (i)} \\[0.2cm]
[\tfrac{3}{2}-{\tt k}]\ \ &  {\rm with}\ \ \ \ \ \ 0<{\tt k}\le 1 \ \ \ {\rm for \ case\ (ii)}
\end{cases}\,.
\eea
\medskip

Denote  the scaling limit of the roots $\zeta_j\equiv\zeta_j^{(+)}$  \eqref{mas9823cas}
as
\be\label{mas9823casB}
\mu_j=\slim_{N\to\infty\atop
j-{\rm fixed}}\,
\bigg(\frac{2N}{r\pi}\bigg)^{\frac{1}{r}}\,\re^{-\frac{\ri\pi(r+1)}{2r}}\ \zeta_j\, .
\ee
It follows from formulae  \eqref{oi89uy21hdsbn},\,\eqref{kdnxcdsjhewio},\,\eqref{ask9812hj32} that 
the set $\{\mu_j\}$ splits into $r$ groups and
\be\label{saj893hjbs}
\mu_m^{(a)}=
-\re^{\frac{2\pi\ri}{r}(a-1)}\,
\big( 2\lambda_a+2{\tt k}+2m_a-1\big)^{\frac{1}{r}}\,.
\ee
For the vacuum state the  integers 
 $m_a$ do not depend on $a$ and $m_a=m^{({\rm vac})}$ with
\be\label{kas8923jhd}
\{m^{({\rm vac})}\}=\{1,2,3,\ldots\}\ .
\ee
The constants $\lambda_a$   are  given by
\be\label{0932jbnfdaA}
\lambda_a=\begin{cases}
\frac{S^z}{2}-\big[\frac{(r-1)\,S^z-{\tt s}}{2r}+\frac{1}{2}\big] & \qquad {\rm for}\ \ \ 
a=1,\ldots, r-{\tt s} \\[0.2cm]
\frac{S^z}{2}-\big[\frac{(r-1)\,S^z-{\tt s}}{2r}+\frac{1}{2}\big]
+\sigma_{\tt k}-1& \qquad {\rm for}\ \ \ a=r-{\tt s}+1,\ldots,r
\end{cases}\,,
\ee
where $\sigma_{\tt k}\in\{0,1\}$ is defined as in \eqref{kjas892jsdbn}.
Note that $ 2\lambda_a+2{\tt k}+1$ may be  a negative number and for taking the 
 $r^{\rm th}$ root appearing in \eqref{saj893hjbs} we will adopt the following convention
\be\label{askjd982yh2g3gsd}
x^{\frac{1}{r}}=|x|^{\frac{1}{r}}\,\times\begin{cases}
+1 & {\rm for}\ \ \ x>0\\
-1 & {\rm for}\ \ \  x<0
\end{cases}\qquad\qquad (r\ {\rm odd})\,,
\ee
where $|x|^{\frac{1}{r}}>0$, i.e., it  is understood as the arithmetic  root.
\medskip

For a low energy excited state,
the scaled Bethe roots $\mu_m^{(a)}$ are still described by the formula \eqref{saj893hjbs}. However, the 
set $\{m_a\}$ differs from the vacuum one by deleting a certain number of entries
(creating ``holes'') and adding some number of
distinct, non-positive entries (creating ``particles''). For given $a$, the exact positions of the holes  in the vacuum distribution 
$m^{({\rm vac})}$ \eqref{kas8923jhd} will be denoted as ${n}_{j,a}^-$, while for the  position of the particles
we use $1-n_{j,a}^{+}$ so that:
\be
\{m_a\}=\{\ldots,1-n_{2,a}^+,1-n_{1,a}^+,1,2,\ldots, \slashed{n}_{1,a}^-,\ldots, \slashed{n}_{2,a}^-,\ldots \}\ .
\ee
Thus, in the sector with given value of $S^z\ge 0$,
the scaled Bethe roots    can be uniquely specified
 by the integer $a=1,2,\ldots, r$ as well as $2r$ ordered sets of  positive integers:
  \bea\label{ka8923jhbdAA}
  1\leq n_{1,a}^{\pm}<  n_{2,a}^{\pm}<\ldots<   n_{M^{\pm}_a,a}^{\pm}\ .
  \eea
Here  $M^{-}_a$ $\big(M^{+}_a\big)$ are the  number of holes (particles)
for the $a^{\rm th}$ group.
\medskip

To take the scaling limit of the $Q$-function we use the l.h.s. of  formula \eqref{jkas8912jhas}  with
$G^{(N/r)}$, therein,  being  set to
\be
G^{(N/r)}\big(- r\mu^r\,\big|\,\tfrac{1}{2}\big)=\bigg(\frac{N}{r \pi}\bigg)^{-\frac{1}{2}\, \mu^r}\ .
\ee
Performing the computation, one arrives at
\bea\label{kjas982jhbsd}
D_+(\mu)&=&\frac{\Gamma(\lambda+\frac{1}{2}+{\tt k})}{\Gamma(\lambda+\frac{1}{2}+{\tt k}+\frac{\mu^r}{2})}\,
\prod_{a=r-{\tt s}+1}^{r}\prod_{m=1}^{1-\sigma_{\tt k}}\frac{%
(2\lambda-1+2{\tt k})^{\frac{1}{r}}+\re^{\frac{2\pi\ri }{r}(1-a)}\,\mu}{%
(2\lambda-1+2{\tt k})^{\frac{1}{r}}}\\[0.2cm]
&\times & \prod_{a=1}^{r}\left[ \frac{ \prod_{j=1}^{M^-_a}\ 
\big( 2\lambda_a+2{\tt k}+2n_{j,a}^{-}-1\big)^{\frac{1}{r}}}
 {\prod_{j=1}^{M^+_a}\ 
\big(2\lambda_a+2{\tt k}+1-2n^+_{j,a}\big)^{\frac{1}{r}}} \ \ 
 \frac{ \prod_{j=1}^{M^+_a}\ 
 \big((2\lambda_a+2{\tt k}+1-2n^+_{j,a})^{\frac{1}{r}}+\re^{\frac{2\pi\ri}{r}(1-a)}\,\mu\big)}
 {\prod_{j=1}^{M^{-}_a}\ 
\big( (2\lambda_a+2 {\tt k}+2n_{j,a}^{-}-1)^{\frac{1}{r}}+
\re^{\frac{2\pi\ri }{r}(1-a)}\,\mu\big)} 
\right]\,,\nonumber
\eea
which, together with the constants $\lambda_a$ \eqref{0932jbnfdaA}, 
involves $\lambda$ that  coincides with $\lambda_a$ for
$a=1,2,\ldots,r-{\tt s}$, i.e.,  
\be
\lambda=\frac{S^z}{2}-\bigg[\frac{(r-1)\,S^z-{\tt s}}{2r}+\frac{1}{2}\bigg]=
\frac{S^z+{\tt s}}{2r}-
\begin{cases}0&  {\rm for\ }  (r-1)S^z- {\tt s}\ \ {\rm even} \\[0.2cm]
\frac{1}{2}&   {\rm for\ }  (r-1)S^z- {\tt s}\ \ {\rm odd} 
\end{cases}\,.
\ee

\medskip

A similar analysis can be carried out for the scaling limit of the Bethe roots accumulating at $\zeta=\infty$.
In this case  the function $\bar{D}_+(\bar{\mu})$ defined through the relation \eqref{jkas8912jhasB}
 would depend on the 
  twist parameter ${\tt k}$, the value of the $z$-projection of the total spin operator
$S^z$  as well as 
the  sets $\{ \bar{n}_{j,a}^{\pm}\}$ such that
\bea\label{ka8923jhbdA}
  1\leq \bar{n}_{1,a}^{\pm}<  \bar{n}_{2,a}^{\pm}<\ldots<   \bar{n}_{\bar{M}^{\pm}_a,a}^{\pm}\ .
  \eea 
The  positive integers $ \bar{n}_{j,a}^{\pm}$ are independent of  $n_{j,a}^{\pm}$ apart from the 
constraint
\be\label{ka8923jhbdB}
\sum_{a=1}^r\big(M^-_a+\bar{M}^-_a\big)= \sum_{a=1}^r\big(M^+_a+\bar{M}^+_a\big)\ .
\ee
The latter expresses that in a given sector with fixed $S^z$, for the low energy states, the  total number of
``left'' and ``right'' holes must be equal to the total number of particles.
For the case $r=1$, i.e., when the lattice model is the usual Heisenberg spin-$\frac{1}{2}$   chain,
the integer $M_1^{-}-M_1^{+}=\bar{M}_1^{+}-\bar{M}_1^{-}$ 
can be identified with the so-called winding number. For $r\ge 1$ we define the winding number ${\tt w}\in\mathbb{Z}$ 
through the relation
\be\label{kasj823jhnbc}
\sum_{a=1}^r \big(M_a^{-}-M_a^{+}\big)=
\sum_{a=1}^r\big(\bar{M}_a^{+}-\bar{M}_a^{-}\big) ={\tt m}+{\tt w} r\,,
\ee
where ${\tt m}=0,1,2,\ldots,r-1$.
\medskip

Thus the CFT states which appear in the scaling limit of the low energy Bethe states of the
lattice model at the free fermion point $\delta=0$ are classified by the (half-)integer  $S^z$  as well as  
the sets $\{{n}_{j,a}^{\pm}\}$  and $\{\bar{n}_{j,a}^{\pm}\}$. It should be pointed out
 that ${\tt s}\in\{0,1,2,\ldots,r-1\}$ \eqref{AAAask8923nbn43}
is not an independent quantum number.
It is specified by $S^z$  as
\be\label{oias9023jknbfdbnjs}
{\tt s}=\begin{cases} -S^z\ \ \ \ \ ({\rm mod}\, r) &\ \ {\rm for}\ \ S^z=0,1,2,\ldots\\[0.2cm]
\frac{r}{2} - S^z\ \ ({\rm mod}\, r) &\ \ {\rm for}\ \ S^z=\frac{1}{2},\frac{3}{2},\frac{5}{2},\ldots
\end{cases}\, .
\ee
\medskip

 For what follows, we will require the expressions for the scaling limit of the
energies of the Bethe states as well as the eigenvalues of the $r$-site lattice translation operator $\mathbb{K}$.
The latter  is a member of the commuting family  and for
its definition and properties see, e.g.,
sec.\,6.1 in ref.\cite{Bazhanov:2020new}.
The eigenvalues of $\mathbb{K}$ are expressed in terms of the Bethe roots as
\be\label{asj21gh}
{\cal K}=\re^{\ri r\pi{\tt k}}\,q^{-r(\frac{N}{2}-S^z)}
\prod_{j=1}^{\frac{N}{2}-S^z}
\frac{\zeta_j^r+q^{+r}}{\zeta_j^r+q^{-r}}\ \ .
\ee
With the above formula at hand, as well as the one for ${\cal E}$  \eqref{kjas8923hjnbxc}, a straightforward computation
shows that for the low energy states, 
\bea\label{kjas8923hjds}
{\cal E}&\asymp& e_{\infty} N+\frac{2\pi r v_{\rm F}}{N}\,\big(I_1+\bar{I}_1\big)+o(N^{-1})\nonumber \\[0.2cm]
{\cal K}&=&
\,(-1)^{n_1}\,\re^{\frac{\ri\pi}{2}n_2}
\,\exp\bigg(\frac{2\pi\ri r }{N}\ \big(\,I_1-\bar{I}_1\,\big)\bigg)\,,
\eea
in agreement with the expectations of scaling in a 1D critical spin chain \cite{Cardy:1986ie}. 
The leading terms contain the specific bulk energy,  $e_{\infty}= -\frac{2r}{\pi}$, the sign factor 
$(-1)^{n_1}=(-1)^{\frac{r-1}{2}(\frac{N}{2}-S^z)+{\tt m}+{\tt w} r}$ and $n_2$, which is an integer defined
modulo four. The latter is given by
\be\label{hsayat}
 n_2=
\begin{cases}
{\tt s}-2\,[1-{\tt k}]\,{\tt s}&({\rm mod}\ 4)\ \ \ \ \ \  {\rm for\ } \  (r-1)S^z- {\tt s}\ \ \  {\rm even} \ \& \ \  {\tt k}\in (-\frac{1}{2},\frac{1}{2}]\\[0.2cm]
{\tt s}-2\,[\frac{3}{2}-{\tt k}]\,{\tt s}-r&({\rm mod}\ 4) \ \ \ \ \ \   {\rm for\ } \  (r-1)S^z- {\tt s}\ \ \ {\rm odd} \ \& \ \ {\tt k}\in(0,1]
\end{cases} \ .
\ee
Notice that, contrary to the specific bulk energy, $(-1)^{n_1}$ and $n_2$  
depend on the state through the quantum numbers
 $S^z$  and
${\tt m}+{\tt w}r$ (${\tt s}$ is specified by $S^z$ according to formula \eqref{oias9023jknbfdbnjs}).
The $1/N$-corrections for the energy contains the state independent Fermi velocity,  $v_{\rm F}=2r$.
\medskip

The coefficients   $I_1$ and $\bar{I}_1$, whose sum and difference are interpreted as the  CFT energy and momentum,
respectively, have the form
\begin{subequations}
\label{kaso23jhnn}
\bea
I_1&=&P^2_{{\tt w},{\tt m}}+\frac{\mathfrak{j}_{\tt m}\,(r-\mathfrak{j}_{\tt m})}{2r} -\frac{r}{24} +
{\tt L}\label{kaso23jhnb}\\[0.2cm]
\bar{I}_1&=&\bar{P}^2_{{\tt w},{\tt m}}+
\frac{\bar{\mathfrak{j}}_{\tt m}\,(r-\bar{\mathfrak{j}}_{\tt m})}{2r}-\frac{r}{24}+\bar{{\tt L}}\, .
\eea
\end{subequations}
Here 
$\mathfrak{j}_{\tt m}$, $\bar{\mathfrak{j}}_{\tt m}$ are integers which take values from the set $\{0,1,\ldots,r-1\}$.
They depend on ${\tt m}$, as is  indicated in their notation, as well as, of course, $S^z$ and are given by
\begin{subequations}
\label{aosiddd9012j23jhjd}
\bea\label{aosiddd9012j23jhjds}
\mathfrak{j}_{\tt m}&=&{\tt m}-(1-\sigma_{\tt k})\,{\tt s} \qquad ({\rm mod}\ r)\\[0.2cm]\label{aosiddd9012j23jhjdBBBs}
\bar{\mathfrak{j}}_{\tt m}&=&{\tt m}+\sigma_{\tt k}\,{\tt s}\qquad\qquad\ \,  ({\rm mod}\ r)\, .
\eea
\end{subequations}
Note that  we use  $\sigma_{\tt k}\in\{0,1\}$ from  eq.\,\eqref{kjas892jsdbn} to make the expression as simple as possible.
Also, $P_{{\tt w},{\tt m}}$ and
 $\bar{P}_{{\tt w},{\tt m}}$  are determined through the relations:
\bea\label{kaso23jhnbAAA}
&&\sqrt{2r}\,\big(P_{{\tt w},{\tt m}}+\bar{P}_{{\tt w},{\tt m}}\big)=S^z   \\[0.2cm]
&&\frac{P_{{\tt w},{\tt m}}-\bar{P}_{{\tt w},{\tt m}}}{\sqrt{2r}}=
{\tt k}+{\tt w}+\frac{{\tt m}}{r} +\frac{(2\sigma_{\tt k}-1)}{2r}\,{\tt s}
-\begin{cases}0&  {\rm for\ }  (r-1)S^z- {\tt s}\ \ {\rm even} \\[0.2cm]
\frac{1}{2}&   {\rm for\ }  (r-1)S^z- {\tt s}\ \ {\rm odd} 
\end{cases}\ .\nonumber
\eea
Finally, in formula \eqref{kaso23jhnn} ${\tt L}$ and $\bar{\tt L}$ are non-negative integers, which are
interpreted  as the chiral levels of the states in the conformal towers.
They  are expressed in terms of
$n^\pm_{j,a}$ and $\bar{n}^\pm_{j,a}$  as
\begin{subequations}\label{aksj902j212309}
\bea\label{aksj902j212309A}
{\tt L}&=&\sum_{a=1}^{r}\bigg(\sum_{j=1}^{M_a^-}\big(n_{j,a}^- - \tfrac{1}{2})+
\sum_{j=1}^{M_a^+}\big(n_{j,a}^+ - \tfrac{1}{2})\bigg)+
(1-\sigma_{\tt k})\sum_{a=r-{\tt s}+1}^{r}\big(M_a^+-M_a^-\big)\nonumber \\[0.2cm]
&-&
\frac{1}{2}\,
\Big(r{\tt w}^2+2{\tt m}{\tt w} -(1-\sigma_{\tt k})\,(2{\tt w}+1)\,{\tt s}+|{\tt m}-(1-\sigma_{\tt k})\,{\tt s}|\Big)
\\[0.2cm]\label{aksj902j212309B}
\bar{{\tt L}}&=&\sum_{a=1}^{r}\bigg(\sum_{j=1}^{\bar{M}_a^-}\big(\bar{n}_{j,a}^- - \tfrac{1}{2})+
\sum_{j=1}^{\bar{M}_a^+}\big(\bar{n}_{j,a}^+ - \tfrac{1}{2})\bigg)-
\sigma_{\tt k}\sum_{a=1}^{r-{\tt s}}\big(\bar{M}_a^+-\bar{M}_a^-\big)\nonumber \\[0.2cm]
&-&
\frac{1}{2}\,
\Big(r{\tt w}^2+2{\tt m}{\tt w}-\sigma_{\tt k}\,(2{\tt w}+1)\,(r-{\tt s})+\big|{\tt m}-\sigma_{\tt k}\,(r-{\tt s})\big|\Big)\,.
\eea
\end{subequations}
\medskip

The above formulae fully describe the 
energy-momentum  spectrum of the conformal field theory
underlying the critical behavior. A practical question concerns 
the degeneracy of the states with given values of the energy, $I_1+\bar{I}_1$, and
momentum, $I_1-\bar{I}_1$.
Since the sets
 $\{n^\pm_{j,a}\}$ \eqref{ka8923jhbdAA}   and $\{\bar{n}^\pm_{j,a}\}$ \eqref{ka8923jhbdA}
labelling the CFT Bethe states are
mostly independent, except for the constraint \eqref{kasj823jhnbc},  one may focus on the former 
set of integers. Thus we arrive at the following combinatorial problem:
for given ${\tt k}$, $S^z$, ${\tt w}$ and ${\tt m}$, how many different
sets $\{n^\pm_{j,a}\}$ such that 
\be
\sum_{a=1}^r \big(M_a^{-}-M_a^{+}\big)={\tt m}+{\tt w}r\,,
\ee
give rise to the same value of $I_1$  \eqref{kaso23jhnb} or, equivalently,
the conformal level ${\tt L}$ \eqref{aksj902j212309A}?
It turns out that this number  depends on ${\tt m}$, $S^z$ and ${\tt k}$ only
through the combination $\mathfrak{j}_{\tt m}$ \eqref{aosiddd9012j23jhjds} and we'll denote it as
${\cal D}_{\mathfrak{j}_{\tt m}}{({\tt L})}$. The generating function  for 
the integers ${\cal D}_{\mathfrak{j}}{({\tt L})}$  
can be recovered from 
the special case ${\tt s}=S^z=0$ and ${\tt k}\to0$. Then the problem 
reduces to
the elementary computation of the character of the fermionic Fock space and 
results in the  relation
\be\label{asjh32h98sd}
\prod_{n=1}^\infty\Big[(1+z {\tt q}^{n-\frac{1}{2}})\,(1+z^{-1}\,{\tt q}^{n-\frac{1}{2}})\Big]^r=
\sum_{{\tt L}=0}^\infty \sum_{{\tt w}\in\mathbb{Z}}\sum_{{\mathfrak j}=0}^{r-1}z^{{\mathfrak j}+{\tt w} r}\ {\tt q}^{{\tt L}+
\frac{{\tt w}}{2}(2\mathfrak{j}+{\tt w} r)+\frac{\mathfrak{j}}{2}}\ 
{\cal D}_{\mathfrak{j}}{({\tt L})}\,.
\ee
Analogously, the number  of sets $\{\bar{n}^\pm_{j,a}\}$, restricted by the condition \eqref{kasj823jhnbc},
which yield the same value of $\bar{I}_1$ coincides with ${{\cal D}}_{\bar{\mathfrak{j}}_{{\tt m}}}{(\bar{{\tt L}})}$.
Hence for fixed twist parameter ${\tt k}$, $z$-projection of the total spin
 $S^z$, winding number ${\tt w}\in\mathbb{Z}$ and  ${\tt m}\in\{0,1,\ldots,r-1\}$
the degeneracy of the energy-momentum
 spectrum  is
given by ${\cal D}_{\mathfrak{j}_{\tt m}}{({\tt L})}\times{{\cal D}}_{\bar{\mathfrak{j}}_{{\tt m}}}{(\bar{{\tt L}})}$.

\subsection{Scaling limit of the Bethe states}
The subject matter of the paper is the scaling limit of the $Q$-functions. 
In the previous subsection this was discussed for the free fermion point.
In fact, for $\delta=0$ it is  possible to go beyond and achieve 
a full description of the space of states arising in the scaling limit of the spin chain
without relying on the Bethe Ansatz. 
This is because of the special form of the Hamiltonian \eqref{kas8932jbncxas}.
Indeed,   define 
the Jordan--Wigner operators
\be\label{kjsa8923jhbs}
\psi_j=\sigma_j^+\,\prod_{i=1}^{j-1}(-\sigma^z_i)\,,\qquad\qquad\psi_j^\dag=\sigma_j^-\,\prod_{i=1}^{j-1}(-\sigma^z_i)\ ,
\ee
which are subject to the  canonical anticommutation relations,
\be
\{\psi_i^\dag,\psi_j\}=\delta_{ij}\,,\qquad\qquad
\{\psi_i,\psi_j\}=\{\psi_i^\dag,\psi_j^\dag\}=0 \,.
\ee
Keeping in mind the quasi-periodic boundary conditions \eqref{BC1b},  the Hamiltonian 
and $z$-projection of the total spin operator are
 expressed in terms of the lattice fermions as
\be\label{akjs892j3hbhas}
\mathbb{H}=(-1)^{\frac{r-1}{2}}r\Bigg[\sum_{j=1}^{N-r}
\big(\psi_j^\dag\psi_{j+r}+\psi_{j+r}^\dag\psi_{j}\big)-
(-1)^{\frac{N}{2}+\mathbb{S}^z}\!\!\!\!\!\sum_{j=N-r+1}^{N}\!\!\!
\Big(\re^{+2\pi\ri {\tt k}}\,
\psi_j^\dag\psi_{j+r-N}+
\re^{-2\pi\ri {\tt k}}\,\psi_{j+r-N}^\dag\psi_{j}\Big)
\Bigg]
\ee
and
\be
\mathbb{S}^z=\sum_{j=1}^{N}\big(\tfrac{1}{2}-\psi_j^\dag\psi_j\big)\ .
\ee
Since these two operators mutually commute, the
  diagonalization problem for $\mathbb{H}$
may be restricted to a sector, where  $\mathbb{S}^z$ takes fixed value.
This way, the Hamiltonian turns out to be a quadratic fermionic form which 
is straightforward to bring to the canonical form.
\medskip

Introduce the notation
\be\label{isa891hj2ds}
\psi_{j,a}=\psi_{a+(j-1)r}\,,\qquad\qquad
\psi_{j,a}^\dag=\psi_{a+(j-1)r}^\dag
\ee
with $a=1,2,\ldots,r$ and $j=1,2,\ldots,\frac{N}{r}$. Then, 
the Hamiltonian  \eqref{akjs892j3hbhas} acting in the sector with given value of 
$\mathbb{S}^z$ can be re-written as
\bea\label{kas932jhasas}
\mathbb{H}\,|_{\mathbb{S}^z=S^z}&=&(-1)^{\frac{r-1}{2}}r\sum_{a=1}^r\Bigg[\sum_{j=1}^{\frac{N}{r}-1}
\big(\psi_{j,a}^\dag\,\psi_{j+1,a}+
\psi_{j+1,a}^\dag\,\psi_{j,a}\big) \\[0.2cm]
&-&
(-1)^{\frac{N}{2}+{S}^z}
\Big(\re^{+2\pi\ri {\tt k}}\,\psi_{\frac{N}{r},a}^\dag\,\psi_{1,a}+
\re^{-2\pi\ri {\tt k}}\,\psi_{1,a}^\dag\,\psi_{\frac{N}{r},a}\Big)\Bigg]\,, \nonumber
\eea
where the fermion operators  carrying different values of the index ``$a$'' are completely decoupled from one another. 
Now we perform a Fourier transform:
\be\label{kasiuh23jbnds}
\psi_{j,a}=\re^{\frac{\ri\pi r j}{2}}\,
\sqrt{\frac{r}{N}}\ \sum_{\ell =1}^{\frac{N}{r}} \mathsf{f}_a(\ell)\,\re^{\ri j p_{\ell}}\,,\qquad\qquad
\psi_{j,a}^\dag=\re^{-\frac{\ri\pi r j}{2}}\,
\sqrt{\frac{r}{N}}\ \sum_{\ell =1}^{\frac{N}{r}} \mathsf{f}_a^\dag(\ell)\,\re^{-\ri j p_{\ell}}
\ee
with
\be
p_\ell=\frac{2\pi r}{N}\Big(\ell+{\tt k}+\frac{S^z-1}{2}\Big) 
\ee
and the Fourier modes  satisfy the  anticommutation relations:
\be\label{oas923iujbnsjd}
\big\{\mathsf{f}_a^\dag(\ell),\mathsf{f}_b(\ell')\big\}=\delta_{a,b}\,\delta_{\ell,\ell'}\ ,\qquad\qquad
\big\{\mathsf{f}_a(\ell),\mathsf{f}_b(\ell')\big\}=\{\mathsf{f}_a^\dag(\ell),\mathsf{f}_b^\dag(\ell')\}=0\, .
\ee
In formulae  \eqref{kasiuh23jbnds} the operators $\psi_{j,a}$ and $\mathsf{f}_a(\ell)$ are understood as intertwiners
acting between the sectors  corresponding to $S^z$ and $S^z+1$, while 
the Hermitian conjugated operators  $\psi_{j,a}^\dag$ and $\mathsf{f}_a^\dag(\ell)$ 
intertwine the sectors with $S^z+1$ and $S^z$. 
Substituting the above expressions for $\psi_{j,a}$ and $\psi_{j,a}^\dag$   into the Hamiltonian
\eqref{kas932jhasas}, one arrives at
\be
\mathbb{H}\,|_{\mathbb{S}^z=S^z}
=\sum_{a=1}^r\sum_{\ell=1}^{\frac{N}{r}}\varepsilon(p_\ell)\, \mathsf{f}_a^\dag(\ell)  \mathsf{f}_a(\ell)\,,
\ee
where 
\be\label{kajs9023jsa}
\varepsilon(p)=-2r\,\cos\big(p-\tfrac{\pi}{2}\big)\, .
\ee
\medskip

The construction of the state with the lowest energy in the given sector $S^z$
follows the usual Dirac procedure. One starts with the pseudovacuum $|\Omega\rangle$, which we take to be the state 
with the maximum possible value of $S^z=\frac{N}{2}$. It  is annihilated by all operators $\psi_j$ or,
equivalently,
\be
|\Omega\rangle\ : \ \ \ \ \ 
\mathsf{f}_a(\ell)\,|\Omega\rangle=0\, .
\ee
The physical vacuum is obtained by acting on $|\Omega\rangle$ with $\mathsf{f}_a^\dag(\ell)$
for which $\varepsilon(p_\ell)$ is negative. In doing so, one should keep in mind that
 we are working in the sector with fixed value of $S^z$ so that the total number of such  operators
is constrained to be $\frac{N}{2}-S^z$.
Thus the vacuum takes the form
\be\label{kjsa8923jjsfd}
|{\rm vac}\rangle=\prod_{a=1}^r\prod_{\ell_a\in\{{\cal I}_m^{(a)}\}}\mathsf{f}_a^\dag({\ell_a})\,|\Omega\rangle\ ,
\ee
where the set $\big\{{\cal I}_m^{(a)}\big\}$ contains $\frac{N}{2}-S^z$ integers.  These integers must be chosen  to minimize the energy
 $\sum\limits_{a=1}^r\sum\limits_{\ell_a} \varepsilon(p_{\ell_a})$. It is easy to see that
the problem of  determining  the  set $\big\{{\cal I}_m^{(a)}\big\}$ is equivalent
to the one considered in the last subsection  when deriving the Bethe root configuration
of the form \eqref{oi89uy21hdsbn} corresponding to the lowest value of the energy \eqref{kjas8923hjnbxc}. Carrying over
the results of our previous analysis, one concludes that the vacuum is $r!/({\tt s}!(r-{\tt s})!)$\,-\,fold degenerate, with
${\tt s}=\frac{N}{2}-S^z$ (mod $r$).
For one representative of these vacua, the corresponding set $\big\{{\cal I}_m^{(a)}\big\}$
 is given by formulae \eqref{ask9812hj32}-\eqref{kjas892jsdbn}.
Note that for the operators $\mathsf{f}_a^\dag({\ell_a})$ in \eqref{kjsa8923jjsfd} with
$\ell_a$  negative we take, by definition,
 $\mathsf{f}_a^\dag(\ell)\equiv \mathsf{f}_a^\dag(\ell+\frac{N}{r})$.
\medskip

For given $a=1,2,\ldots,r$, let us denote by  ${\cal I}_{\rm min}^{(a)}$ and ${\cal I}_{\rm max}^{(a)}$ 
the smallest and greatest integers, respectively, belonging to the set $\{{\cal I}_m^{(a)}\}$
\eqref{ask9812hj32}-\eqref{kjas892jsdbn}. 
The oscillator modes 
\be\label{aks092jk3hjsd}\arraycolsep=0.7cm
\begin{array}{ll}
\mathsf{c}_a(\nu)=\mathsf{f}_a\big({\cal I}_{\rm min}^{(a)}-\tfrac{1}{2}-\nu\big)\,, &
\bar{\mathsf{c}}_a(\nu) =\mathsf{f}_a\big({\cal I}_{\rm max}^{(a)}+\tfrac{1}{2}+\nu\big) \\[0.4cm]
\mathsf{d}_a(\nu)=\mathsf{f}_a^\dag\big({\cal I}_{\rm min}^{(a)}-\tfrac{1}{2}+\nu\big)\,, &
\bar{\mathsf{d}}_a(\nu) =\mathsf{f}_a^\dag\big({\cal I}_{\rm max}^{(a)}+\tfrac{1}{2}-\nu\big)
\end{array}\qquad\qquad \big(\nu\in\mathbb{Z}+\tfrac{1}{2}\big)\, ,
\ee
labelled by the half-integer $\nu$,
create/annihilate the low-lying particle-hole excitations
close to the Fermi surface. The latter, in the case under consideration,
consists of two points for each fermionic flavor.  The commutation relations of the
modes \eqref{aks092jk3hjsd} follows from \eqref{oas923iujbnsjd}:
\be\label{jkas892jhbnd}
\big\{\mathsf{c}_a(\nu),\mathsf{d}_b(\nu')\big\}=\big\{\bar{\mathsf{c}}_a(\nu),\bar{\mathsf{d}}_b(\nu')\big\}=
\delta_{a,b}\,\delta_{\nu+\nu',0}
\ee
with all other anticommutators vanishing.
Linearizing the dispersion relation \eqref{kajs9023jsa} in the vicinity of the Fermi points
we  take the scaling limit $N\to\infty$ keeping fixed $S^z$  (i.e.,  zero magnetization) 
as well as the macroscopic distance
 $x=\frac{2\pi r}{N} j$. In this limit the lattice operators $\psi_{j,a}$ are 
approximated by complex fermion fields that vary slowly on the lattice scale,
\be\label{askj32jdj}
\Psi_a(x)=\re^{\ri   (\lambda_a+{\tt k}) x}
\sum_{\nu\in\mathbb{Z}+\frac{1}{2}}\mathsf{c}_a(\nu)\,\re^{-\ri \nu x}\,,\qquad\qquad 
\bar{\Psi}_a(x)=
\re^{\ri  (\bar{\lambda}_a+{\tt k}) x}\sum_{\nu\in\mathbb{Z}+\frac{1}{2}}\bar{\mathsf{c}}_a(\nu)\,
\re^{\ri  \nu x}\,.
\ee
More precisely, one has the asymptotic formula
\be\label{kajs9823hjsd}
\psi_{j,a}\asymp
(-1)^{\frac{r+1}{2}j}\ \sqrt{\frac{r}{N}}\ \Big(
\re^{-\frac{\ri\pi j}{2}}\,\Psi_a(x)+\re^{\frac{\ri\pi j}{2}}\,
\bar{\Psi}_a(x)\Big)+o\big(N^{-\frac{1}{2}}\big)\ .
\ee
Above we use the notation $\lambda_a$ from eq.\,\eqref{0932jbnfdaA}
together with
\be
\bar{\lambda}_a=\begin{cases}
\frac{S^z}{2}-\frac{S^z+{\tt s}}{r}-\big[\frac{(r-1)\,S^z-{\tt s}}{2r}+\frac{1}{2}\big]& \qquad {\rm for}\ \ \ 
a=1,\ldots, r-{\tt s} \\[0.2cm]
\frac{S^z}{2}-\frac{S^z+{\tt s}}{r}-\big[\frac{(r-1)\,S^z-{\tt s}}{2r}+\frac{1}{2}\big]+\sigma_{{\tt k}}& 
\qquad {\rm for}\ \ \ a=r-{\tt s}+1,\ldots,r
\end{cases}\,.
\ee
The  relation  for $\psi_{j,a}^\dag$
 is obtained from \eqref{kajs9823hjsd} via the formal Hermitian conjugation with
\be\label{askj32jdjB}
\Psi_a^\dag(x)=\re^{-\ri   (\lambda_a+{\tt k})x}
\sum_{\nu\in\mathbb{Z}+\frac{1}{2}}\mathsf{d}_a(\nu)\,\re^{-\ri \nu x}\,,\qquad\qquad 
\bar{\Psi}_a^\dag(x)=
\re^{-\ri  (\bar{\lambda}_a+{\tt k}) x}\sum_{\nu\in\mathbb{Z}+\frac{1}{2}}\bar{\mathsf{d}}_a(\nu)\,
\re^{\ri  \nu x}\,.
\ee
\medskip

The scaling limit of the Hamiltonian \eqref{kas932jhasas} is described as
\be
\frac{1}{2\pi r v_{\rm F}}\ \slim_{N\to\infty} N\,\Big(\mathbb{H}\,|_{\mathbb{S}^z=S^z}-e_\infty N\Big)={\bf I}_1+\bar{{\bf I}}_1\, .
\ee
Here  $e_\infty=-\frac{2r}{\pi},\, v_{\rm F}=2r $, while
 ${\bf I}_1$ and $\bar{\bf{I}}_1$ are operators built from the continuous fermionic fields:
\bea\label{aksj9823hjja}
{\bf I}_1&=&+\,\frac{1}{2}\sum_{a=1}^r\int_0^{2\pi}\frac{\rd x}{2\pi}\ 
\big(\Psi_a^\dag\,\ri\partial_x\Psi_a+\Psi_a\,\ri\partial_x\Psi_a^\dag\big) \nonumber \\[0.1cm]
\bar{\bf{I}}_1&=&
-\,\frac{1}{2}\sum_{a=1}^r\int_0^{2\pi}\frac{\rd x}{2\pi}\ 
\big(\bar{\Psi}_a^\dag\,\ri\partial_x\bar{\Psi}_a+\bar{\Psi}_a\,\ri\partial_x\bar{\Psi}_a^\dag\big)\ .
\eea
They act in the CFT space of states that arises in the scaling limit of the low-energy sector of the lattice model with a fixed value of the
 $z$-projection of the total spin operator. The latter
quantum number  can be identified  with
the ${\rm U}(1)$ charge
\bea
S^z=- \sum_{a=1}^r\int_0^{2\pi}\frac{\rd x}{2\pi}\ 
\big(\Psi^\dagger_a \Psi_a+{\bar \Psi}^\dagger_a {\bar \Psi}_a \big)\ .
\eea
The CFT vacuum, which will be denoted by the same symbol as its lattice counterpart \eqref{kjsa8923jjsfd},
obeys the conditions
\be
\mathsf{c}_a(\nu)|{\rm vac}\rangle=\mathsf{d}_a(\nu)|{\rm vac}\rangle=
\bar{\mathsf{c}}_a(\nu)|{\rm vac}\rangle=\bar{\mathsf{d}}_a(\nu)|{\rm vac}\rangle=0\qquad\quad {\rm for}
\qquad\quad \nu=\tfrac{1}{2}\,,\,\tfrac{3}{2}\,,\,\tfrac{5}{2}\,,\,\ldots\ .
\ee
Notice that  the above formulae do not contain a dependence on $S^z$, though the latter goes into the 
construction
of the vacuum of the spin chain and affects the position of the  Fermi points.
However, the eigenvalues of ${\bf I}_1$ and $\bar{\bf{I}}_1$ do depend on $S^z$ since
it  enters into the Dirac fields \eqref{askj32jdj},\,\eqref{askj32jdjB} through $\lambda_a$ and ${\bar \lambda}_a$.

\medskip

The low energy Bethe states in the scaling limit become the states of the form
\be\label{k983hjbhjcx}
\prod_{a=1}^r 
\bigg[\prod_{j=1}^{M^-_a}\mathsf{c}_a(\tfrac{1}{2}-n_{j,a}^-)\bigg]
\bigg[\prod_{j=1}^{M^+_a}\mathsf{d}_a(\tfrac{1}{2}-n_{j,a}^+)\bigg]
\bigg[\prod_{j=1}^{\bar{M}^-_a}\bar{\mathsf{c}}_a(\tfrac{1}{2}-\bar{n}_{j,a}^-)\bigg]
\bigg[\prod_{j=1}^{\bar{M}^+_a}\bar{\mathsf{d}}_a(\tfrac{1}{2}-\bar{n}_{j,a}^+)\bigg]
|{\rm vac}\rangle\,. %
\ee
In what follows we will refer to these as the CFT Bethe states.
Recall that  the $2r$ sets of positive integers  $\{{n}_{j,a}^{\pm}\}$ \eqref{ka8923jhbdAA}
and $\{\bar{n}_{j,a}^{\pm}\}$ \eqref{ka8923jhbdA}
are subject to the constraint $\sum_{a=1}^r\big(M^-_a+\bar{M}^-_a\big)= \sum_{a=1}^r\big(M^+_a+\bar{M}^+_a\big)$,
see eq.\,\eqref{ka8923jhbdB}. The CFT Bethe states 
 form a basis for the field theory  space of states  and turn out to be
eigenstates of the operators ${\bf I}_1$ and $\bar{\bf{I}}_1$. 
A straightforward computation shows
that the corresponding eigenvalues
coincide with $I_1$ and
$\bar{I}_1$, respectively, defined in eqs.\,\eqref{kaso23jhnn}-\eqref{aksj902j212309}.
\medskip

The original spin chain Hamiltonian \eqref{kas8932jbncxas}  belongs to a family of commuting operators generated by the 
 $Q$-operators.
Similarly, ${\bf I}_1$ and $\bar{{\bf I}}_1$ from \eqref{aksj9823hjja} 
are members of an infinite set of operators
 $\{{\bf I}_m,\bar{{\bf I}}_m\}_{m=1}^\infty$,  all 
of which mutually commute:
\be
[{\bf I}_m,{\bf I}_{m'}]=[\bar{{\bf I}}_m,\bar{{\bf I}}_{m'}]=[{\bf I}_m,\bar{{\bf I}}_{m'}]=0\, .
\ee
They are given by
\begin{subequations}
\bea\label{kjas9023hjjhd}
{\bf I}_m&=&\frac{1}{2}\sum_{a=1}^r\int_0^{2\pi}\frac{\rd x}{2\pi}\ 
\Big(\Psi_a^\dag\,(\ri\partial_x)^m\,\Psi_a-\Psi_a\,(-\ri\partial_x)^m\,\Psi_a^\dag\Big) \\[0.2cm]
\label{kjas9023hjjhdA}
\bar{{\bf I}}_m&=&\frac{1}{2}\sum_{a=1}^r\int_0^{2\pi}\frac{\rd x}{2\pi}\ 
\Big(\bar{\Psi}_a^\dag\,(-\ri\partial_x)^m\,\bar{\Psi}_a-\bar{\Psi}_a\,(\ri\partial_x)^m\,\bar{\Psi}_a^\dag\Big)\ .
\eea
\end{subequations}
A common eigenbasis for 
all these operators is provided by the CFT Bethe states \eqref{k983hjbhjcx}.
Combining the definition \eqref{kjas9023hjjhd}, together with the mode expansions
of the continuous fermionic fields \eqref{askj32jdj},\,\eqref{askj32jdjB}, one can compute the eigenvalue  of ${\bf I}_m$
for the CFT Bethe states.
It turns out that the same expression appears as a certain coefficient in the  large-$\mu$  expansion of the 
scaling limit of the $Q$-functions.
Indeed, as  $\mu\to+\infty$  the asymptotic expansion of  $D_+(\mu)$  \eqref{kjas982jhbsd}  reads as
\be\label{kjnbbnxcjsdas}
\log D_+\asymp -\frac{\mu^r}{2}\,\log\Big(\frac{\mu^r}{2{\rm e}}\Big)-\sqrt{2r}\, P\log(\mu)+\log \mathfrak{C}_++
\sum_{j=1}^\infty \frac{(-1)^{j}}{j}\, 2^{\frac{j}{r}} 
\  I_{\frac{j}{r}}\ \mu^{-j}+O(\mu^{-\infty})\ ,
\ee
where
\be
\sqrt{2r}\, P=r\,(\lambda+{\tt k})+  {\mathfrak m} +{\tt w} r-(1-\sigma_{\tt k})\ {\tt s}\ ,
\ee
while
\bea\label{9032jksdbnbsaaaaa}
I_{\frac{j}{r}}&=&r\, \delta_{j,0\,({\rm mod}\,r)}\ \tfrac{1}{1+\frac{j}{r}}\,
\ B_{\frac{j}{r}+1}(\lambda+\tfrac{1}{2}+{\tt k})-(1-\sigma_{\tt k})\,
\sum_{a=r-{\tt s}+1}^{r}\,\re^{\frac{2\pi\ri}{r} (a-1)j}\,\big(\lambda+{\tt k}-\tfrac{1}{2}\big)^{\frac{j}{r}}
\\[0.1cm]
&+&\sum_{a=1}^r
\Bigg[\sum_{i=1}^{M_a^-}\re^{\frac{2\pi\ri}{r} (a-1)j}\,
\big(n_{i,a}^{-}-\tfrac{1}{2}+\lambda_a+{\tt k}\big)^{\frac{j}{r}}-
\sum_{i=1}^{M_a^+}\re^{\frac{2\pi\ri}{r} (a-1)j}\,\big(
\lambda_a+{\tt k}+\tfrac{1}{2}-n^+_{i,a}\big)^{\frac{j}{r}}\Bigg]\nonumber
\eea
and the conventional notation $B_{m+1}(z)$ for the Bernoulli polynomial is being used.
If the integer $j$  is divisible by $r$, i.e., $j=mr$ with $m=1,2,3,\ldots\ $
then  the Kronecker symbol $\delta_{j,0\,({\rm mod}\,r)}$ is non-zero and
a straightforward computation shows that the 
  coefficient
$I_{m}$ in the asymptotic expansion  coincides with the eigenvalue of the operator \eqref{kjas9023hjjhd}.
For  $\frac{j}{r}$  a fractional number,  one may expect that $I_{\frac{j}{r}}$ can also be interpreted as
 eigenvalues of certain
operators belonging to the commuting family. Formally, they would correspond to the case when 
a fractional power of the derivative occurs in
the r.h.s. of \eqref{kjas9023hjjhd}. Such operators
would be  non-local in the sense that they can not be expressed as  integrals over  local densities
built from $\Psi_a$, $\Psi^\dag_a$ and their derivatives. Notice that the formula \eqref{9032jksdbnbsaaaaa} for $I_{\frac{j}{r}}$
involves  combinations like $\lambda_a+{\tt k}+\tfrac{1}{2}-n^+_{i,a}$, which may be negative,  raised to a fractional power.
In this case we use the convention \eqref{askjd982yh2g3gsd}.
Together with $I_{\frac{j}{r}}$ the asymptotic expansion \eqref{kjnbbnxcjsdas} also involves the constant term
$\log(\mathfrak{C}_+)$ and, explicitly,
\be
\label{kjas982jhbsdaaaaa}
\mathfrak{C}_+=\frac{\re^{\frac{\ri\pi}{r} (1-\sigma_{\tt k}){\tt s}({\tt s}+1)}\,2^{\lambda+{\tt k}}\,
\Gamma(\lambda+\frac{1}{2}+{\tt k})}{%
\sqrt{2\pi}\,
(2\lambda+2{\tt k}-1)^{\frac{{\tt s}}{r}(1-\sigma_{\tt k})}}\,
\ 
\prod_{a=1}^{r}\left[ \frac{ \prod_{j=1}^{M^-_a}\ 
\big( 2\lambda_a+2{\tt k}+ 2n_{j,a}^{-}-1\big)^{\frac{1}{r}}\re^{\frac{2\pi\ri}{r}(a-1)}}
 {\prod_{j=1}^{M^+_a}\ 
 \big(2\lambda_a+2{\tt k}+1-2n^+_{j,a}\big)^{\frac{1}{r}}\re^{\frac{2\pi\ri}{r}(a-1)}}
\right]\,.
\ee
This is  expected to coincide with the
eigenvalue of a certain non-local integral of motion.
\medskip

All the above can be repeated for the operators $\{\bar{{\bf I}}_m\}$ \eqref{kjas9023hjjhdA}. Their eigenvalues
would appear in the large-$\bar{\mu}$ asymptotic expansion of  the scaling function $\bar{D}_+(\bar{\mu})$ defined by 
\eqref{jkas8912jhasB} with $\delta=0$.

\section{CFT space of states for $|\delta|<\frac{1}{r}$\label{sec4}}
As we saw, the classification of the  states occurring in the scaling limit 
of the ${\cal Z}_r$ invariant inhomogeneous XXZ spin-$\frac{1}{2}$ chain
 is relatively easy to achieve at the free fermion point.
Drawing on the  experience with the Heisenberg XXZ  spin-$\frac{1}{2}$  chain, one may expect that the same
description remains valid in the vicinity of $\delta=0$. By this we mean that the field theory space of states possesses a special
basis of the CFT Bethe states, which are labelled by the quantum number $S^z$ along with the sets of integers 
$\{{n}_{j,a}^{\pm}\}$ \eqref{ka8923jhbdAA}
and $\{\bar{n}_{j,a}^{\pm}\}$ \eqref{ka8923jhbdA} subject to the constraint \eqref{ka8923jhbdB}. While a
 rigorous proof of the statement is beyond our reach, we have accumulated a large amount of supporting numerical evidence.
It is based on the investigation of RG trajectories of the low energy Bethe states on the finite lattice
along the lines of ref.\cite{Bazhanov:2019xvy}.
\medskip

For  $\delta=0$ numerous RG trajectories were constructed in a variety of sectors with $S^z\ge0$ and
corresponding to different  choices of the sets $\{{n}_{j,a}^{\pm}\}$, $\{\bar{n}_{j,a}^{\pm}\}$. It was checked
that these trajectories could be  unambiguously deformed away from the free fermion point provided $|\delta|$ is sufficiently small.
 In addition, we found that
the  large-$N$ behavior of the eigenvalues of the 
 Hamiltonian $\mathbb{H}$ and $r$-site lattice translation operator $\mathbb{K}$  for all the trajectories 
are  described by  relations of the form \eqref{kjas8923hjds},\,\eqref{hsayat}.
However, for  general $|\delta|<\frac{1}{r}$ the specific bulk energy and Fermi velocity read as
 \bea\label{uassaysa}
e_{\infty}= -\frac{2v_{{\rm F}}}{\pi}\,\int_0^\infty{\rm d}t\ \frac{\sinh\big(\frac{1- r\delta}{1+ r\delta}\,t\big)}
{\sinh\big(\frac{2t}{1+r\delta}\big)\,\cosh(t)}\,,\qquad \qquad\qquad
v_{{\rm F}}=\frac{2r}{1+r\delta}\ .
\eea
The  coefficients $I_1$ and $\bar{I}_1$ are still given by formula \eqref{kaso23jhnn}. Furthermore, the values of 
the integers $\mathfrak{j}_{\tt m}$ and $\bar{\mathfrak{j}}_{\tt m}$, therein, are the same as in
 eq.\,\eqref{aosiddd9012j23jhjd}, i.e.,
\be\label{ajks8723hj}
\mathfrak{j}_{\tt m}=\begin{cases}
{\tt m}-(1-[1-{\tt k}])\,{\tt s}\ \, \, ({\rm mod}\ r)\ \  & {\rm case\ (i)}\\[0.2cm]
{\tt m}-(1-[\frac{3}{2}-{\tt k}])\,{\tt s}\ \, \, ({\rm mod}\ r)\ \  & {\rm case\ (ii)}
\end{cases}\,,\qquad
\bar{\mathfrak{j}}_{\tt m}=\begin{cases}
{\tt m}+[1-{\tt k}]\,{\tt s}\ \, \, ({\rm mod}\ r) \ \ & {\rm case\ (i)}\\[0.2cm]
{\tt m}+[\frac{3}{2}-{\tt k}]\,{\tt s}\
\, ({\rm mod}\ r)\  \ & {\rm case\ (ii)}
\end{cases}\, ,
\ee
where we take into account the definition \eqref{kjas892jsdbn} of $\sigma_{\tt k}$.
Recall that the two cases stand for
\bea
&&{\rm case}\ {\rm (i)}\ \ :\ \ \ \ \ (r-1)S^z- {\tt s}\ \ {\rm even}\,,\ \ \, {\rm and\ we\ take}\ \  
 {\tt k}\in\big(-\tfrac{1}{2},\tfrac{1}{2}\big] \nonumber  \\[-0.25cm]
&&\qquad\qquad\qquad\qquad\qquad\qquad\qquad\qquad\qquad\qquad\qquad\qquad\qquad\ \ \ \ \ \  .\\[-0.25cm]
&&{\rm case}\ {\rm (ii)}\ :\ \ \ \  \  (r-1)S^z-{\tt s}\ \ {\rm odd}\,,\ \ \, \ {\rm and\ we\ take}\ \  
{\tt k}\in(0,1]\nonumber
\eea
The   $\delta$-dependence in the subleading corrections are traced only in a minor modification 
in the expressions \eqref{kaso23jhnbAAA} for $P_{{\tt w},{\tt m}}$  and $\bar{P}_{{\tt w},{\tt m}}$:
\bea\label{ksajijh2389jidsAA}
&&\!\!\!\!\!\!\!\!\!\!   \sqrt{\frac{2r}{1- r\delta}}\,\big(P_{{\tt w},{\tt m}}+\bar{P}_{{\tt w},{\tt m}}\big)=S^z   \\[0.2cm]
&&\!\!\!\!\!\!\!\!\!\!  \sqrt{\frac{1- r\delta}{2r}}\,\big(P_{{\tt w},{\tt m}}-\bar{P}_{{\tt w},{\tt m}}\big)=
{\tt k}+{\tt w}+\frac{{\tt m}}{r} -\frac{{\tt s}}{2r}
+\begin{cases} 
[1-{\tt k}]\,\frac{{\tt s}}{r}&  {\rm case\ (i)} \\[0.2cm]
[\frac{3}{2}-{\tt k}] \,\frac{{\tt s}}{r}-\frac{1}{2}&   {\rm case\ (ii)}
\end{cases}\ .\nonumber
\eea
The non-negative integers ${\tt L}$ and $\bar{\tt L}$  
do not depend on the continuous parameter $\delta$
similar to the integers $\mathfrak{j}_{\tt m}$ and  $\bar{\mathfrak{j}}_{\tt m}$.
They are
 still given in terms of the sets $\{{n}_{j,a}^{\pm}\}$, $\{\bar{n}_{j,a}^{\pm}\}$  as in eq.\,\eqref{aksj902j212309}.
\medskip

It should be pointed out that the explicit expressions for the coefficients
appearing in formula \eqref{kjas8923hjds}
describing the large-$N$ behavior of the eigenvalues of $\mathbb{H}$ and $\mathbb{K}$ 
for the ${\cal Z}_r$ invariant inhomogeneous XXZ spin-$\frac{1}{2}$ chain
in the domain of the anisotropy parameter $\gamma\in\big(\tfrac{\pi}{2}(1-\tfrac{1}{r}),\tfrac{\pi}{2}(1+\tfrac{1}{r})\big)$
were already presented in a different, but equivalent form in sec.\,2.5 of ref.\cite{Gehrmann:2024tue}.
\medskip

The CFT Bethe states form a special basis in the field theory space of states. Their construction relies essentially  on the
integrable structure, which is inherited from that of the lattice system.  At the same time, the CFT space of states
can be described without any reference to integrability,   in the spirit of the seminal works \cite{BPZ, Knizhnik:1984nr,Zamolodchikov:1985wn}.
In this case, the central object of interest is the algebra of (extended) conformal symmetry and its representation theory.
The extended conformal symmetry algebra is generated by a set of chiral and antichiral currents. The latter include
$T$ and $\bar{T}$ --- 
the Lorentz spin $+2$ and $-2$
components of the energy-momentum tensor.
In the usual interpretation, $I_1$ and $\bar{I}_1$ from formula \eqref{kaso23jhnn} are the eigenvalues of the integrals of motion
${\bf I}_1$ and $\bar{\bf I}_1$, which are given by 
\be\label{kja09jh21jhjh}
{\bf I}_1=\int_0^{2\pi}\frac{\rd x}{2\pi} \ T(x)\,,\qquad\qquad
\bar{{\bf I}}_1=\int_0^{2\pi}\frac{\rd x}{2\pi}\  \bar{T}(x)
\ee
so that the CFT Hamiltonian coincides with
\be\label{kjasdasdsa8932b21}
\hat{H}_{\rm CFT}={\bf I}_1+\bar{\bf{I}}_1\, .
\ee
\medskip

In the case of the free fermion point,  in view of eqs.\,\eqref{aksj9823hjja},\,\eqref{askj32jdj} and \eqref{askj32jdjB},
the CFT Hamiltonian  can be
expressed in terms of the oscillator modes  \eqref{jkas892jhbnd} as
\bea
\hat{H}_{\rm CFT}|_{\delta=0}&=&\sum_{a=1}^r\sum_{\nu\in \mathbb{Z}+\frac{1}{2}}
\Big(\big(\nu-\lambda_a-{\tt k}\big)\,:\!\mathsf{d}_a(-\nu)\,\mathsf{c}_a(\nu)\!:+\,\big(\nu+\bar{\lambda}_a+{\tt k}\big)
:\!\bar{\mathsf{d}}_a(-\nu)\,\bar{\mathsf{c}}_a(\nu)\!:\Big)\nonumber \\[0.1cm]
&-&\frac{r}{12}+\frac{1}{2}\,\sum_{a=1}^r\Big(
\big(\lambda_a+{\tt k}\big)^2+\big(\bar{\lambda}_a+{\tt k}\big)^2\Big)\,,
\eea
where the symbol $:\!(\ldots)\!:$ stands for the normal ordering.
It is then straightforward to show that the Heisenberg equations of motion for $\Psi_a$ and $\bar{\Psi}_a$ imply that
these are, respectively, chiral and antichiral fields,  i.e.,
\be
\Psi_a=\Psi_a(u)\,,\qquad\qquad\bar{\Psi}_a=\bar{\Psi}_a(\bar{u})
\ee
with 
\be
u=t+x\,,\qquad\qquad \bar{u}=t-x\  .
\ee
Furthermore, they satisfy the following operator product expansions
\be
\Psi_a^\dag(u)\Psi_b(u')=-\frac{\ri\,\delta_{ab}}{u-u'}+O(1)\,,\qquad\qquad \Psi_a^\dag(u)\Psi_b^\dag(u')=O(1)\,,\qquad\qquad
\Psi_a(u)\Psi_b(u')=O(1)
\ee
and similarly for the antichiral fields. Thus, the algebra of conformal symmetry for the CFT underlying the critical behavior
of the lattice model for $\delta=0$ is a tensor product of two  algebras ---
 a chiral one generated by $\Psi_a$, $\Psi_a^\dag$ and an antichiral one generated by $\bar{\Psi}_a$, $\bar{\Psi}_a^\dag$.
In what follows, we mainly focus on the chiral algebra.
\medskip

Any composite local field built from  $\Psi_a$, $\Psi_a^\dag$ and their derivatives belongs to the algebra
of extended conformal symmetry. Among them are the quadratic currents \cite{Witten:1983ar}:
\begin{subequations}
\bea
{\cal J}^A&=&\bm{\Psi}^{\dag} {\tt t}^A \bm{\Psi} \\[0.2cm]
{\cal J}&=&-\frac{1}{\sqrt{2r}}\,\bm{\Psi}^\dag\bm{\Psi}\, .
\eea
\end{subequations}
Here  we use the notation $\bm{\Psi}$ and $\bm{\Psi}^\dag$ for the matrix column and row, respectively, built from the
basic fermionic fields:
\bea
{\boldsymbol \Psi}=\big(\Psi_1,\Psi_2,\ldots,\Psi_r\big)^{\rm T}
\ \ \ \ \ \ \ \ \ \ \ \ \qquad \ \ {\boldsymbol \Psi}^\dagger=\big(\Psi^\dagger_1,\Psi^\dagger_2,\ldots,\Psi^\dagger_r\big)\, .
\eea
Also, ${\tt t}^A$ with $A=1,2,\ldots r^2-1$ stand for the $r\times r$ traceless matrices,  which form the defining representation
of the   $\mathfrak{su}(r)$ 
Lie algebra and satisfy the commutation relations
\be
\big[{\tt t}^A,{\tt t}^B\big]=\ri {f^{AB}}_C\,{\tt t}^C\, .
\ee
The currents ${\cal J}^A$, ${\cal J}$ possess some special properties. Apart from being chiral,
\be\label{asoki92jjn12}
\partial_{\bar{u}}{\cal J}^A=\partial_{\bar{u}}{\cal J}=0\,,
\ee
they turn out to form a closed infinite dimensional Lie algebra. Since,
\bea\label{asj9812ja}
&&{\cal J}^A(u)\,{\cal J}^B(u')=-\frac{q^{AB}}{(u-u')^2}+\frac{{f^{AB}}_C\,{\cal J}^C(u)}{u-u'}+O(1)\\[0.2cm]
&&{\cal J}(u)\,{\cal J}(u')=-\frac{1}{2(u-u')^2}\,,\qquad\qquad  \qquad\qquad  {\cal J}^A(u)\, {\cal J}(u')=O(1)\,,\nonumber
\eea
where
\be
q^{AB}={\rm Tr}\big({\tt t}^A{\tt t}^B\big)
\ee
this algebra is the direct sum  of the $\mathfrak{su}(r)$ Kac-Moody algebra at level one, $\widehat{\mathfrak{su}}_1(r)$, and the $\widehat{\mathfrak{u}}(1)$ algebra.
In addition, the chiral component of the energy momentum tensor can be
expressed in terms of the currents as
\be\label{kaui32jnbas}
T={\cal J}^2+\frac{ q_{AB}}{2(r+1)}\ {\cal J}^A{\cal J}^B
\ee
with $q_{AB}$ being the inverse of $q^{AB}$, i.e., 
$
q_{AB}\, q^{BC}=\delta_A^C
$.
\medskip

All the above concerned the free fermion point. Of course, the lattice fermion operators can still be defined
via the Jordan--Wigner transformation as in \eqref{kjsa8923jhbs} for any value of $\delta$. However, 
having in mind the Heisenberg XXZ spin-$\tfrac{1}{2}$  chain corresponding to  the case $r=1$,
 there is no reason to expect that their scaling limit
results in the (anti)chiral fermionic fields.
Nevertheless, we  conjecture
that the algebra of extended conformal symmetry underlying the critical behavior of the lattice model for $|\delta|<\frac{1}{r}$
is generated by ${\cal J}^A$, ${\cal J}$, $\bar{{\cal J}}^A$, $\bar{{\cal J}}$
 for which all of the relations \eqref{asoki92jjn12}-\eqref{kaui32jnbas} along with their barred counterparts remain valid.
To motivate this hypothesis, let's recall some simple facts from the representation theory of $\widehat{\mathfrak{su}}_1(r)$ and $\widehat{\mathfrak{u}}(1)$. 
The affine Lie algebra $\widehat{\mathfrak{su}}_1(r)$ admits $r$ integrable irreps, 
${\cal V}_{\mathfrak{j}}$ with $\mathfrak{j}=0,1,2,\ldots,r-1$,
which are associated with the fundamental  irreps of the finite dimensional Lie algebra 
$\mathfrak{su}(r)$. In particular, $\mathfrak{j}=0,1$ and $\mathfrak{j}=r-1$ correspond
to the trivial, defining and complex conjugate to the defining irrep of $\mathfrak{su}(r)$, 
respectively. 
An irrep for $\widehat{\mathfrak{u}}(1)$ 
is provided by the bosonic Fock space ${\cal F}_P$, labelled by the value of the charge
\be
\hat{P}=\int_0^{2\pi}\frac{\rd x}{2\pi}\ {\cal J}(x)\,,
\ee
which commutes with all of the currents. Then, the space
\be
{\cal H}_{\mathfrak{j},P}={\cal V}_{\mathfrak{j}}\otimes {\cal F}_P
\ee
is an irreducible representation of the direct sum $\widehat{\mathfrak{su}}_1(r)\oplus\widehat{\mathfrak{u}}(1)$.
Similarly, one can introduce the ``right'' space
\be
\bar{{\cal H}}_{\bar{\mathfrak{j}},\bar{P}}=\bar{{\cal V}}_{\bar{\mathfrak{j}}}\otimes \bar{{\cal F}}_{\bar{P}}\, ,
\ee
which is an irrep of the antichiral algebra formed by the currents $\bar{{\cal J}}^A$ and $\bar{\cal J}$.
Then  the operator ${\bf I}_1$
defined in \eqref{kja09jh21jhjh} with $T$ as in \eqref{kaui32jnbas} 
and  $\bar{\bf I}_1$, which may be introduced by the analogous formulae, act
in the space ${\cal H}_{\mathfrak{j},P}\otimes\bar{{\cal H}}_{\bar{\mathfrak{j}},\bar{P}}$
and their eigenvalues are given by
\begin{subequations}\label{kjsa8932jhb21AAA}
\bea\label{kjsa8932jhb21}
I_1&=&P^2+\frac{\mathfrak{j}\,(r-\mathfrak{j})}{2r} -\frac{r}{24}+{\tt L} \\[0.2cm]
\bar{I}_1&=&\bar{P}^2+\frac{\bar{\mathfrak{j}}\,(r-\bar{\mathfrak{j}})}{2r} -\frac{r}{24}+\bar{{\tt L}}
\eea
\end{subequations}
with ${\tt L},\bar{\tt L}=0,1,2,\ldots\ $. A comparison with eq.\eqref{kaso23jhnn}
suggests that the space of states occurring in the scaling limit of the lattice model
organizes into the direct sum of the spaces 
${\cal H}_{\mathfrak{j},P}\otimes\bar{{\cal H}}_{\bar{\mathfrak{j}},\bar{P}}$.
In particular, the sector  with fixed value of $S^z$, denoted as ${\cal H}_{S^z}$,
admits the decomposition
\be\label{sakj832jhsd}
{\cal H}_{S^z}=\bigoplus_{{\tt w}\in\mathbb{Z}}\bigoplus_{{\tt m}=0}^{r-1}
{\cal H}_{\mathfrak{j}_{\tt m},P_{{\tt w},{\tt m}}}\otimes\bar{{\cal H}}_{\bar{\mathfrak{j}}_{\tt m},\bar{P}_{{\tt w},{\tt m}}}\,.
\ee
Here $\mathfrak{j}_{\tt m}$, $\bar{\mathfrak{j}}_{\tt m}$ and $P_{{\tt w},{\tt m}}$, $\bar{P}_{{\tt w},{\tt m}}$
 are given by eqs.\,\eqref{ajks8723hj}
and \eqref{ksajijh2389jidsAA}.\footnote{%
Strictly speaking, a comparison of formulae \eqref{kjsa8932jhb21AAA} and \eqref{kaso23jhnn} is insufficient 
to identify the value of the highest weights labelling  the irreps in the decomposition \eqref{sakj832jhsd}.
However, any ambiguities can be resolved by considering the case $\delta=0$  for which the space of states
${\cal H}_{S^z}$ may be classified w.r.t. the extended conformal symmetry algebra generated by
the basic chiral and antichiral  fermion fields.}
\medskip

Formula \eqref{sakj832jhsd} describes the decomposition of the CFT space of states  
into the irreps 
${\cal H}_{\mathfrak{j},P}\otimes
\bar{{\cal H}}_{\bar{\mathfrak{j}},\bar{P}}$
of the algebra of extended conformal symmetry.  
Let's focus on the space ${\cal H}_{\mathfrak{j},P}$, which was introduced as 
 an irrep of the Lie algebra $\widehat{\mathfrak{su}}_1(r)\oplus\widehat{\mathfrak{u}}(1)$
but turns out to be an irrep of the  chiral algebra of  local fields built from ${\cal J}^A$ and ${\cal J}$.
It is naturally decomposed into
eigenspaces of the operator ${\bf{I}}_1$, the chiral part of the CFT Hamiltonian \eqref{kjasdasdsa8932b21}, as
\be
{\cal H}_{\mathfrak{j},P}=\bigoplus_{{\tt L}=0}^\infty {\cal H}_{\mathfrak{j},P}^{({\tt L})}\,.
\ee
Here the integer ${\tt L}$ is the conformal level.
Each of the level subspaces ${\cal H}_{\mathfrak{j},P}^{({\tt L})}$ is finite dimensional
so that
\be\label{kjas9823jsa}
{\rm Tr}_{{\cal H}_{\mathfrak{j},P}}\big({\tt q}^{{\bf I}_1}\big)=
{\tt q}^{P^2+\frac{\mathfrak{j}\,(r-\mathfrak{j})}{2r} -\frac{r}{24}}
\sum_{{\tt L}=0}^\infty \dim\big({\cal H}_{\mathfrak{j},P}^{({\tt L})}\big) \,{\tt q}^{{\tt L}}\, .
\ee
Since ${\cal H}_{\mathfrak{j},P} ={\cal V}_{\mathfrak{j}}\otimes {\cal F}_P$, the l.h.s. is given by
\be
{\rm Tr}_{{\cal H}_{\mathfrak{j},P}}\big({\tt q}^{{\bf I}_1}\big)=
\frac{{\tt q}^{P^2-\frac{1}{24}}}{\prod_{n=1}^\infty(1-{\tt q}^{n})}
\ \ \chi^{(\rm KM)}_{\mathfrak{j}}({\tt q}) \,,
\ee
where the first factor coincides with the character of  the algebra $\widehat{\mathfrak{u}}(1)$ represented in
 the Fock space ${\cal F}_P$,
while $\chi^{(\rm KM)}_{\mathfrak{j}}$ with $\mathfrak{j}=0,1,\ldots,r-1$ is the  character of the integrable
 irrep of the Kac-Moody algebra $\widehat{\mathfrak{su}}_1(r)$.
A closed-form expression  for $\chi^{(\rm KM)}_{\mathfrak{j}}$ can be found in ref.\cite{Itzykson}.
However, for our purposes,  
it is  convenient 
to use the generating function for all the characters of the integrable representations:
\be\label{aks32jnjds}
{\tt q}^{-\frac{r-1}{12}}\prod_{n=1}^\infty(1-{\tt q}^{n})\Big[
(1+z {\tt q}^{n-\frac{1}{2}})\,(1+z^{-1}\,{\tt q}^{n-\frac{1}{2}})\Big]^r=\sum_{\mathfrak{j}=0}^{r-1}\sum_{{\tt w}\in\mathbb{Z}}
z^{\mathfrak{j}+{\tt w}r } \, {\tt q}^{\frac{1}{2r}(\mathfrak{j}+{\tt w} r)^2}\,
\chi^{(\rm KM)}_{\mathfrak{j}}({\tt q})\ .
\ee
From the above formulae \eqref{kjas9823jsa}-\eqref{aks32jnjds} it follows that the dimensions
of the level subspaces ${\cal H}_{\mathfrak{j},P}^{({\tt L})}$ do not depend on $P$
and coincide with the  integers ${\cal D}_{\mathfrak{j}}{({\tt L})}$ defined through the relation
\eqref{asjh32h98sd}:
\be\label{kjas892jhbnkxsasfkj3}
\dim\big({\cal H}_{\mathfrak{j},P}^{({\tt L})}\big)={\cal D}_{\mathfrak{j}}{({\tt L})}\ .
\ee

\section{Identification of parameters\label{sec5}}
The main ingredients are now in place and we are ready
to formulate the ODE/IQFT correspondence for the problem at hand.
\medskip

The analysis of the previous sections, in particular, eq.\,\eqref{jkas8912jhas}
 suggests that the lattice $Q$-operators admit the scaling limit of the form
\be\label{jkas8912jhasA}
{\mathlarger{\mathlarger{\mathlarger {\boldsymbol  a}}}}_\pm(\mu)=
\slim_{N\to\infty}G^{(N/r)}\Big(-\big( \tfrac{r}{1-r \delta}\big)^{1+r\delta}\mu^r\,\big|\,\tfrac{1}{2}\, (1 - r\delta)\Big)\,
\mathbb{A}_\pm\bigg(\Big(\tfrac{r^2N_0}{(1-r\delta)N}\Big)^{\frac{1}{r}+\delta}\,\re^{\frac{\ri\pi(r+1)}{2r}}\,\mu\bigg)\,.
\ee
Here
${\mathlarger{\mathlarger{\mathlarger {\boldsymbol  a}}}}_\pm(\mu)$  are operators acting
in the space ${\cal H}_{\mathfrak{j},P}$ --- the building block for the chiral space of states. Furthermore, 
these operators
  must commute with   
${\bf I}_1$, the chiral part of the CFT Hamiltonian,
 so that they act invariantly in the finite dimensional level subspaces:
\be\label{kas092jkjd}
{\mathlarger{\mathlarger{\mathlarger {\boldsymbol  a}}}}_\pm(\mu)\,:\ \ {\cal H}_{\mathfrak{j},P}^{({\tt L})}
\mapsto {\cal H}_{\mathfrak{j},P}^{({\tt L})}\ .
\ee
The values of
$P$ labelling the irreps appearing in the scaling limit of the lattice model 
 are determined through the relations \eqref{ksajijh2389jidsAA}. 
For given value of the twist parameter ${\tt k}$ they form the discrete set.
However, since ${\tt k}$  is a continuous parameter one may expect that the action of 
${\mathlarger{\mathlarger{\mathlarger {\boldsymbol  a}}}}_\pm(\mu)$ 
is well
 defined in ${\cal H}_{\mathfrak{j},P}^{({\tt L})}$
at least for generic values of $P$. Within the ODE/IQFT correspondence,
the eigenvalues of   ${\mathlarger{\mathlarger{\mathlarger {\boldsymbol  a}}}}_\pm(\mu)$   are related to 
 the spectral determinants $D_\pm(\mu)$ of the ODEs of the form \eqref{askjdjh21A}-\eqref{kasj892jhsa}. 
The complete formulation
of the correspondence assumes, among other things, an explicit construction of
the operators \eqref{kas092jkjd} and an effective procedure for
obtaining their joint eigenbasis
at least for the first few lowest levels ${\tt L}$, in the spirit of the works \cite{Bazhanov:1994ft,Bazhanov:1996dr,Bazhanov:1998dq}.
This problem lies beyond our current study.
Here we discuss how the parameters 
on the ODE side such as ${\tt K}$ and the number of apparent singularities
${\tt J}$ are related to $\mathfrak{j}$, $P$ and ${\tt L}$, which specify
the level subspaces ${\cal H}_{\mathfrak{j},P}^{({\tt L})}$. 
\medskip

The operators ${\mathlarger{\mathlarger{\mathlarger {\boldsymbol  a}}}}_\pm(\mu)$  obey the commutativity conditions
\be
\big[{\mathlarger{\mathlarger{\mathlarger {\boldsymbol  a}}}}_\pm(\mu),
{\mathlarger{\mathlarger{\mathlarger {\boldsymbol  a}}}}_\pm(\mu')\big]=
\big[{\mathlarger{\mathlarger{\mathlarger {\boldsymbol  a}}}}_+(\mu),
{\mathlarger{\mathlarger{\mathlarger {\boldsymbol  a}}}}_-(\mu')\big]= 0\, .
\ee
By considering their expansions about different points in the complex $\mu$-plane 
one may generate the family of mutually commuting Integrals of Motion (IM) in the CFT (so-called local IM, 
non-local IM, dual non-local IM, etc.).
Among these is the local IM ${\bf I}_1$ 
defined by eqs.\,\eqref{kja09jh21jhjh} and \eqref{kaui32jnbas}. It is expected to occur in the large-$\mu$ 
asymptotic expansion
of ${\mathlarger{\mathlarger{\mathlarger {\boldsymbol  a}}}}_+(\mu)$
and, hence, its eigenvalue $I_1$ should appear in the same expansion of the spectral determinant
$D_+(\mu)$  of the ODE \eqref{askjdjh21A}-\eqref{kasj892jhsa}. 
This was already shown to be the case for the free fermion point, see  eq.\,\eqref{kjnbbnxcjsdas}.
\medskip

The zeroes of  $D_+(\mu)$ accumulate along the rays 
\be
\arg(\mu)=\frac{\pi}{r}\,(2\ell-1)\ \ 
({\rm mod}\ 2\pi)\qquad {\rm with}\qquad \ell=1,2,\ldots,r\ .
\ee
 These split the complex $\mu$-plane into $r$ wedges in each of which
the large-$\mu$ asymptotic of the spectral determinant is described differently. Namely,
if inside the $\ell^{\rm th}$
wedge we swap $\mu$ for $\theta$ such that
\be
\mu=\re^{\frac{2\pi\ri}{r}\,\ell}\,\re^{\frac{1+r\delta}{r}\theta}
\ee
then a  WKB analysis of the ODE   yields\footnote{%
Generalizing  the terminology of ref.\cite{Bazhanov:1996dr}   concerning 
the quantum KdV theory (the case $r=1$), 
we will call the coefficients appearing in the Taylor expansion  of
${\mathlarger{\mathlarger{\mathlarger {\boldsymbol  a}}}}_+(\mu)$ or its logarithm as  the non-local IM.
The asymptotic coefficients $\mathfrak{C}_+$, $\tilde{H}_{m}$  from \eqref{aysayasy}
 are  the eigenvalues of the dual non-local IM, while 
$Q_{1},\,Q_{3},\,Q_{5},\ldots $  give the eigenvalues of the local IM. There is no analogue of $Q_{\frac{2m-1}{r}}$
with $2m-1$ not divisible by $r$ in \cite{Bazhanov:1996dr}. We will refer to the corresponding operators as the semi-local IM.}
\bea\label{aysayasy}
\log D_+(\mu)&\asymp&\frac{\Gamma(\frac{r\delta }{1+r\delta})\Gamma(-\frac{1}{2}+\frac{1}{1+r\delta})}
{2\sqrt{\pi}}\   \re^\theta -\Big({\tt K}+\frac{{\tt M}}{r}\,\Big)\frac{1+r\delta}{1-r\delta}\  \theta
+\log\Big(\re^{-\frac{2\pi\ri}{r}{\tt M}\ell}\ {\mathfrak C}_+\Big) 
\\[0.1in]
&- &\sum_{m=1}^{\infty}
   \re^{-\frac{2\pi\ri}{r}(2m-1)\ell}\,Q_{\frac{2m-1}{r}}\   \re^{-\frac{2m-1}{r}\theta}-
\sum_{m=1}^\infty
  \re^{-\frac{4\pi\ri}{r}m\ell}\, \tilde{H}_{m}\  \re^{-\frac{2m}{r}\frac{1+r\delta}{1-r\delta}\theta} 
+O\Big(\big(\,\re^{\theta}\,\big)^{-\infty} \Big)
   \nonumber
 \eea
as
\be
|\theta|\to+\infty\qquad \qquad {\rm and}\qquad\qquad  |\arg(\theta)|<\frac{\pi}{1+r\delta}\  .
\ee
The asymptotic coefficients
are mostly unavailable apart from $\mathfrak{C}_+$ and
$Q_{\frac{2m-1}{r}}$ with $m=1,2,\ldots,\frac{3r+1}{2}$, whose   expressions 
 are presented in Appendix \ref{AppA}. For now, we need only
\bea\label{ajksu23jnbas}
Q_1=\frac{2\Gamma(\frac{1}{2}+\frac{r\delta }{1+r\delta})\Gamma(\frac{1}{1+r\delta})}
{\sqrt{\pi}  r}\ \bigg[\frac{r}{2(1-r\delta)}\ \Big({\tt K}+\frac{{\tt M}}{r}\Big)^2-
\frac{ {\tt M}^2 }{2 r} +\frac{{\tt J}}{2}-\frac{r}{24}\bigg]\, .
\eea
Note that the expansion \eqref{aysayasy} is not literally applicable for 
$\delta\ne -\tfrac{k-1}{rk}$  with $k=1,2,3,\ldots\ $.
In particular, as $\delta=0$  the numerical coefficient in the leading 
term possesses a simple pole. In this case it is replaced by 
$-\frac{\mu^r}{2}\,\log\big(\frac{\mu^r}{2 {\rm e}}\big)$ with $\mu^r=\re^{\theta}$. All other terms admit the limit 
$\delta\to0$ and reduce to those presented in eqs.\eqref{kjnbbnxcjsdas}-\eqref{kjas982jhbsdaaaaa}.
As was already pointed out at the beginning of sec.\,\ref{sec2}, formally setting $\delta=0$  in the ODE 
 \eqref{askjdjh21A}-\eqref{kasj892jhsa}  leads to the vacuum differential equation, 
which can not be used to compute the spectral determinant apart from the case when the
apparent singularities are absent. 
\medskip

The following comment is in order here.
To derive the asymptotic formula \eqref{aysayasy} we used an important property of the 
algebraic system \eqref{kasj892jhsa}, which was already  mentioned  in sec.\,\ref{sec3}. Namely,
for all of its solutions $\{(\varpi_i,c_i)\}_{i=1}^{{\tt J}}$,
the sum of $c_i$ may only take a certain set of values, 
\be \label{9023jdjnkjsd}
\sum_{i=1}^{\tt{J}} c_i={\tt M}\,{\tt K}+\frac{{\tt M}^2}{2}\,\delta\,,
\ee
where ${\tt M}$ is an integer. This was confirmed by numerically solving the system \eqref{kasj892jhsa}
for small values of ${\tt J}$. Furthermore, 
we found that the integer ${\tt M}$ has the same parity as ${\tt J}$,
i.e., 
\be\label{kas891jbdaskjkds}
{\tt M}={\tt J}\ \ ({\rm mod}\ 2)\,,
\ee
and is  bounded by the condition
\be\label{askj9ddd832jsdas}
|{\tt M}|\le {\tt J}\ .
\ee

\medskip

One should expect that the asymptotic coefficient
$Q_1$ \eqref{ajksu23jnbas} coincides with $I_1$ up to an overall state-independent  factor.
The latter may be fixed by considering the case of the vacuum   in the sector with $S^z=0$,
where $I_1=\frac{r}{2(1-r\delta)}\ {\tt k}^2-\frac{r}{24}$.
On the other hand, the state corresponds to the differential equation with ${\tt K}={\tt k}$,  ${\tt J}=0$ and
hence  the integer ${\tt M}$ also vanishes. 
Thus, one concludes that the eigenvalue of the local IM ${\bf I}_1$ coincides with the expression
in the square brackets in \eqref{ajksu23jnbas}, i.e.,
\be
I_1=\frac{r}{2(1-r\delta)}\ \Big({\tt K}+\frac{{\tt M}}{r}\Big)^2-
\frac{ {\tt M}^2 }{2 r} +\frac{{\tt J}}{2}-\frac{r}{24}\ .
\ee
A comparison with the formula \eqref{kjsa8932jhb21} suggests the following identifications
\be\label{aksj8923jhsd}
P=\sqrt{\frac{r}{2(1-r\delta)}}\,\Big({\tt K}+\frac{{\tt M}}{r}\,\Big)\,,\qquad\qquad\qquad 
 {\tt L}=\frac{{\tt J}}{2}-\frac{{\tt M^2}}{2r}-\frac{\mathfrak{j}(r-\mathfrak{j})}{2r}\, .
\ee
An additional relation is needed to express $\mathfrak{j}$ in terms of  ${\tt J}$ and ${\tt M}$.
It can be obtained  from the condition that the number of solutions  ${\cal N}_{{\tt J},{\tt M}}$
  of the algebraic system  \eqref{kasj892jhsa} supplemented by  \eqref{9023jdjnkjsd}
coincides with the dimension of the level subspace
 ${\cal H}_{\mathfrak{j},P}^{({\tt L})}$, i.e.,
\be\label{0932jjhjdhgsa}
{\cal N}_{{\tt J},{\tt M}}={\cal D}_{\mathfrak{j}}{({\tt L})}\,.
\ee
This is required for there to be a one-to-one correspondence between the states in the level subspaces 
and the solution sets $\{(\varpi_i,c_i)\}$. A numerical
study shows that \eqref{0932jjhjdhgsa} holds true
provided the pairs $({\tt J},{\tt M})$ and $({\tt L},\mathfrak{j})$ are related as in eq.\,\eqref{aksj8923jhsd} and
\be\label{9023kj54b}
\mathfrak{j}={\tt M}\ ({\rm mod}\ r)\, .
\ee
Notice that, as an immediate consequence, 
$\frac{{\tt J}}{2}-\frac{{\tt M^2}}{2r}-\frac{\mathfrak{j}(r-\mathfrak{j})}{2r}$ with
$\mathfrak{j}$ as above must be a non-negative number, 
which yields a stronger bound on $|{\tt M}|$
than \eqref{askj9ddd832jsdas}. 
Moreover, the relation \eqref{kas891jbdaskjkds} simply follows from  \eqref{aksj8923jhsd} and \eqref{9023kj54b}.
\medskip

Given ${\tt K}$, ${\tt J}$ and ${\tt M}$, formulae
  \eqref{aksj8923jhsd},\,\eqref{9023kj54b} allow one to uniquely determine $(P,{\tt L},\mathfrak{j})$.
However, it is easy to see that the group of transformations generated by
\be\label{kjas23jkji90i32}
{\tt K}\mapsto {\tt K}+1\,,\qquad\qquad 
{\tt J}\mapsto {\tt J}+r-2{\tt M}\,,\qquad\qquad
{\tt M}\mapsto {\tt M}- r
\ee
leaves the r.h.s. of those relations invariant as well as the condition ${\tt J}\ge 0$.
As a result, infinitely many triples  $({\tt K},{\tt J},{\tt M})$ correspond to the same values of $(P,{\tt L},\mathfrak{j})$.
This ambiguity can be traced
at the level of the differential equations as follows.
\medskip

Consider two ODEs.
The first one is given by \eqref{askjdjh21A},\,\eqref{askjdjh21B}  and is
characterized by the parameters ${\tt K},$ ${\tt J}$ and the set $\{(\varpi_i,c_i)\}_{i=1}^{{\tt J}}$. The latter
satisfies the algebraic system \eqref{kasj892jhsa}  as well as the condition \eqref{9023jdjnkjsd}  with some integer ${\tt M}$.
The second ODE takes the similar form:
\be\label{askjui23jhdsn}
\big(-\partial_v^2+\tilde{U}(v)\big)\tilde{\psi}=0
\ee
with 
\be\label{askjui23jhdsnA}
\tilde{U}(v)=\tilde{{\tt K}}^2+\re^{2v}+\epsilon^r\,\re^v+\sum_{i=1}^{\tilde{{\tt J}}}\bigg(
\frac{3\delta^2\tilde{\varpi}_i^2\epsilon^2}{(\epsilon^2-\tilde{\varpi}_i^2)^2}+
\frac{2\delta}{\epsilon^2-\tilde{\varpi}_i^2}\,\big(\tilde{c}_i\epsilon^2+
\tilde{\varpi}_i\epsilon\,\re^v\big)\bigg)
\ee
and
\be
\tilde{{\tt K}}={\tt K}+1\,,\qquad\qquad \tilde{{\tt J}}={\tt J}+r-2{\tt M}\ .
\ee  
Also $\{(\tilde{\varpi}_i, \tilde{c}_i)\}_{i=1}^{\tilde{\tt J}}$  is a solution of 
the  algebraic system, which is obtained from \eqref{kasj892jhsa} through the substitutions ${\tt K}\mapsto \tilde{{\tt K}}$,
 ${\tt J}\mapsto \tilde{{\tt J}}$ and $\{(\varpi_i,c_i)\}\mapsto \{(\tilde{\varpi}_i,\tilde{c}_i)\}$, such that
\be\label{asjhj3298u32hj}
\sum_{i=1}^{\tilde{{\tt J}}}\tilde{c}_i=\tilde{{\tt M}}\tilde{{\tt K}}+\frac{\tilde{{\tt M}}^2}{2}\,\delta
\ee
with $\tilde{{\tt M}}={\tt M}-r$.
It turns out that  $\{(\tilde{\varpi}_i,\tilde{c}_i)\}$ can be chosen in such a way so that 
if $\tilde{\psi}$  satisfies the  ODE \eqref{askjui23jhdsn},\,\eqref{askjui23jhdsnA},
then
\bea\label{sak90kj21jh3290A}
\psi=F_1(v)\,\partial_v \tilde{\psi}(v)+F_2(v)\,\tilde{\psi}(v)
\eea
with certain functions $F_1$ and $F_2$ obeys  the original differential equation \eqref{askjdjh21A}-\eqref{kasj892jhsa}.
The relation between the two solution
 sets $\{(\varpi_i,c_i)\}$ and $\{(\tilde{\varpi}_i,\tilde{c}_i)\}$, as well as the explicit form for the functions,
are  given in Appendix \ref{AppB} (see 
eqs.\eqref{ashbdhgv32198},\,\eqref{asldjh3298hi},\,\eqref{asldjh3298hiA},\,\eqref{kasjhd8ihfds}, therein).
It follows from these formulae
 that the spectral determinants for the two ODEs, whose solutions are related via \eqref{sak90kj21jh3290A},
 are the same.
\medskip

The symmetry transformation \eqref{kjas23jkji90i32} allows one to always 
restrict to the case where 
the integer ${\tt M}$ satisfies the condition
\be
0\le {\tt M}\leq  r-1\ .
\ee
Then, the relations between $P$, ${\tt L},\mathfrak{j}$ and  ${\tt K}$, ${\tt J},{\tt M}$ 
may be unambiguously inverted yielding
\be
{\tt K}=\sqrt{\frac{2(1-r\delta)}{r}}\ P-\frac{\mathfrak{j}}{r}\,,\qquad\qquad
{\tt J}= 2{\tt L}+\mathfrak{j}\, ,\qquad\qquad
{\tt M} =\mathfrak{j}\ .
\ee

\medskip
The differential equations \eqref{askjdjh21A}-\eqref{kasj892jhsa}
  describe the eigenvalues of the commuting family of operators
acting in the space ${\cal H}_{\mathfrak{j},P}$. The  ODEs which encode the eigenvalues 
of   $\bar{\mathlarger{\mathlarger{\mathlarger {\boldsymbol  a}}}}_\pm(\mu)
\in{\rm End}\big(\bar{\cal{H}}_{\bar{\mathfrak{j}},\bar{P}}\big)$ appearing in the scaling limit of the operators 
$\bar{\mathbb{A}}_\pm(\zeta)$ \eqref{kjsajh1289},  differ only in notation.
They depend 
on $\bar{\mu}$, $\bar{{\tt K}}$, $\bar{{\tt J}}$ as well as $\{(\bar{\varpi}_i, \bar{c}_i)\}_{i=1}^{\bar{\tt J}}$, which obeys
an algebraic system similar to \eqref{kasj892jhsa}. Such solutions can be assigned the integer $\bar{{\tt M}}$ 
according to
\be \label{aasdas0923jdjh}
\sum_{i=1}^{\bar{{\tt J}}} \bar{c}_i=\bar{{\tt M}}\,\bar{{\tt K}}+\frac{\bar{{\tt M}}^2}{2}\,\delta\,. 
\ee
The parameters 
 $\bar{{\tt K}}$, $\bar{{\tt M}}$, $\bar{{\tt J}}$ are given in terms of
 those labeling the level $\bar{{\tt L}}$ subspace of $\bar{\cal{H}}_{\bar{\mathfrak{j}},\bar{P}}$
by the formulae 
\be
\bar{P}=\sqrt{\frac{r}{2(1-r\delta)}}\,\Big(\bar{\tt K}+\frac{\bar{{\tt M}}}{r}\,\Big)\,,\qquad\qquad 
 \bar{{\tt L}}=\frac{\bar{{\tt J}}}{2}-\frac{\bar{{\tt M}}^2}{2r}-\frac{\bar{\mathfrak{j}}(r-\bar{\mathfrak{j}})}{2r}\,,
\qquad\qquad \bar{\mathfrak{j}}=-\bar{\tt M}\ ({\rm mod}\ r)\, 
\ee
that  are analogous to eqs.\,\eqref{aksj8923jhsd} and \eqref{9023kj54b}. 
One can always restrict to the case where $\bar{{\tt M}}=0,1,\ldots,r-1$ 
after which the above relations may be inverted. Keeping in mind that $\bar{\mathfrak{j}}$ 
also takes the values from $\{0,1,\ldots,r-1\}$, one arrives at
\begin{subequations}
\be
\bar{{\tt K}}=\sqrt{\frac{2(1-r\delta)}{r}}\ {\bar P}-\frac{r-\bar{\mathfrak{j}}}{r}\,,\qquad\quad
\bar{{\tt J}}= 2\bar{{\tt L}}+r-\bar{\mathfrak{j}}\, ,\qquad\quad
\bar{{\tt M}} =r-\bar{\mathfrak{j}}\qquad {\rm for}\qquad \bar{\mathfrak{j}}=1,2,\ldots,r-1
\ee
and
\be
\bar{{\tt K}}=\sqrt{\frac{2(1-r\delta)}{r}}\ {\bar P}\,,\qquad\quad
\bar{{\tt J}}= 2\bar{{\tt L}}\, ,\qquad\quad
\bar{{\tt M}} =0
\qquad \!\!{\rm for}\qquad \bar{\mathfrak{j}}=0\, .
\ee
\end{subequations}
\section{Numerical checks\label{sec6}}
The differential equations 
 \eqref{askjdjh21A}-\eqref{kasj892jhsa} together with
 their barred counterparts describe the states from the infinite dimensional space
which occurs in the scaling limit of the
${\cal Z}_r$ invariant inhomogeneous  XXZ spin-$\frac{1}{2}$ chain in the vicinity of the free fermion point.
The results obtained so far  allow one to isolate the correspondence to a finite set of ODEs
 and the CFT Bethe states belonging to the 
finite dimensional level $({\tt L},\bar{\tt L})$ subspace
\be\label{ais903284jh}
{\cal H}_{\mathfrak{j}_{\tt m},P_{{\tt w},{\tt m}}}^{({\tt L})}
\otimes\bar{{\cal H}}_{\bar{\mathfrak{j}}_{\tt m},\bar{P}_{{\tt w},{\tt m}}}^{(\bar{{\tt L}})}\,,
\ee
where the  integers $\mathfrak{j}_{\tt m},\,\bar{\mathfrak{j}}_{\tt m}\in\{0,1,\ldots,r-1\}$ and
$P_{{\tt w},{\tt m}}$, $\bar{P}_{{\tt w},{\tt m}}$ are  given by formulae 
\eqref{ajks8723hj} and \eqref{ksajijh2389jidsAA}, respectively.
For the reader's convenience, we summarize here how the parameters 
from \eqref{ais903284jh} are related to those entering into the ODEs.
The latter include ${\tt K}$, $\tt{J}$ and ${\tt M}$ from  \eqref{9023jdjnkjsd} 
as well as $\bar{\tt K}$, $\bar{{\tt J}}$ and $\bar{{\tt M}}$ defined through
\eqref{aasdas0923jdjh}.
The relation is described as
\be\label{9jijjdso23}
{\tt K}=\sqrt{\frac{2(1-r\delta)}{r}}\ P_{{\tt w},{\tt m}}-\frac{\mathfrak{j}_{\tt m}}{r}\,,\qquad\qquad
{\tt J}= 2{{\tt L}}+\mathfrak{j}_{\tt m}\,,\qquad\qquad {\tt M}=\mathfrak{j}_{\tt m}\ \ 
\ee
and
\be\arraycolsep=0.2cm
\begin{array}{llll}
\bar{{\tt K}}=\sqrt{\dfrac{2(1-r\delta)}{r}}\ {\bar P}_{{\tt w},{\tt m}}-\dfrac{r-\bar{\mathfrak{j}}_{\tt m}}{r}\,,&
\bar{{\tt J}}= 2\bar{{\tt L}}+r-\bar{\mathfrak{j}}_{\tt m}\, ,&
\bar{{\tt M}} =r-\bar{\mathfrak{j}}_{\tt m}& {\rm for} \qquad \bar{\mathfrak{j}}_{\tt m}=1,2,\ldots,r-1\\[0.4cm]
\bar{{\tt K}}=\sqrt{\dfrac{2(1-r\delta)}{r}}\ {\bar P}_{{\tt w},{\tt m}}\,,&
\bar{{\tt J}}= 2\bar{{\tt L}}\,, &
\bar{{\tt M}} =0
&{\rm for}  \qquad \bar{\mathfrak{j}}_{\tt m}=0
\end{array}\,.
\ee
Given a
 CFT Bethe state from the finite dimensional subspace \eqref{ais903284jh}, its precise identification with a particular 
pair of ODEs may be achieved based on formulae \eqref{jkas8912jhas} and \eqref{jkas8912jhasB},
which express  the scaling limit of the $Q$-functions 
in terms of the spectral determinants of the ODEs. Their l.h.s. can be estimated
from   the numerical solution of the Bethe Ansatz equations for large, but finite $N$.
This should be matched with the numerical data coming from  the integration of the differential equations.
In practice, there is no need to numerically determine the full functions 
$D_\pm(\mu)$ and $\bar{D}_\pm(\bar{\mu})$. Instead, it turns out to be
sufficient to focus on the first few  coefficients that occur 
in their expansions at small and/or large values of the spectral parameters. 

\subsection{Sum rules}
The spectral determinants are entire functions of the spectral parameters.
As such, their  logarithms admit the Taylor
series expansions with a finite radius of convergence. For $D_+(\mu)$ and $\bar{D}_+(\bar{\mu})$,
 in view of the normalization condition $D_+(0)=\bar{D}_+(0)=1$, one has:
\be
\log D_+(\mu)=-\sum_{s=1}^\infty H_s\,\mu^s\,, \qquad\qquad
\log \bar{D}_+(\bar{\mu})=-\sum_{s=1}^\infty \bar{H}_s\,\bar{\mu}^s\, .
\ee
Then, the formulae describing the scaling limit of $A_+(\zeta)$
\eqref{jkas8912jhas} and $\bar{A}_+({\zeta})$ \eqref{jkas8912jhasB}
can be equivalently
written as an infinite number of relations:
\begin{subequations}\label{kjas9832jh12oi}
\bea\label{ksai23jnnbjsas}
&&\slim_{N\to\infty}\bigg(\frac{r^2N_0}{(1-r\delta)N}\bigg)^{(\delta+\frac{1}{r})s}\,\re^{\frac{\ri\pi(r+1)}{2r}s}\,
 h_s^{(N,{\rm reg})}=H_s \\[0.2cm]\label{ksai23jnnbjsasA}
&&\slim_{N\to\infty}\bigg(\frac{r^2N_0}{(1-r\delta)N}\bigg)^{(\delta+\frac{1}{r})s}\,\re^{\frac{\ri\pi(r+1)}{2r}s}\,
 \bar{h}_s^{(N,{\rm reg})}=\bar{H}_s\, ,
\eea
\end{subequations}
where $s=1,2,3,\ldots\ $.
We'll refer to these as the sum rules. The reason is because
$ h_s^{(N,{\rm reg})}$ and $ \bar{h}_s^{(N,{\rm reg})}$ are defined in terms of
  sums over the integers powers of the Bethe roots as
\begin{subequations}
\bea
 h_s^{(N,{\rm reg})}&=&s^{-1}\sum_{j=1}^{\frac{N}{2}-S^z}\big(\zeta_j\big)^{-s}-
\frac{(-1)^{\frac{r+1}{2}\frac{s}{r}}\,N}{2s\cos\big(\frac{\pi s}{2r}(1-r\delta)\big)}\ \,\delta_{s,0\,({\rm mod}\,r)}\\[0.2cm]
 \bar{h}_s^{(N,{\rm reg})}&=&s^{-1}\sum_{j=1}^{\frac{N}{2}-S^z}\big(\zeta_j\big)^{+s} -
\frac{(-1)^{\frac{r+1}{2}\frac{s}{r}}\,N}{2s\cos\big(\frac{\pi s}{2r}(1-r\delta)\big)}
\ \, \delta_{s,0\,({\rm mod}\,r)}\ .
\eea
\end{subequations}
Notice that the counterterms $\propto N$ do not affect the limit
\eqref{kjas9832jh12oi} as $s>\frac{r}{1+r\delta}$ and may be dropped in the definition of $ h_s^{(N,{\rm reg})}$
and $\bar{h}_s^{(N,{\rm reg})}$. In particular, as $\delta>0$ they can always be ignored.
Nevertheless, we found that  the inclusion of such terms
 significantly improves the rate of convergence of the  limits in \eqref{kjas9832jh12oi}
for any  $|\delta|<\frac{1}{r}$.
\medskip

The advantage of rewriting formulae \eqref{jkas8912jhas},\,\eqref{jkas8912jhasB}
as an infinite number of scaling relations \eqref{kjas9832jh12oi} 
 is that $H_s$ and $\bar{H}_s$, entering therein, for small $s$ admit
 analytic expressions in terms of the parameters of the ODEs. They are
  obtained by means of perturbation
theory applied to the differential equations with the spectral parameters assumed to be small.
In particular, one can show that
\bea\label{kjsa89jhjh12}
H_1&=&\frac{\delta}{2^{\delta}\,\Gamma^2(\frac{1+\delta}{2})}\ 
\bigg(\sum_{i=1}^{{\tt J}}\varpi_i^{-1} \bigg)\ 
f_1\Big(\frac{{\tt K}}{2},\frac{1-\delta}{2}\Big)\\[0.2cm]
 H_2&=&
\frac{\delta }{2^{2\delta+2}\,\Gamma^2(1+\delta)}\bigg(
\sum_{i=1}^{{\tt J}}\frac{2c_i-3\delta}{\varpi_i^2}\bigg)\ 
f_1\Big(\frac{{\tt K}}{2},-\delta\Big)+\frac{\delta^2}{2^{2\delta}\,\Gamma^4(\frac{1+\delta}{2})}\ 
\bigg(\sum_{i=1}^{{\tt J}}\varpi_i^{-1} \bigg)^2\ 
f_2\Big(\frac{{\tt K}}{2},\frac{1-\delta}{2}\Big)\ . \nonumber
\eea
The function $f_1$ is given by
\be
f_1(h,g)=\frac{\pi\Gamma(1-2g)}{\sin(\pi g)}\ \frac{\Gamma(g+2h)}{\Gamma(1-g+2 h)}\, ,
\ee
while $f_2$ admits an integral representation.
We do not give it here as the formula is rather cumbersome to describe and instead refer the reader to
 Appendix A of 
ref.\cite{Gehrmann:2024tue}.
There is no need to present the expressions for  $\bar{H}_s$, since they are only notationally 
different to those of $H_s$.
\medskip

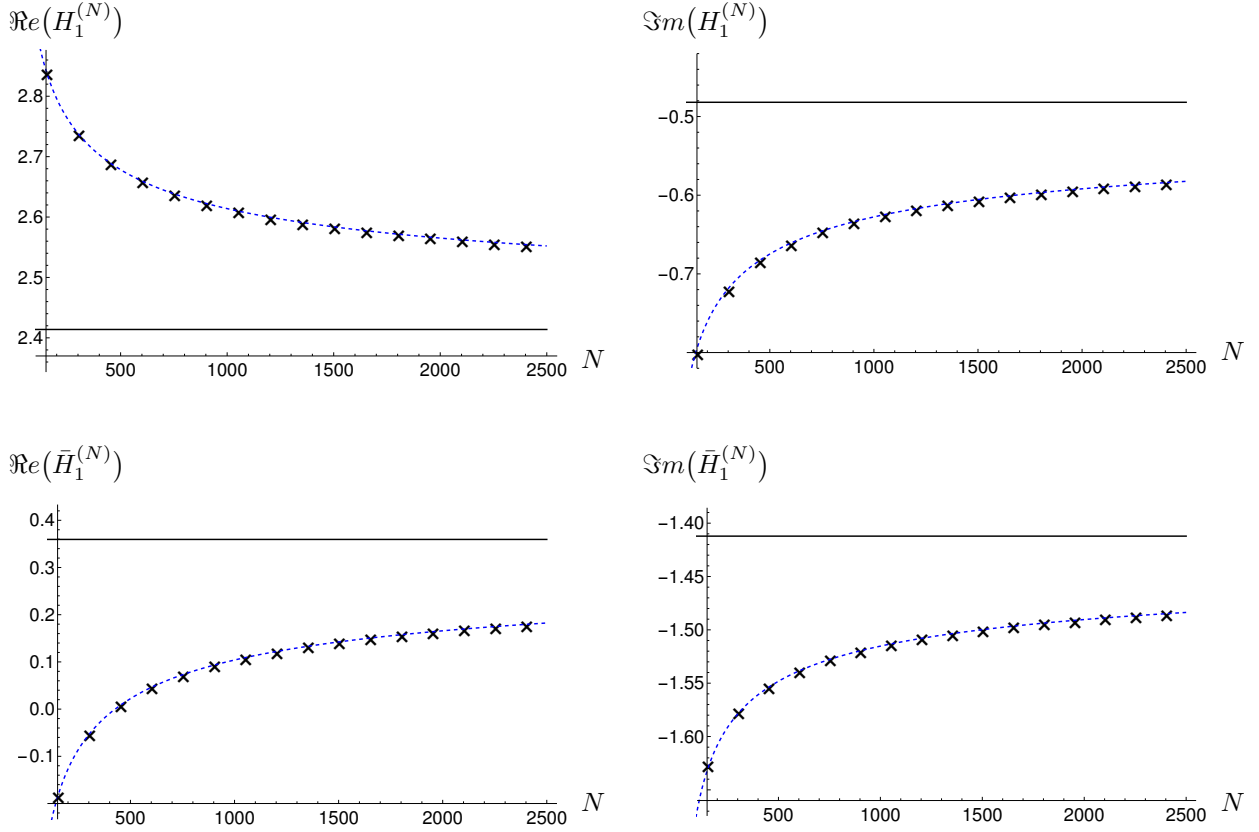
\begin{figure}
\begin{center}
\scalebox{1}{
\begin{tikzpicture}
\node at (0,0) {\includegraphics[width=7.3cm]{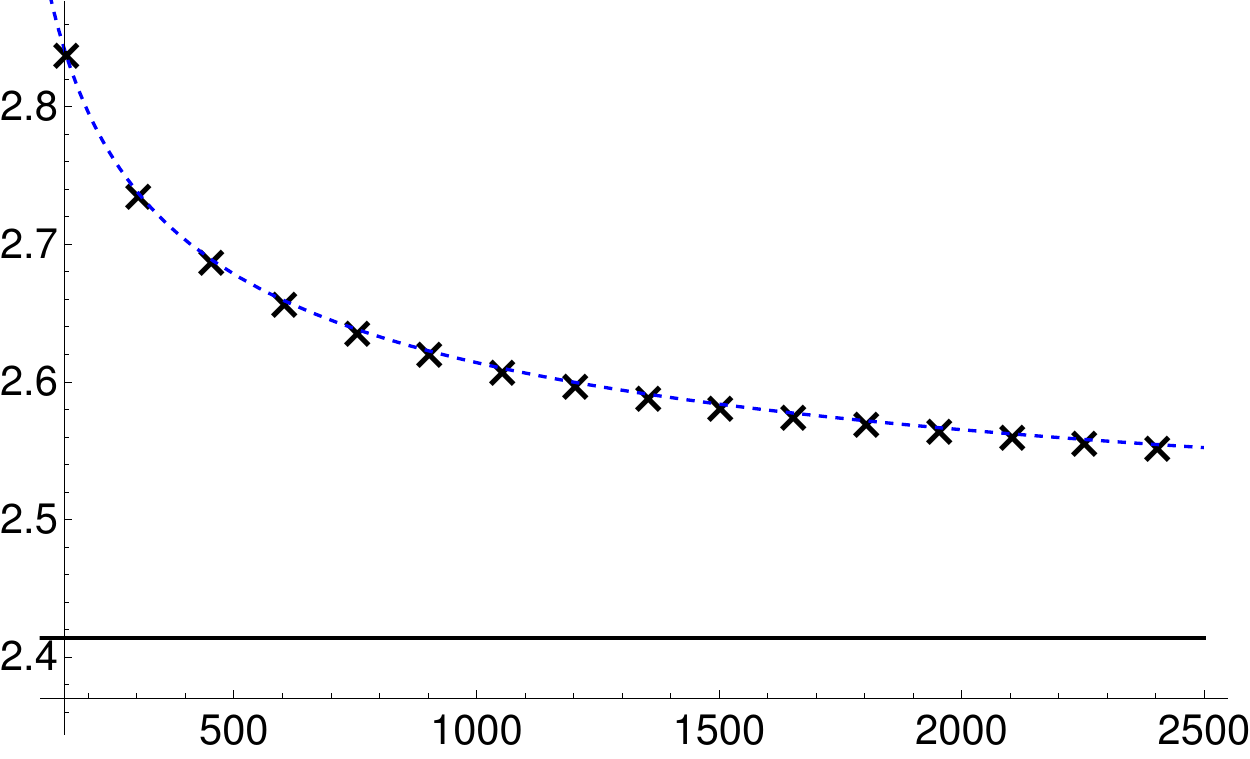}};
\node at (8.5,0) {\includegraphics[width=7.3cm]{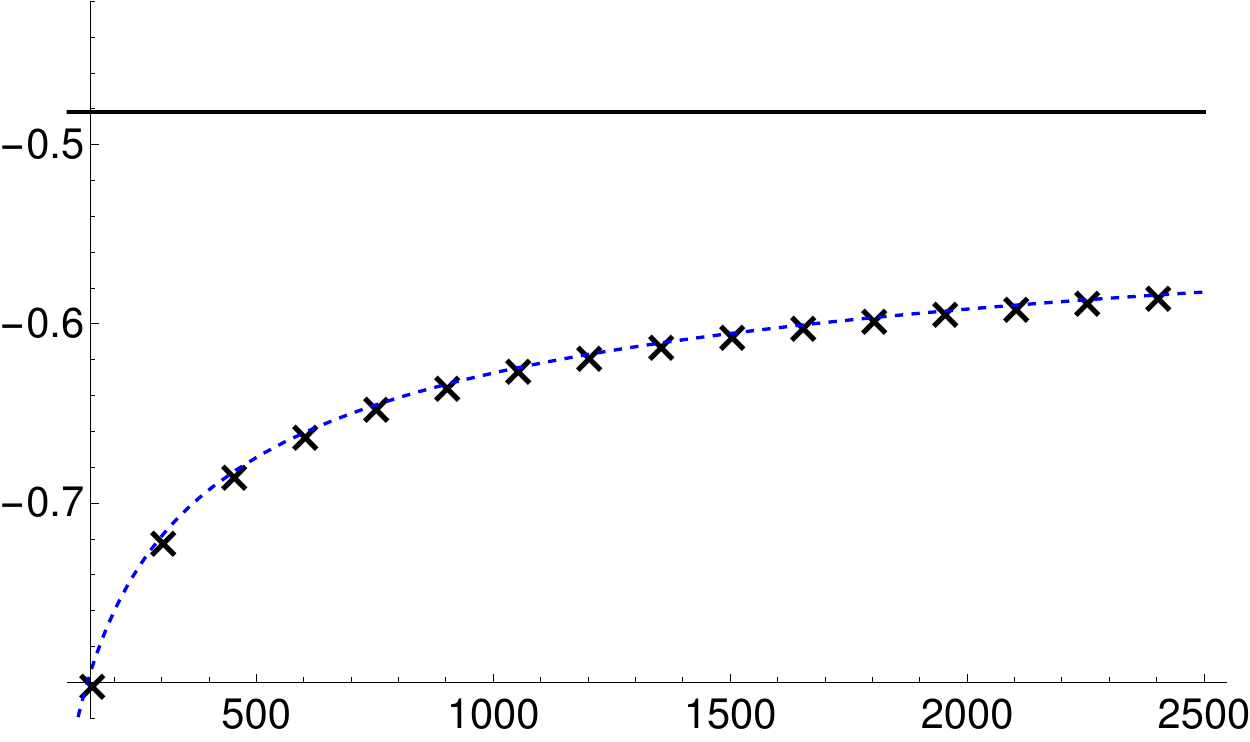}};
\node at (0,-6) {\includegraphics[width=7.3cm]{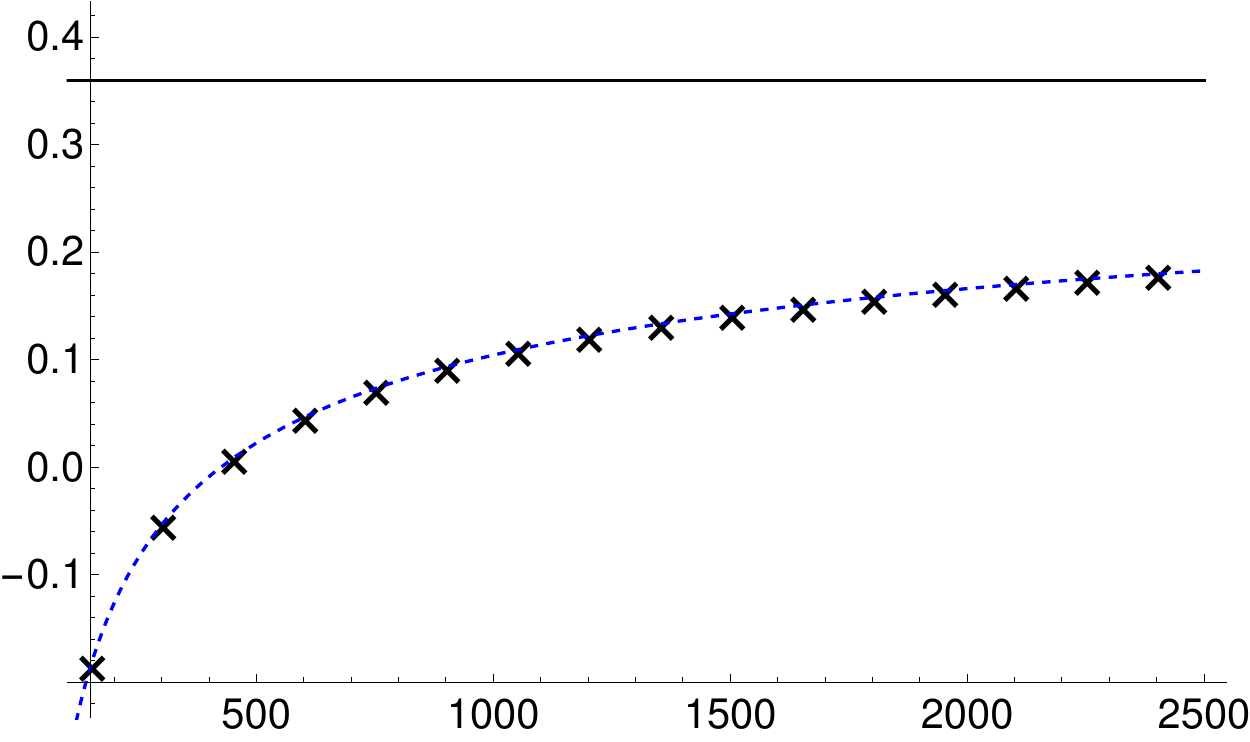}};
\node at (8.5,-6) {\includegraphics[width=7.3cm]{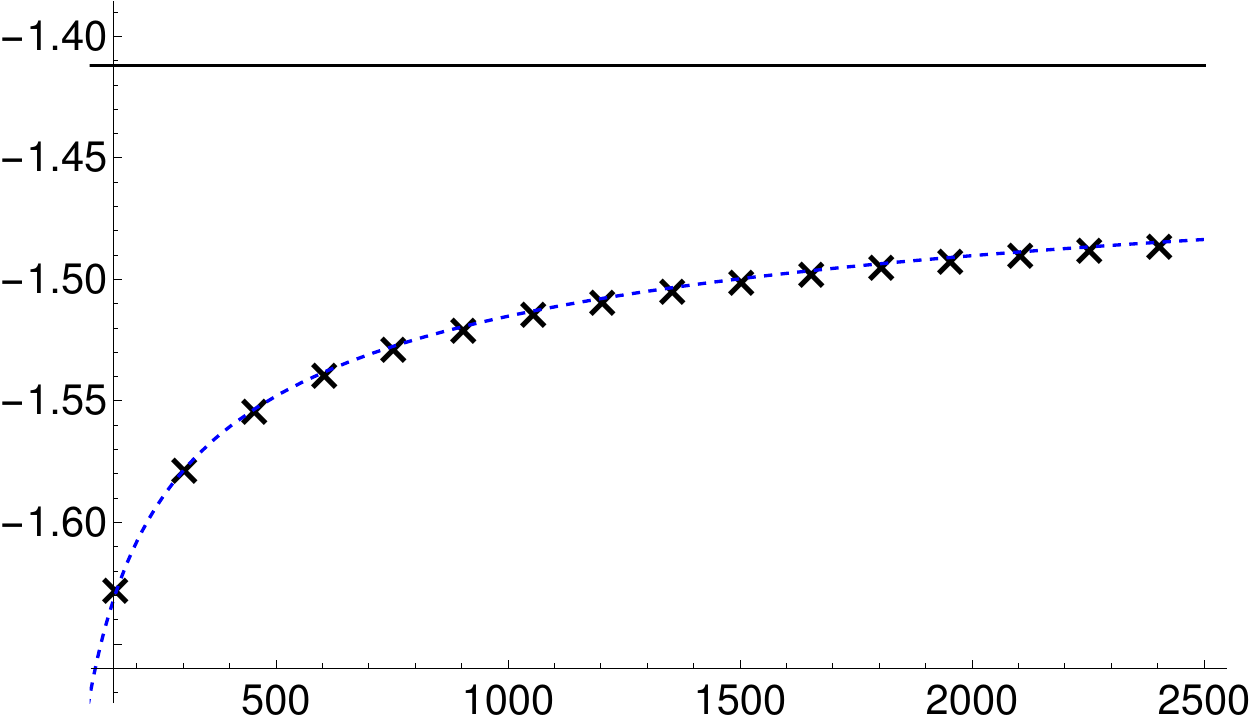}};
\node at (4,-1.9) {\small$N$};
\node at (12.5,-1.8) {\small$N$};
\node at (4,-7.8) {\small$N$};
\node at (12.5,-7.8) {\small$N$};
\node at (-3,2.6) {\small $\Re e\big(H_1^{(N)}\big)$};
\node at (5.5,2.6) {\small $\Im m\big(H_1^{(N)}\big)$};
\node at (-3,-3.3) {\small $\Re e\big(\bar{H}_1^{(N)}\big)$};
\node at (5.5,-3.3) {\small $\Im m\big(\bar{H}_1^{(N)}\big)$};
\end{tikzpicture}
}
\end{center}
\bigskip

\caption{\small%
Presented is numerical data for a certain RG trajectory of the
${\cal Z}_r$ invariant inhomogeneous XXZ spin-$\frac{1}{2}$ chain with $r=5$,
$\delta=\frac{1}{15}$ and ${\tt k}=\frac{1}{50}$.
The corresponding quantum numbers are 
 $S^z=\frac{3}{2}$ $({\tt s}=1)$, ${\tt w}=0$, ${\tt m}=1$,
${\tt L}=2$, $\bar{\tt L}=1$.  It follows from formula \eqref{ajks8723hj}
that $\mathfrak{j}_{{\tt m}}=1$ and $\bar{\mathfrak{j}}_{{\tt m}}=2$.
The sets of Bethe roots corresponding to the RG trajectory for different $N$ 
can be found in the online repository \cite{rep}.  From this data, it was computed
$H_1^{(N)}$  and $\bar{H}_1^{(N)}$, which are  defined as the l.h.s. of 
\eqref{ksai23jnnbjsas} and \eqref{ksai23jnnbjsasA}, respectively, considered at finite fixed $N$ without taking the 
limit $N\to\infty$.  The obtained values are depicted  by the black crosses in the plots.
 The black solid lines represent the predicted  limiting values  
$H_1=2.41386-0.48188\,\ri$ and $\bar{H}_1=0.35948-1.41215\,\ri$.
This was computed by means of  eq.\,\eqref{kjsa89jhjh12}, where 
${\tt K}=-\frac{7}{25}$, ${\tt J}=5$ and using a certain solution set of the algebraic 
system \eqref{kasj892jhsa} satisfying  \eqref{9023jdjnkjsd} with ${\tt M}=1$, while for the barred counterpart
$\bar{\tt K}=-\frac{8}{25}$, $\bar{\tt J}=5$ and $\bar{\tt M}=3$.
The solution sets $\{(c_i,\varpi_i)\}_{i=1}^{5}$ and $\{(\bar{c}_i,\bar{\varpi}_i)\}_{i=1}^{5}$ are also stored in the repository
\cite{rep}.
 The dashed blue lines 
are plots of the real and imaginary parts of
  $2.41257-0.48021\,\ri+(3.19121-2.33556\,\ri)\,N^{-0.4}$  and 
$0.35959-1.41237\,\ri-(4.05034+1.63060\,\ri)\,N^{-0.4}$, which were obtained via a fit of the last three data points
for $H_1^{(N)}$ and $\bar{H}_1^{(N)}$, respectively.\label{fig1}}
\end{figure}

The sum rules \eqref{kjas9832jh12oi}, combined with the  explicit formula  for $H_s$ and $\bar{H}_s$,
provides an effective way of 
identifying the pair of differential equations corresponding to a given CFT Bethe state at least for small values of
the conformal levels. The procedure goes along the following lines.
Having at hand
 $P_{{\tt w},{\tt m}}$, ${{\tt L}}$ and $\mathfrak{j}_{\tt m}$  one extracts ${\tt K}$, $\tt{J}$ and
${\tt M}$ by means of \eqref{9jijjdso23}. Then one numerically solves the system 
 \eqref{kasj892jhsa}, supplemented by the additional constraint \eqref{9023jdjnkjsd}, and finds all
${\cal D}_{\mathfrak{j}_{\tt m}}({\tt L})$ solutions $\{(\varpi_i,c_i)\}$.
For each solution, eq.\eqref{kjsa89jhjh12} allows one to 
 calculate the corresponding $H_1$ and $H_2$.
This gives the r.h.s. of \eqref{ksai23jnnbjsas} for $s=1,2$.
 On the other hand, by solving the Bethe Ansatz equations for increasing values of $N$ for 
the RG trajectory that  flows to the CFT Bethe state, 
one can  numerically estimate the 
limits in the l.h.s. of the sum rules 
(the notion of a RG trajectory for the low energy states of
 a spin chain is discussed in ref.\cite{Bazhanov:2019xvy}). 
Through a matching of the numerical data for the left and right hand sides of  \eqref{ksai23jnnbjsas}
the set of parameters $\{(\varpi_i, c_i)\}$ and hence, the differential equation, is unambiguously determined.
Repeating the procedure one recovers the barred differential equation. In fig.\,\ref{fig1},  numerical data for the
sum rules with $s=1$ is presented
for a specific CFT Bethe state, where $r=5$, ${\tt L}=2$ and $\bar{\tt L}=1$.
\medskip

The scaling relation \eqref{jkas8912jhas} also implies another
set of sum rules involving the 
Taylor coefficients for the spectral determinant $D_-(\mu)$
and  sums over the integer powers of the  zeroes of the $Q$-function $A_-(\zeta)$.
However,  such relations turn out to be functionally dependent on \eqref{ksai23jnnbjsas}.
This follows from the quantum Wronskian relations for the spectral determinants 
\eqref{asd9h12jhbsa},\,\eqref{asd9h12jhbsaAA}  and  for
the lattice operators $\mathbb{A}_\pm(\zeta)$ \eqref{qwron}.

\subsection{Quasi-shift operators\label{sec62}}
The  mutually commuting family, which includes the spin chain Hamiltonian,
possesses a certain class of operators known as the quasi-shift operators.
Their importance for the study of the critical behavior was first pointed out 
in  ref.\cite{Ikhlef:2011ay} in the context of the
${\cal Z}_2$
invariant inhomogeneous XXZ spin-$\frac{1}{2}$ chain in the critical regime with $\gamma\in(0,\frac{\pi}{2})$.
  The  construction of the quasi-shift operators
as well as some  formal algebraic relations that they satisfy are described  in secs.\,6 and 7 of ref.\cite{Bazhanov:2020new},
where they are denoted as $\mathbb{K}^{(\ell)}$. For our purposes 
we just need the expression for their 
 eigenvalues,  ${\cal K}^{(\ell)}$, in terms of the solutions to the Bethe Ansatz equations:
\be\label{kjsa9823hhg4332}
{\cal K}^{(\ell)}=\re^{\ri\pi{\tt k}}\,q^{S^z-\frac{N}{2}}\prod\limits_{j=1}^{\frac{N}{2}-S^z}\frac{\zeta_j+\eta_{\ell}\,q}{\zeta_j+\eta_{\ell}\,q^{-1}}\qquad\qquad (\ell=1,2,\ldots,r)\,.
\ee
The above formula is written in a way that makes it valid for the general
 inhomogeneous XXZ spin-$\frac{1}{2}$ chain depending on the inhomogeneity parameters $\{\eta_J\}_{J=1}^N$
subject to the $r$-site periodicity condition $\eta_{J+r}=\eta_J$ (see, again, the paper \cite{Bazhanov:2020new} for details).
The ${\cal Z}_r$  invariant lattice system, which is the subject matter of this work, corresponds to setting
\be\label{ajk8932h}
\eta_\ell=(-1)^r\,\re^{\frac{\ri\pi}{r}\,(2\ell-1)}\ \ \ \qquad\qquad\qquad (\ell=1,2,\ldots,r)\ .
\ee
Notice that the product of ${\cal K}^{(\ell)}$ over all $\ell$ yields an expression, which
coincides with ${\cal K}$ from \eqref{asj21gh} --- the eigenvalue of the $r$-site lattice translation operator.
This, of course, holds true as an operator relation $\mathbb{K}=\mathbb{K}^{(1)}\mathbb{K}^{(2)}\ldots\,\mathbb{K}^{(r)}$.

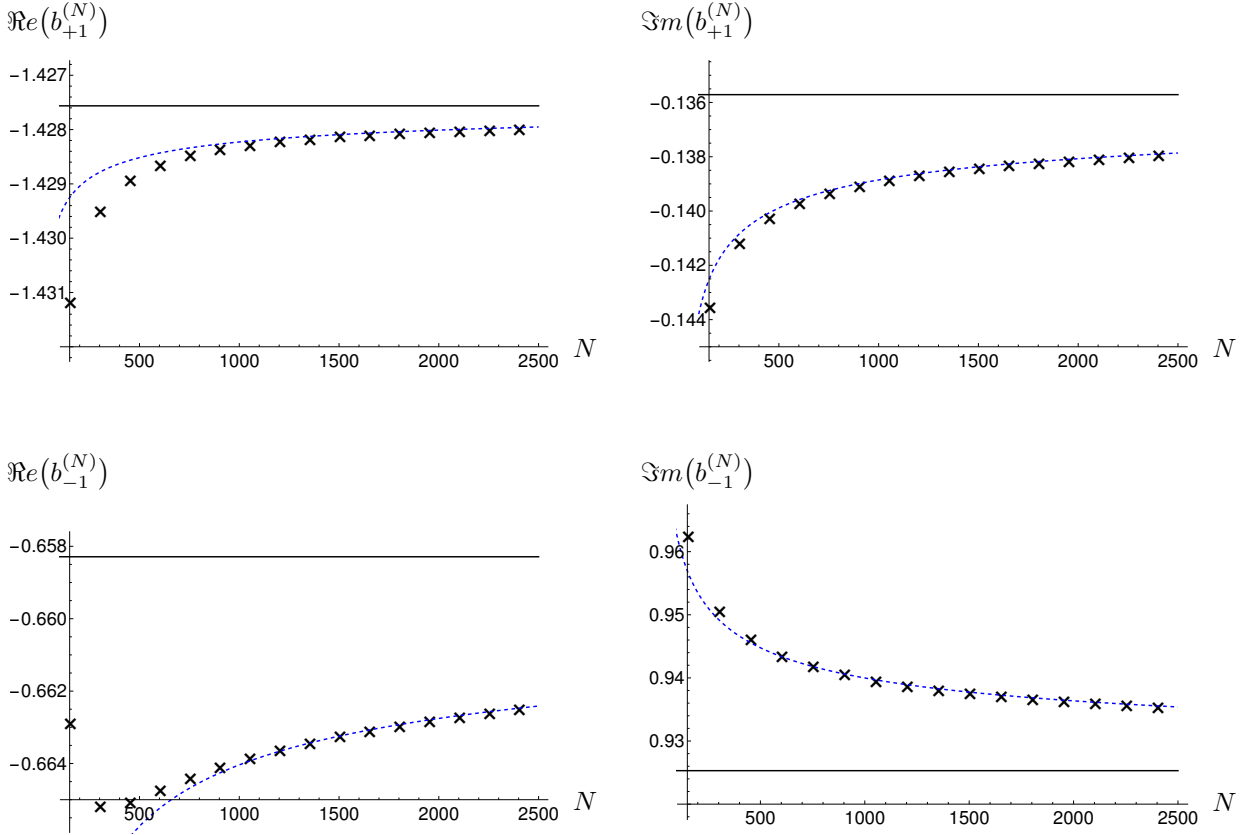
\begin{figure}
\begin{center}
\scalebox{1}{
\begin{tikzpicture}
\node at (0,0) {\includegraphics[width=7.3cm]{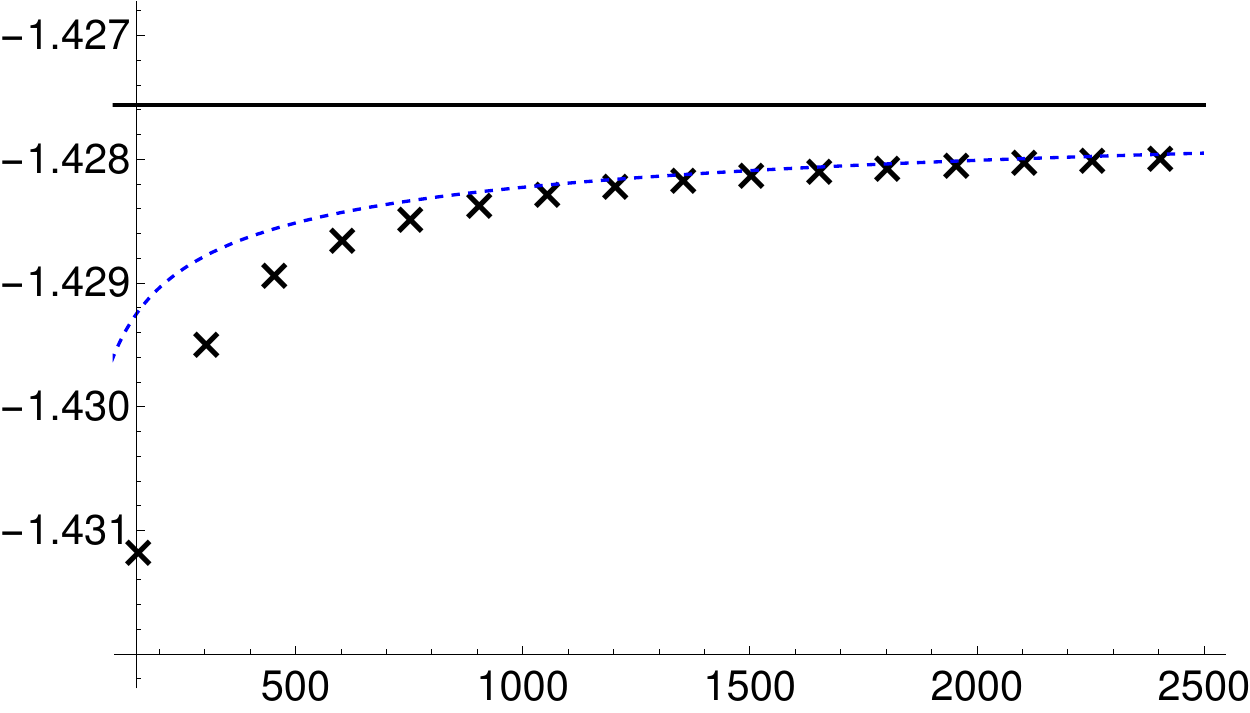}};
\node at (8.5,0) {\includegraphics[width=7.3cm]{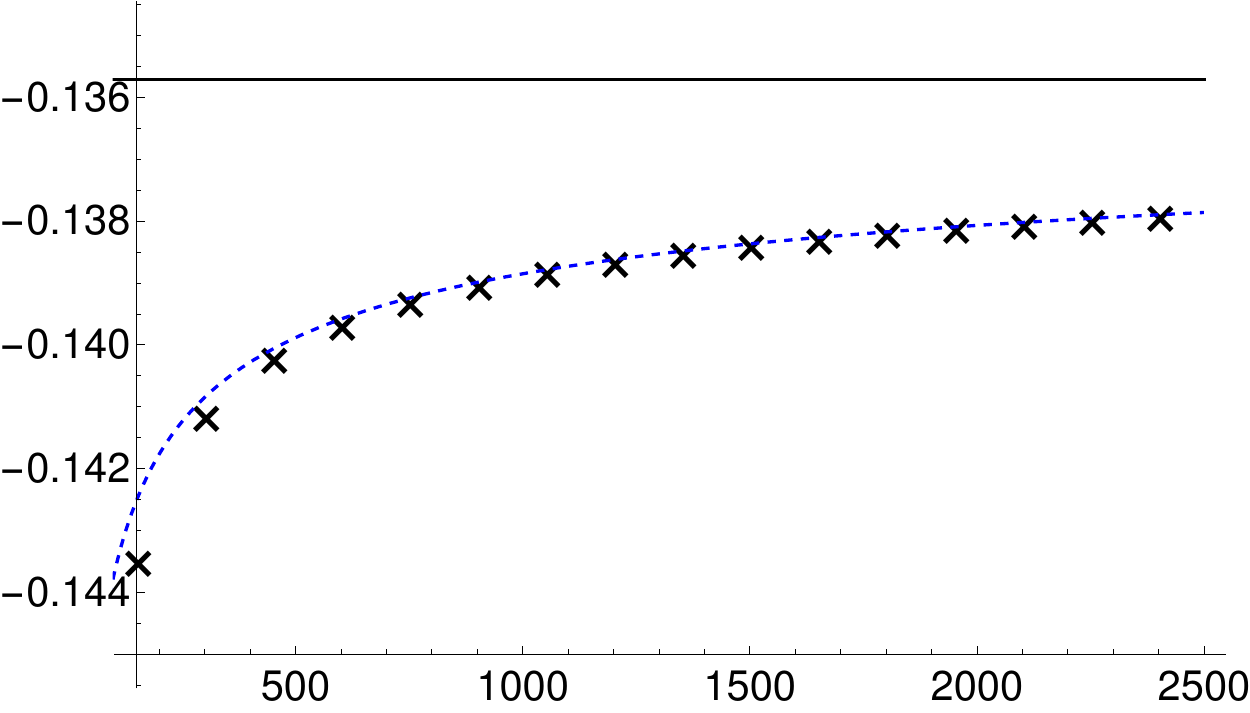}};
\node at (0,-6.2) {\includegraphics[width=7.3cm]{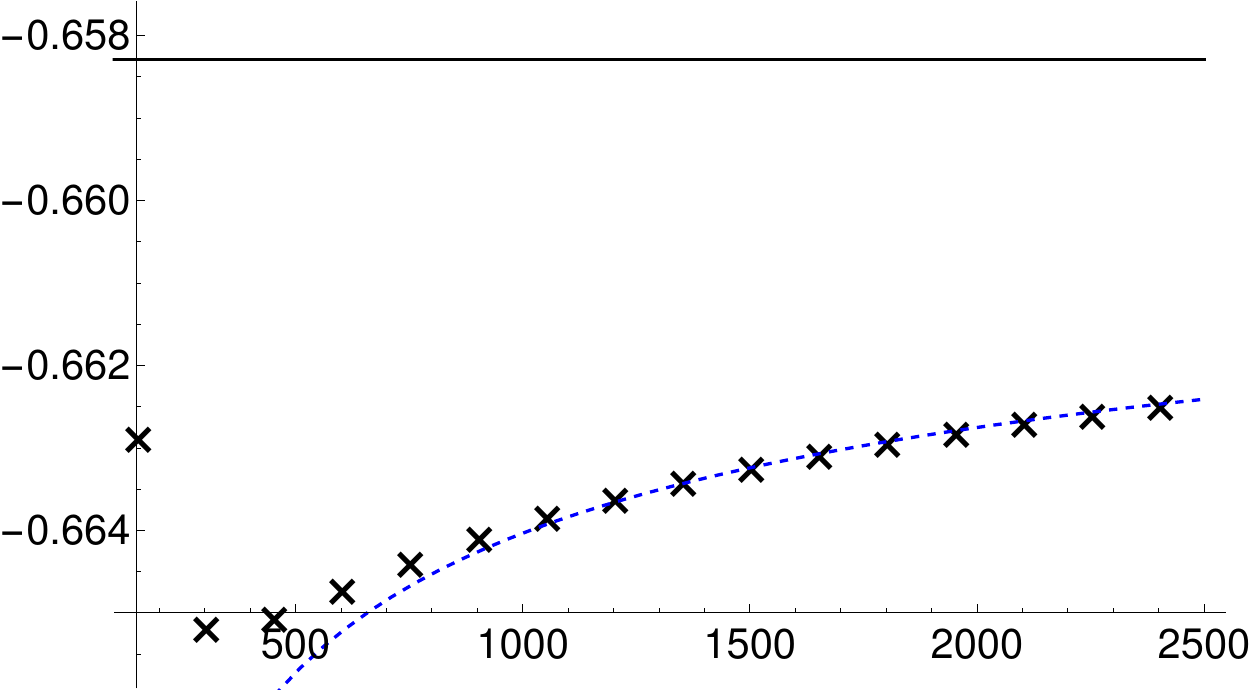}};
\node at (8.5,-6) {\includegraphics[width=7.3cm]{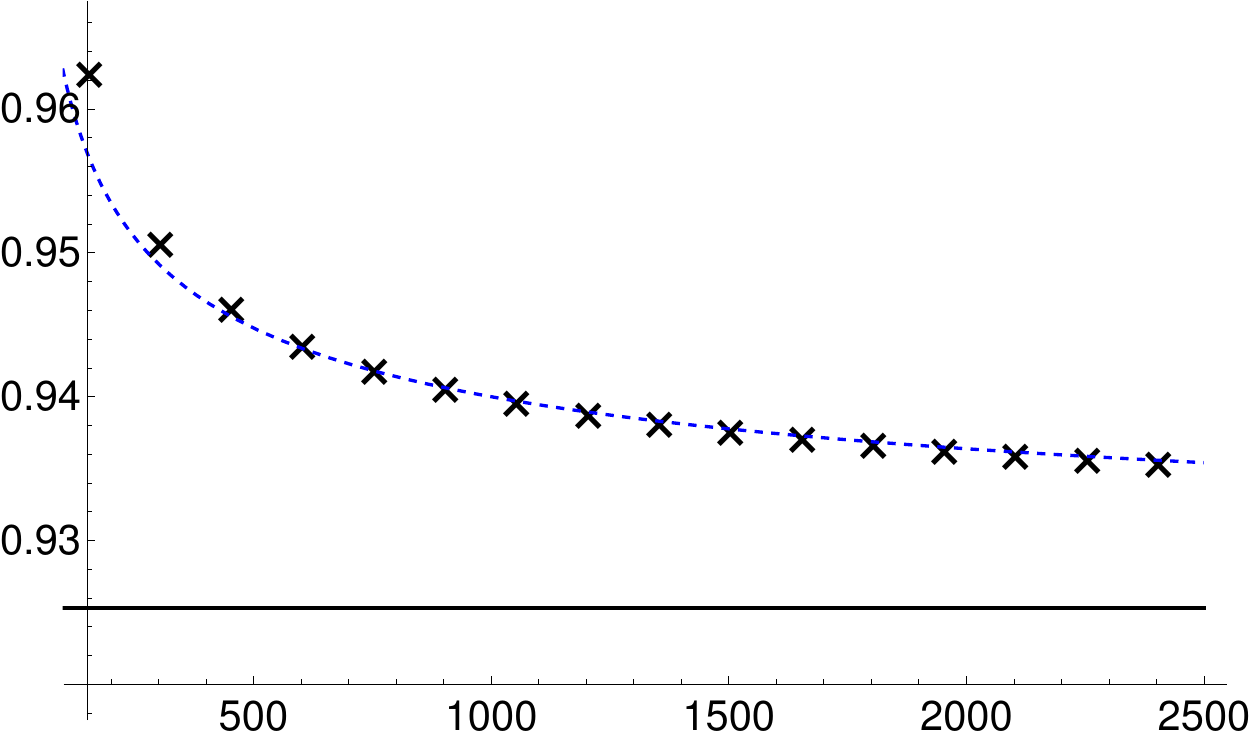}};
\node at (4,-1.75) {\small$N$};
\node at (12.5,-1.75) {\small$N$};
\node at (4,-7.78) {\small$N$};
\node at (12.5,-7.8) {\small$N$};
\node at (-3,2.6) {\small $\Re e\big(b_{+1}^{(N)}\big)$};
\node at (5.5,2.6) {\small $\Im m\big(b_{+1}^{(N)}\big)$};
\node at (-3,-3.4) {\small $\Re e\big(b_{-1}^{(N)}\big)$};
\node at (5.5,-3.4) {\small $\Im m\big(b_{-1}^{(N)}\big)$};
\end{tikzpicture}
}
\end{center}
\bigskip

\caption{\small%
Presented is numerical data for $b_{\pm1}^{(N)}$, defined as the r.h.s. of \eqref{kjsa23j12io} with $m=1$,
where the number of lattice sites $N$ is kept fixed. The data corresponds   
 to the same RG trajectory as in fig.\ref{fig1}, i.e., $r=5$, $\delta=\frac{1}{15}$, 
${\tt k}=\frac{1}{50}$, $S^z=\frac{3}{2}$ $({\tt s}=1)$,
${\tt w}=0$, ${\tt m}=1$, ${\tt L}=2$ and $\bar{\tt L}=1$. Additional details can be found in the  caption to that figure.
The crosses come from the solutions to the Bethe Ansatz equations. The black solid lines represent the limiting values
$b_{+1}=-1.42756-0.13571\,\ri$ and $b_{-1}=-0.65829+0.92530\,\ri$
calculated by means of eq.\,\eqref{aksjhdsfi234} with ${\tt K}=-\frac{7}{25}$, $\bar{\tt K}=-\frac{8}{25}$ as well as the same
 sets $\{(c_i,\varpi_i)\}_{i=1}^5$ and $\{(\bar{c}_i,\bar{\varpi}_i)\}_{i=1}^5$ 
that were used to compute $H_1$ and $\bar{H}_1$ for fig.\,\ref{fig1}. The dashed blue lines are plots of the real/imaginary parts of
$-1.42733-0.13562\,\ri-(0.01428+0.05120\,\ri)\,N^{-0.4}$ for $b_{+1}^{(N)}$ and
$-0.65874+0.92504\,\ri-(0.08395-0.23694\,\ri)\,N^{-0.4}$ for $b_{-1}^{(N)}$, which were obtained 
via a  fit of  the numerical data.
The fit was performed in a rather rough way, using  only the last three points with the largest values of $N$.
\label{fig2}}
\end{figure}

\medskip
The eigenvalues of the quasi-shift operators   for the low energy Bethe states
develop a certain scaling behavior. Their scaling limits are described in terms of the coefficients
that occur in the asymptotic expansion of the spectral determinants $D_+(\mu)$ and  $\bar{D}_+(\bar{\mu})$ 
at large $\mu$ and $\bar{\mu}$, respectively. The reason for this can be traced to the fact that 
$\mathbb{K}^{(\ell)}$
admit a simple expression in terms of the lattice $Q$-operators:
\be
\mathbb{K}^{(\ell)}=\re^{\ri\pi{\tt k}}\,q^{\mathbb{S}^z-\frac{N}{2}}\,
\mathbb{A}_+(-q^{+1}\eta_\ell)\big(\mathbb{A}_+(-q^{-1}\eta_\ell)\big)^{-1}\, ,
\ee
which is just the operator form of eq.\,\eqref{kjsa9823hhg4332}.
 For the ${\cal Z}_r$ invariant  inhomogeneous XXZ spin-$\frac{1}{2}$ chain in the
regime with $\gamma\in(0,\frac{\pi}{r})$, the scaling behavior of ${\cal K}^{(\ell)}$ was
described  in ref.\cite{Kotousov:2023zps}.
In the case under consideration where $r$ is odd and 
$\gamma\in\big(\frac{\pi}{2}(1-\frac{1}{r}),\frac{\pi}{2}(1+\frac{1}{r})\big)$, the 
scaling limit is expressed in terms of the asymptotic coefficients 
$Q_{\frac{2m-1}{r}}$  from the large-$\mu$ expansion \eqref{aysayasy}
and the similar coefficients $\bar{Q}_{\frac{2m-1}{r}}$ from the expansion of  $\bar{D}_+(\bar{\mu})$.
In what follows, we will need their explicit forms:
\begin{subequations}\label{asj23bncjasd}
\bea\label{asj23bncjasdAAA}
Q_{\frac{2m-1}{r}}=
 \frac{\Gamma(1-\frac{(2m-1)\delta}{1+r \delta}) \Gamma(\frac{1}{2}+\frac{(2m-1)\delta}{1+r \delta})}{(2m-1) \sqrt{\pi}}\ \sum_{i=1}^{\tt J}\varpi^{2m-1}_i
\eea
and
\bea
{\bar Q}_{\frac{2m-1}{r}}=
 \frac{\Gamma(1-\frac{(2m-1)\delta}{1+r \delta}) \Gamma(\frac{1}{2}+\frac{(2m-1)\delta}{1+r \delta})}{(2m-1) \sqrt{\pi}}\ \sum_{i=1}^{\tt J}\bar{\varpi}^{2m-1}_i\ ,
\eea
\end{subequations}
where
$m=1,\ldots,\tfrac{r-1}{2}$ and
 $\varpi_i$, $\bar{\varpi}_i$ specify the positions of the  singularities in the function $U(v)$ \eqref{askjdjh21B}
and its barred counterpart (see 
eqs.\,\eqref{askjru92jhsfdh3yyuyA} and 
\eqref{askjru92jhsfdh3yyuyB} from Appendix \ref{AppAb}).
\medskip

Under the ${\cal Z}_r$  transformation the operator $\mathbb{K}^{(\ell)}$ is mapped to $\mathbb{K}^{(\ell+1)}$,
where the superscript $\ell+1$ is understood modulo $r$ (see ref.~\cite{Bazhanov:2020new}
for details).
This provides a hint to consider, instead of the eigenvalues themselves, the Fourier transform of the logarithm of 
 ${\cal K}^{(\ell)}$.
Numerical work shows that the following limits exist:\footnote{%
For the low energy states $\slim_{N\to\infty}{\cal K}^{(\ell)}= (-1)^{\frac{r-1}{2}(\frac{N}{2}-S^z)+{\tt m}+{\tt w} r}
\re^{\frac{\ri\pi }{2}\tilde{ n}}$  with
$\tilde{n}$ being an integer that is related to $n_2$
from eqs.\,\eqref{kjas8923hjds},\,\eqref{hsayat} as
$r\tilde{n}=n_2$ (mod 4).
 However, since 
the leading term in the asymptotics for $\log ({\cal K}^{(\ell)})$  does not depend on $\ell$
(provided the branch of the logarithm is suitably chosen), it
gets cancelled when the sum over $\ell$ in \eqref{kjsa23j12io} is taken and 
gives no contribution to the definition of $b_{\pm m}$.}
\bea\label{kjsa23j12io}
b_{\pm m}&=&\slim_{N\to\infty}\ 
 \bigg(\frac{(1-r\delta)\,N}{r^2N_0}\bigg)^{\frac{2m-1}{r}}
\frac{1}{2r}\ \sum_{\ell= 1}^r\re^{\pm\frac{\ri\pi}{r} (2m-1) (2\ell-1)}\,\log\big({\cal K}^{(\ell)}\big)\,,
\eea
where $m=1,2,\ldots,\frac{r-1}{2}$.
Note that as $m=\frac{r+1}{2}$  the sum in the r.h.s. is
proportional to  the logarithm of the eigenvalue of the $r$-site lattice translation operator, whose 
scaling behavior was already described in formula \eqref{kjas8923hjds}.
We found that the following relations hold true:
\bea\label{aksjhdsfi234}
b_{+m}&=&-\,\re^{\frac{\ri\pi}{2r}(2m-1)}\ 
\cos\big(\tfrac{\pi (2m-1)\delta}{1+r\delta}\big)\ 
Q_{\frac{2m-1}{r}}
\nonumber\\[-0.2cm]
&& \qquad\qquad\qquad\qquad\qquad 
\qquad\qquad\qquad\qquad\qquad \qquad \big(m=1,2,\ldots,\tfrac{r-1}{2}\big)\ .\\[-0.2cm]
b_{-m}&=&+\,\re^{\frac{\ri\pi}{2r}(2m-1)}\ 
\cos\big(\tfrac{\pi (2m-1)\delta}{1+r\delta}\big)\  {\bar Q}_{\frac{2m-1}{r}}\nonumber
\eea
In fig.\,\ref{fig2} we  illustrate the above formulae as well as the rate of convergence of the limit in \eqref{kjsa23j12io}
for the same Bethe state that was used in fig.\,\ref{fig1}.
\medskip

The relations  \eqref{aksjhdsfi234} provide an alternative procedure to the  one discussed in the previous subsection for
identifying the pair of ODEs 
corresponding to a given CFT Bethe state, at least for small values of the conformal levels ${\tt L}$ and $\bar{\tt L}$.

\subsection{Product rules\label{sec63}}
The lowest subleading terms in the large-$\mu$ asymptotic expansion of $D_+(\mu)$ 
includes the coefficient $\mathfrak{C}_+$. Its closed-form expression as a function of the
parameters of the ordinary differential equation is presented and discussed in Appendix \ref{AppAa}.
The asymptotic coefficient $\mathfrak{C}_+$, together with its barred counterpart
$\bar{\mathfrak{C}}_+$, appears in the scaling limit of certain products over the Bethe roots,
such as the eigenvalues of the operator $\mathbb{A}_+^{(\infty)}$ \eqref{8932io90218932}. 
For the case $r=1$, such scaling relations were originally presented in the work \cite{Kotousov:2019ygw}. 
We found that the similar relations hold true for any $r=1,3,5,\ldots\ $, namely,
\bea\label{asj9hj21gds}
\prod_{m=1}^{\frac{N}{2}-S^z}\re^{-\frac{\ri\pi}{2r}(r-1)}\,\zeta_m=
\re^{\frac{\ri\pi}{r}(r-1)\,\mathfrak{j}_{{\tt m}}}\ 
\frac{{\bar  {\mathfrak C}}_+}
{{\mathfrak C}_+}\ 
\bigg(\frac{(1-r\delta)N}{r^2N_0}\bigg)^{ 
 (\frac{1}{r}+\delta)\sqrt{\frac{2r}{1-r\delta}}\,(P_{{\tt w},{\tt m}}-\bar{P}_{{\tt w},{\tt m}})}
\ \big(1+o(1)\big)
\eea
and
\bea\label{asidaa8912hsd}
\prod_{m=1}^{\frac{N}{2}-S^z}
\big(\re^{-\frac{\ri\pi}{2r}(r-1)}\,q\big)
\,\big(\zeta_m+\eta_\ell\, q^{-1}\big)\big(\zeta^{-1}_m+\eta_{\ell'}^{-1} \, q^{-1}\big)&=&
(\eta_\ell)^{-{\mathfrak{j}}_{\tt m}}\,(\eta_{\ell'})^{-\bar{\mathfrak{j}}_{\tt m}}\ 
{\mathfrak C}_+\,{\bar  {\mathfrak C}}_+
 \\[0.2cm]
&\times&
\bigg(\frac{(1-r\delta)N}{r^2N_0}\bigg)^{-(\frac{1}{r}+\delta)\,S^z}
\,\big(2+2r\delta\big)^{\frac{N}{r}}
 \big(1+o(1)\big)\,.\nonumber
\eea
Here $\eta_\ell=(-1)^r\re^{\frac{\ri\pi}{r}\,(2\ell-1)}$, 
$\eta_{\ell'}=(-1)^r\re^{\frac{\ri\pi}{r}\,(2\ell'-1)}$ with arbitrary integers $\ell,\ell'$.
\medskip

The r.h.s of the above relations  involve both asymptotic coefficients $\mathfrak{C}_+$ and 
$\bar{\mathfrak{C}}_+$. 
By taking the ratio/product of \eqref{asidaa8912hsd} and \eqref{asj9hj21gds}
one arrives at
\bea
\prod_{m=1}^{\frac{N}{2}-S^z}
\big(q+\eta_\ell\,\zeta_m^{-1}\big)\big(\zeta^{-1}_m+\eta_{\ell'}^{-1} \, q^{-1}\big)&=&
\re^{-\frac{\ri\pi}{r}(r-1)\,\mathfrak{j}_{{\tt m}}}\,(\eta_\ell)^{-{\mathfrak{j}}_{\tt m}}\,(\eta_{\ell'})^{-\bar{\mathfrak{j}}_{\tt m}}\ 
({\mathfrak C}_+)^2
 \\[0.2cm]
&\times&
\bigg(\frac{(1-r\delta)N}{r^2N_0}\bigg)^{ 
 -2(\frac{1}{r}+\delta)\sqrt{\frac{2r}{1-r\delta}}\,P_{{\tt w},{\tt m}}}
\,\big(2+2r\delta\big)^{\frac{N}{r}}
 \big(1+o(1)\big)\nonumber
\eea
and
\bea
\prod_{m=1}^{\frac{N}{2}-S^z}
\big(q+\eta_{\ell'}^{-1} \, \zeta_m\big)
\,\big(\zeta_m+\eta_\ell\, q^{-1}\big)&=&
\re^{\frac{\ri\pi}{r}(r-1)\,\bar{\mathfrak{j}}_{{\tt m}}}\,(\eta_\ell)^{-{\mathfrak{j}}_{\tt m}}\,(\eta_{\ell'})^{-\bar{\mathfrak{j}}_{\tt m}}\ 
(\bar{{\mathfrak C}}_+)^2
 \\[0.2cm]
&\times&
\bigg(\frac{(1-r\delta)N}{r^2N_0}\bigg)^{ 
 -2(\frac{1}{r}+\delta)\sqrt{\frac{2r}{1-r\delta}}\,\bar{P}_{{\tt w},{\tt m}}}
\,\big(2+2r\delta\big)^{\frac{N}{r}}
 \big(1+o(1)\big)\,.\nonumber
\eea
These last two formulae are convenient for numerical checks since they involve the asymptotic coefficients for the
spectral determinants $D_+(\mu)$ and $\bar{D}_+(\bar{\mu})$, separately.
\medskip

The product rules   were verified numerically 
for many examples. This provides further  evidence in support of the conjecture regarding the scaling
limits \eqref{jkas8912jhas} and \eqref{jkas8912jhasB} of the lattice $Q$-functions. 

\section{Enumerating   solutions of the algebraic system \label{sec7}}

This section is devoted to a  discussion of some  mathematical properties
of the solutions to the 
algebraic system that determines the position of the apparent singularities and their
residues entering   the ODE. 
In order to formulate them without referring the reader to the opening parts of the paper
we recall the system here:
\bea \label{ask9823jhjhpsa}
2c_j&=&\varpi_j^r-1+2\delta\sum_{i\ne j}\frac{\varpi_i\varpi_j}{\varpi_j^2-\varpi_i^2}\\[0.2cm]
c_j^2&=& {\tt K}^2+\delta c_j-\frac{\delta^2}{4}+\sum_{i\ne j}\bigg(
\frac{3\delta^2\varpi_i^2\varpi_j^2}{(\varpi_i^2-\varpi_j^2)^2}+\frac{2\delta\,c_i\varpi_j^2}{\varpi_j^2-\varpi_i^2}\bigg)
\qquad\qquad (j=1,2,\ldots,{\tt J})\, .\nonumber
\eea
Throughout this discussion, we do not distinguish between the solution
set $\{(\varpi_i,c_i)\}_{i=1}^{{\tt J}}$ and the one obtained via any
 permutation $\sigma$ of its elements: $(\varpi_i,c_i)\mapsto  (\varpi_{\sigma(i)},c_{\sigma(i)})$.
Then our conjecture regarding the ODE/IQFT correspondence supposes that
  the total number of solutions of  \eqref{ask9823jhjhpsa}, subject to the additional constraint
\be \label{9023jdjnkjsdd}
\sum_{i=1}^{{\tt J}} c_i=\mathfrak{j}\,{\tt K}+\frac{\mathfrak{j}^2}{2}\ \delta
\qquad\qquad
{\rm with}
\qquad\qquad
\mathfrak{j}=0,1,\ldots, r-1\ ,
\ee
is given by ${\cal D}_{\mathfrak{j}}{(\frac{{\tt J}-\mathfrak{j}}{2})}$, where the integers
${\cal D}_{\mathfrak{j}}{({\tt L})}$  coincide 
with the dimensions of the level ${\tt L}$ subspace of ${\cal H}_{\mathfrak{j},P}$ --- the irrep  of the 
 Kac-Moody algebra  $\widehat{\mathfrak{su}}_1(r)\oplus\widehat{\mathfrak{u}}(1)$.
\medskip


The integers ${\cal D}_{\mathfrak{j}}{(\frac{{\tt J}-\mathfrak{j}}{2})}$
  admit a simple combinatorial interpretation. They coincide with the number
of ordered sets
\be\label{kajs89jdhj}
\{\mathfrak{n}^\pm_{i,a}\}_{i=1}^{{M}^\pm_a}\, :\qquad\qquad
1\le \mathfrak{n}^\pm_{1,a}<\mathfrak{n}^\pm_{2,a}<\ldots<\mathfrak{n}^\pm_{M^\pm_a,a}\,,
\ee
satisfying the constraints:
\be\label{kjas23jbhjsiudaas}
\sum_{a=1}^r (M_a^--M_a^+)=\mathfrak{j}\,,\qquad \qquad \qquad
\sum_{a=1}^{r}\bigg(\sum_{j=1}^{M_a^-}\big(2\mathfrak{n}_{j,a}^- -1)+
\sum_{j=1}^{M_a^+}\big(2\mathfrak{n}_{j,a}^+ - 1)\bigg)
={\tt J}\ .
\ee
Notice that  ${\cal D}_{\mathfrak{j}}{(\frac{{\tt J}-\mathfrak{j}}{2})}$ does not vanish only if 
$\mathfrak{j}$ obeys the conditions
\be\label{aj8923hjhgh21}
\mathfrak{j}={\tt J}\ \ ({\rm mod}\ 2)\,,\qquad\qquad\qquad
0\le \mathfrak{j}\le {\tt J}\,.
\ee
The combinatorial structure suggests that there is a natural way of assigning to each solution 
$\{(\varpi_i,c_i)\}$
certain  sets $\{\mathfrak{n}^+_{i,a}\}$ and $\{\mathfrak{n}^-_{i,a}\}$. 
The key idea behind this identification is 
the assumption that any solution $\{(\varpi_i,c_i)\}$ with  $|\delta|<\frac{1}{r}$
can be obtained  from a continuous deformation in the parameter $\delta$ 
of a solution
at  $\delta=0$. In the latter case the algebraic system simplifies considerably,   and the  correspondence with the sets
$\{\mathfrak{n}^\pm_{i,a}\}$ becomes straightforward to establish.
Before outlining  the general  procedure, we first illustrate it on a few simple examples, specifically for the cases 
 ${\tt J}=1,2$,  where  the 
solutions of  \eqref{ask9823jhjhpsa}
admit an especially transparent    description.
\medskip

\subsection{Solutions of the algebraic system for ${\tt J}=1,2$\label{sec71}}
The simplest case is when ${\tt J}=1$. Then,
as follows from 
the conditions \eqref{aj8923hjhgh21},
 only $\mathfrak{j}=1$ is possible 
and the number of corresponding solutions of \eqref{ask9823jhjhpsa} is ${\cal D}_{1}{(0)}=r$. The latter coincides
with the dimensions of the  first fundamental  (defining) irrep of the Lie algebra
$\mathfrak{su}(r)$.
The algebraic system \eqref{ask9823jhjhpsa} together with \eqref{9023jdjnkjsdd} 
reduces to  elementary equations for $\varpi\equiv \varpi_1$ and $c\equiv c_1$,
whose solutions, labeled by $a=1,2,\ldots,r$,
are given by
 \be\label{0923uhdfjj32}
\varpi=\re^{\frac{2\pi\ri}{r}a}\ \big(1+2 {\tt K}+\delta\big)^{\frac{1}{r}}\ ,\ \ \ \ \ \
\qquad  \ \ c= {\tt K}+\frac{\delta}{2}\ .
 \ee
It follows from \eqref{kjas23jbhjsiudaas} that $\{\mathfrak{n}^+_{j,b}\}=\emptyset$ for $b=1,2,\ldots,r$,
while $\{\mathfrak{n}^-_{j,b}\}$ with $b\ne a $ are empty and $\{\mathfrak{n}^-_{j,a}\}=\{1\}$.
\medskip

For the case ${\tt J}=2$ the integer $\mathfrak{j}$ may take the values zero or two. 
Let's first set ${\mathfrak{j}}=2$. The number of solutions of eqs.\,\eqref{ask9823jhjhpsa} and 
\eqref{9023jdjnkjsdd} should coincide with ${\cal D}_{2}(0)=\frac{r(r-1)}{2}$, which matches
the dimension of the second fundamental irrep of $\mathfrak{su}(r)$. A straightforward analysis of the system
leads to the equation for
$w\equiv \frac{\varpi_1^2+\varpi_2^2}{2\varpi_1\varpi_2}$:
  \bea\label{kjsa98hhhg23}
(1-\kappa)\ U_{\frac{r-1}{2}}(w)+ (1+\kappa)\ U_{\frac{r-3}{2}}(w)=0\ .
 \eea
Here we also use the shortcut notation 
\bea
\kappa= \frac{\delta}{1+2{\tt K}+2\delta}
\eea
and $U_n(w)$ stands for the Chebyshev polynomials of the second kind
(for their definition see eq.\,\eqref{8932haajho3}, below).
Provided $\kappa$ is sufficiently small, eq.\,\eqref{kjsa98hhhg23} possesses $\frac{r-1}{2}$
distinct solutions 
$w=w_b$, with $b=1,2,\ldots,\frac{r-1}{2}$, distinguished by the condition
\be
\lim_{\kappa\to 0} w_b=\cos\Big(\frac{2\pi b}{r}\Big) \ .
\ee
Given  $w_b$  one reconstructs  $\varpi_1,\varpi_2$  and $c_1,c_2$  as
 \be
\arraycolsep=0.5cm
\begin{array}{ll}
  \varpi_1=\re^{\frac{2\pi\ri}{r} a_1}\ \big( 1+2{\tt K}+2\delta+\ri   f_{a_1-a_2}\,\delta\, \big)^{\frac{1}{r}}\ ,& 
  \varpi_2= \re^{\frac{2\pi\ri}{r} a_2}\ \big( 1+2{\tt K}+2\delta-\ri   f_{a_1-a_2}\,\delta\, \big)^{\frac{1}{r}}
   \\[0.3cm]
  c_1= 
{\tt K}+\delta-\frac{\ri}{2}\, g_{a_1-a_2}\,\delta \ ,&
 c_2=
{\tt K}+\delta+\frac{\ri}{2}\,  g_{a_1-a_2}\,\delta
\end{array}\!\!\!\!\!\!\!\!\!\!\,,
  \ee
where 
\bea
f_b= \sqrt{\frac{1-w_{b}}{1+w_{b}}}=\tan\Big(\frac{\pi b}{r}\Big)+O(\delta)\ ,\ \ \ \ \ \ \ \ \qquad
g_b=\frac{w_b}{\sqrt{1-w^2_b}}=\cot\Big(\frac{2\pi b}{r}\Big)+O(\delta)\  ,
\eea
while the integers $a_1$ and $a_2$ satisfy
\bea
1\leq a_2<a_1\leq r-1\ .
\eea
For these solutions, all sets $\{\mathfrak{n}^+_{j,c}\}$ are empty. Among the sets
$\{\mathfrak{n}^-_{j,c}\}$, only those with $c=a_1$ and $c=a_2$ are non-empty  
and  $\{\mathfrak{n}^-_{j,a_1}\}=\{\mathfrak{n}^-_{j,a_2}\}=\{1\}$.
\medskip

For ${\tt J}=2$ and $\mathfrak{j}=0$ there should be 
${\cal D}_0(1)=r^2$ solutions. 
This  coincides with the  number of spin-$1$ local fields in the chiral component 
of the algebra of extended conformal symmetry.
A perturbative analysis of the algebraic system  yields
\bea\label{ahsysayas}
 &&\varpi_1=\re^{\frac{2\pi\ri }{r} a}\ \big(1+2 {\tt K}+f\, \delta+O(\delta^2)\big)^{\frac{1}{r}} \ ,\ \ \ \ \
 \varpi_2=\re^{\frac{2\pi\ri }{r} b}\ \big(1-2 {\tt K}- f\,\delta+O(\delta^2)\big)^{\frac{1}{r}} \nonumber\\[0.1in]
&& c_1=-c_2={\tt K}+\tfrac{\ri}{2}\,f\, \delta +O(\delta^2)
 \eea
 with  $a,b=1,2,\ldots, r$ and
 \bea
 f=\frac{(1-2 {\tt K})^{\frac{1}{r}}\, \re^{\frac{2\pi\ri }{r} (b-a)}+(1+2 {\tt K})^{\frac{1}{r}} \,\re^{\frac{2\pi\ri }{r} (a-b)}}
 {(1-2 {\tt K})^{\frac{1}{r}} \,\re^{\frac{2 \pi \ri  }{r} (b-a)}-(1+2 {\tt K})^{\frac{1}{r}}\, \re^{\frac{2\pi\ri }{r}(a- b)}}\ .
 \eea
The   solutions involving infinitesimal $\delta$ can be continued to small but finite $\delta$ via  the following expressions:
 \bea\label{gasasar}
 &&\varpi_1=\rho \, \varpi_2\ ,
 \ \ \ \ \ \ 
 \varpi_2= \Big(\frac{2}{1+\rho^{r}}\Big)^{\frac{1}{r}}\\[0.1in]
 &&c_1=-c_2=\frac{1}{2}\ \frac{\rho^r-1}{\rho^r+1}+\delta\ \frac{\rho}{\rho^2-1}\ .\nonumber
 \eea
 Here
 \bea
 \rho=\big(\sqrt{t_*}\pm\sqrt{t_*-1}\big)^{2}
 \eea
and  $t_*$ is  a  root of the polynomial equation in $t$ of order $r$,
\bea\label{aisaiisa}
 (t-1)\ U^2_{r-1}\big(\sqrt{t}\big)+2\delta\ \sqrt{t}\, T_r(\sqrt{t})\,  U_{r-1}\big(\sqrt{t}\big)=
 (4 {\tt K}^2-\delta^2)\ T_r^2\big(\sqrt{t}\big)-\delta^2\ \frac{T^2_r\big(\sqrt{t}\big)}{t}\,,
 \eea
written using the  Chebyshev polynomials of the first and second kind
 \bea\label{8932haajho3}
 T_n\big(\cos(\theta)\big)=\cos(n\theta)\ ,\ \ \ \ \ \qquad
\ \ \ \ U_n\big(\cos(\theta)\big)=\frac{\sin\big((n+1)\theta\big)}{\sin(\theta)}\ .
 \eea
For generic values of the parameters, equation \eqref{aisaiisa} possesses  $r$ distinct  roots and the   fractional power  $(\ldots)^{\frac{1}{r}}$ in \eqref{gasasar} defines an $r$-valued function.
 Thus, there are  indeed $r^2$ solutions, which  can be  labeled by integers $a,b=1,2,\ldots, r$,  as in \eqref{ahsysayas}.
For these solutions
\be\label{sak9823jh899021}
\lim_{\delta\to 0}\varpi_1=\re^{\frac{2\pi\ri }{r} a}\ \big(1+2 {\tt K}\big)^{\frac{1}{r}}\,,\qquad\qquad
\lim_{\delta\to 0}\varpi_2=\re^{\frac{2\pi\ri }{r} b}\ \big(1-2 {\tt K}\big)^{\frac{1}{r}}\ .
\ee
We will assign $\{\mathfrak{n}^-_{j,a}\}= \{\mathfrak{n}^+_{j,b}\}=\{1\}$, while all the other
sets $\{\mathfrak{n}^\pm_{j,c}\}$ are empty.

\subsection{Solutions of the algebraic system in the vicinity of $\delta=0$}
When the number of apparent singularities ${\tt J}$ exceeds two, 
the solutions to the system of equations \eqref{ask9823jhjhpsa}  do not 
possess a tractable analytic description so we resorted to a numerical study.
It was observed that the  solutions can be distinguished by the limiting values of
$\varpi_i$ at $\delta=0$. For given $\varpi_i$, this always takes the form
 \bea\label{sajjhghhdsv}
\lim_{\delta\to 0} \varpi_i=\sigma \,\re^{\frac{2\pi\ri}{r}  a}\ \big(2\mathfrak{n}^{\pm}_{j,a}-1\mp 2{\tt K}\big)^{\frac{1}{r}}\,,
\eea
where the sets of integers $\{\mathfrak{n}_{j,a}^\pm\}_{j=1}^{M^\pm_a}$  \eqref{kajs89jdhj} obey the conditions 
\eqref{kjas23jbhjsiudaas} and $\sigma=\pm$ is a sign factor. 
For the cases ${\tt J}=1,2$, discussed above, the sets $\{\mathfrak{n}_{j,a}^\pm\}$ 
were either
always empty or coincided with $\{1\}$ and $\sigma=+$ 
(see eq.\,\eqref{sak9823jh899021} and the $\delta\to 0$ limit of \eqref{0923uhdfjj32} and \eqref{ahsysayas}). 
In general,  $\varpi_{i_1}^2$ and $\varpi_{i_2}^2$
with $i_1\ne i_2$ from the solution $\{(\varpi_i,c_i)\}$, though they should be distinct at finite $\delta$
(otherwise eq.\,\eqref{ask9823jhjhpsa} becomes singular\footnote{%
In the derivation of the algebraic system \eqref{ask9823jhjhpsa} as the conditions specifying
the location of the apparent singularities of the ODE \eqref{askjdjh21A}, \eqref{askjdjh21B}, it is assumed that all $\varpi_i^2$
are distinct.}), may nevertheless converge to  the same limit as $\delta\to 0$.
The number of $\varpi_i$ with given limiting value \eqref{sajjhghhdsv} 
turns out to coincide with $\mathfrak{n}_{j,a}^\pm$ for $\sigma=+$ and 
$\mathfrak{n}_{j,a}^\pm-1$ for $\sigma=-$. This way, we expect that for sufficiently small
$\delta$ every solution   $\{(\varpi_i,c_i)\}$ can be assigned the  integers  $\{\mathfrak{n}_{j,a}^\pm\}$
according to \eqref{sajjhghhdsv}.
Moreover, for a given solution,
$\varpi_i$ and $c_i$ may be labelled as 
\be\label{akjs98rr232323jhhjdsA}
\varpi_i\equiv\varpi_{j,a|\beta}^\pm\,,\qquad\qquad\qquad \qquad  c_i\equiv c_{j,a|\beta}^\pm
\ee
so that
 \bea\label{akjs98rr232323jhhjdsB}
\lim_{\delta\to 0} \varpi_{j,a|\beta}^\pm=(-1)^{\beta-1}\  \re^{\frac{2\pi\ri}{r}  a}\ \big(2\mathfrak{n}^{\pm}_{j,a}-1\mp 2{\tt K}\big)^{\frac{1}{r}}\ \ \ \ \ \  \ \ \ \ \ {\rm and}\ \ \ \ \ \ \ 
\ \ \ \ \ \beta=1,\,\ldots,\,  2\mathfrak{n}_{j,a}^\pm-1 \ ,
\eea
where the integer $\beta$ is used to distinguish between the $\varpi_i$ belonging to each degenerate group.
 For example, in the notation \eqref{akjs98rr232323jhhjdsA},\,\eqref{akjs98rr232323jhhjdsB},  $\varpi_1$ and $\varpi_2$ 
from  formula \eqref{sak9823jh899021}  become
$\varpi_{1,a|1}^-$ and $\varpi_{1,b|1}^+$,
respectively.

\medskip

So long as all the limiting values of $\varpi_i^2$ for $\delta\to 0$ are different, 
there is no issue in solving the system of equations \eqref{ask9823jhjhpsa}
as a formal series in $\delta$.
Then one may expect that the series possesses a finite radius of convergence so that 
$\{(\varpi_i,c_i)\}$ admits an unambiguous continuation to small and finite $\delta$,
as is the case for the examples with ${\tt J}=1,2$ discussed above. In the presence of
degeneracies of $\lim_{\delta\to 0}\varpi_i^2$ the analysis is more subtle, akin to degenerate perturbation theory in quantum mechanics.
Below we argue that the degeneracy occurring in the $\delta\to0$ limit
is completely lifted at the lowest perturbative order.
\medskip

A brief inspection of the algebraic system \eqref{ask9823jhjhpsa} leads one to conclude that, as 
in quantum mechanics, at  the  lowest  perturbative order one can focus on $\varpi^{\pm}_{j,a|\beta}$
belonging to the same degenerate group. This way, the indices $j$ and $a$ will be kept fixed, while the 
different members are to be distinguished by the integer $\beta$. Focusing on, say, $\varpi^{-}_{j,a|\beta}$
 we assume the following perturbative structure
 \bea\label{asjk89jhhjd3}
 \varpi^{-}_{j,a|\beta}&=&(-1)^{\beta-1}\  \re^{\frac{2\pi\ri}{r}a}\ \big(2\mathfrak{n}-1+2{\tt K}\big)^{\frac{1}{r}}\ \big(1+
 \varepsilon_\beta\,\delta+O(\delta^2)\big)\nonumber\\[0.1in]
 c^{-}_{j,a|\beta}&=&C_\beta+O(\delta)\ \ \ \ \ \ \ \ \  \ \ \ \ \ \ \ \ \ (\beta=1,2,\ldots,2\mathfrak{n}-1)\ ,
 \eea
 where, to lighten notation, we use $\mathfrak{n}\equiv\mathfrak{n}_{j,a}^{-}$. Substituting this into 
eq.\eqref{ask9823jhjhpsa} yields
 \bea
 2C_\beta&=&(-1)^{\beta-1}\ (2 \mathfrak{n}-1+2{\tt K})-1+  \sum_{\alpha\not=\beta}\frac{(-1)^{\alpha+\beta}}{\varepsilon_\beta-\varepsilon_\alpha}\\
 C_\beta^2&=&{\tt K}^2+\sum_{\alpha\not=\beta}\bigg(\frac{3}{4(\varepsilon_\beta-\varepsilon_\alpha)^2}+\frac{C_\alpha}{\varepsilon_\beta-\varepsilon_\alpha}\bigg)\ \ \ \ \ \ \ \ \ \qquad \big(\alpha,\beta=1,2,\ldots,2\mathfrak{n} -1)\ .\nonumber
 \eea
Excluding $C_\beta$  one arrives at a closed system of equations for $\varepsilon_\beta$:
\bea\label{asgsatsata}
 \sum_{\alpha\not=\beta}\sigma_{\alpha+\beta}\ 
 \bigg(\frac{1}{(\varepsilon_\beta-\varepsilon_\alpha)^2}-\frac{1}{\varepsilon_\beta-\varepsilon_\alpha}\bigg)
 =\big(\mathfrak{n}-\sigma_\beta\big)^2+2\, \big(\mathfrak{n}-\sigma_\beta\big) \, {\tt K}
 \eea
 with
 \bea
 \sigma_\alpha=\begin{cases}
 1\ \ \ \  \ \ &{\rm for}\ \ \ \ \alpha-{\rm odd}\\
 0\ \ \ \  \ \ &{\rm for}\ \ \ \ \alpha-{\rm even}\
 \end{cases}\,.
 \eea
Note that, working at the lowest perturbative order, we can only recover the values of the differences
 $\varepsilon_\beta-\varepsilon_\alpha$. The determination of  $\varepsilon_\alpha$ themselves requires
an analysis at the next order in $\delta$. The computation becomes more subtle since not only
$\varpi^{\pm}_{j,a|\beta}$ within the same degenerate group ``interact'', but  also those labelled by 
different $j$, $a$ and superscript $\pm$ are coupled with one another. 
We do not go into this here.
\medskip

Together with the invariance w.r.t. an additive shift 
$\varepsilon_\alpha\mapsto \varepsilon_\alpha+{\rm const}$,
the system \eqref{asgsatsata} is unchanged under the permutations of the elements 
within each of the sets 
$\{\varepsilon_{2m}\}_{m=1}^{\mathfrak{n}-1}$ and
$\{\varepsilon_{2l-1}\}_{l=1}^{\mathfrak{n}}$.
A numerical analysis shows that for ${\tt K}>0$ it
admits a unique solution, up to the symmetry transformations. It can be chosen such that 
$\varepsilon_{\alpha+1}-\varepsilon_\alpha$ with $\alpha=1,2,\ldots,2\mathfrak{n}-2$ turn out to be positive.
For example, as $\mathfrak{n}=2$ one has
 \bea
 \varepsilon_2-\varepsilon_1=\frac{1}{2(1+2{\tt K})}\ \big(\sqrt{5+8{\tt K}}+1\big)\ ,\ \ \ \ \ \ \ \ \ \ 
 \varepsilon_3-\varepsilon_2=\frac{1}{2(1+2{\tt K})}\ \big(\sqrt{5+8{\tt K}}-1\big)\ .
 \eea
For generic ${\tt K}$ we expect that the solutions of the system \eqref{asgsatsata} 
are such that all the differences $\varepsilon_\beta-\varepsilon_\alpha$ for $\beta\ne\alpha$ are non-vanishing.
Furthermore, all solutions are related to each other via permutations and the additive overall shift of $\varepsilon_\alpha$. 
In view of the equation \eqref{asjk89jhhjd3} for $\varpi^{\pm}_{j,a|\beta}$, this would imply that
the degeneracy is  resolved at the lowest perturbative order.

\section{The case $r=1$\label{sec8}}
Despite that the differential equation \eqref{askjdjh21A}-\eqref{kasj892jhsa} with $r=1$ 
looks different to the Schr\"{o}dinger equation with monster potential \eqref{sa903hjbdf32},\,\eqref{jssysys}, 
the  corresponding spectral determinants turn out to be the same. This suggests that the two classes of ODEs are related via 
changes of variables, which do not affect their monodromy properties. Here we briefly discuss  these
transformations.

\medskip
For the vacuum case, the change of variables that brings  the Schr\"{o}dinger equation 
for the anharmonic oscillator \eqref{mnasghuiqw}   to the ODE \eqref{as9021jjbnsd}  (i.e., \eqref{askjdjh21A},\,\eqref{askjdjh21B}
with ${\tt J}=0$)
was the starting point of our discussion. The explicit formulae are given by
eqs.\,\eqref{sa0923jdsjhA},\,\eqref{sa0923jdsjhB} and \eqref{askj892h2aaa3gh8721}. In particular, for $r=1$ the parameters 
of the differential equations are related as 
\be\label{ks8djhwebjjdAA}
\alpha=\frac{1+\delta}{1-\delta}\,,\,\qquad\qquad \ell+\frac{1}{2}=\frac{2{\tt K}}{1-\delta}\,
\qquad\qquad E=-\Big(\frac{2}{1-\delta}\Big)^{1+\delta}\mu\ .
\ee
It follows from our discussion in sec.\,\ref{sec5}
that as  ${\tt J}=1$   the ODE 
\eqref{askjdjh21A}-\eqref{kasj892jhsa} can be reduced  to the vacuum equation. Indeed, the
integer ${\tt M}$  \eqref{9023jdjnkjsd}  may take only the values $\pm1$ and, in view of eq.\,\eqref{kjas23jkji90i32} 
specialized to $r=1$,
 the ODE with ${\tt J}=1$, ${\tt M}=\pm1$ and ${\tt K}$
is  mapped to the one with ${\tt J}={\tt M}=0$ and ${\tt K}$  substituted by ${\tt K}\pm 1$.
The  transformation has the form \eqref{sak90kj21jh3290A}, where for ${\tt M}=+1$
the functions $F_1$ and $F_2$ entering therein follow from the results presented in Appendix \ref{AppB}. 
\medskip

The map \eqref{kjas23jkji90i32} allows one to bring the general ODE \eqref{askjdjh21A}-\eqref{kasj892jhsa}
to the equation of the same form, but with ${\tt J}$ an even number and ${\tt M}=0$, i.e.,
\be\label{as8923hjshghsd}
\sum_{i=1}^{{\tt J}}c_i=0\,,\qquad\qquad 
{\tt J}=2{\tt L}\ \qquad\qquad  ({\tt L}=0,1,2,\ldots)\ .
\ee
In this case, the number of solutions to the algebraic system  \eqref{kasj892jhsa}  coincides with 
the integer ${\cal D}_0({\tt L})$, which is defined via its generating function in eq.\,\eqref{asjh32h98sd}.
As $r=1$ the sum over ${\tt w}$ in the r.h.s. of that formula
may be explicitly taken resulting in a Jacobi  theta function. This, combined  with the Jacobi triple product formula,
yields
\be
\prod_{n=1}^\infty\frac{1}{1-{\tt q}^n}=
\sum_{{\tt L}=0}^\infty
 {\tt q}^{{\tt L}}\ 
{\cal D}_{0}{({\tt L})}\qquad\qquad\qquad (r=1)\, .
\ee
Hence the number of  solutions to the algebraic system \eqref{kasj892jhsa}, subject to the additional constraint
\eqref{as8923hjshghsd},   is given by
the number of partitions of ${\tt L}=\frac{{\tt J}}{2}$ into positive integers. This coincides with the number of solutions
$\{z_j\}_{j=1}^{{\tt L}}$ 
of  \eqref{jssysys} for the monster potentials  \cite{Bazhanov:2003ni,Masoero}. We conjecture that  
there is a one-to-one correspondence between the solutions of  the algebraic systems \eqref{jssysys} and 
 \eqref{kasj892jhsa},\,\eqref{as8923hjshghsd} such that
the ODE 
 \eqref{askjdjh21A}-\eqref{kasj892jhsa}  can be brought to the form of the Schr\"{o}dinger equation with monster potential
 \eqref{sa903hjbdf32} via a change of variables that preserves the monodromy properties.
While the general relation between the solution sets $\{z_j\}$ and $\{(\varpi_i,c_i)\}$
along with a closed-form expression for the change of variables is unavailable,
we have verified the conjecture for  ${\tt J}=2,4,6$. 
\medskip

For ${\tt J}=2$ the algebraic system   \eqref{kasj892jhsa} 
 supplemented by the condition $c_1+c_2=0$ 
possesses  a unique  solution (up to permutations) $\varpi_1=\varpi_+$, $\varpi_2=\varpi_-$ and 
$c_1=-c_2=c$, where
\be
\varpi_{\pm}=1\pm\sqrt{\frac{4{\tt K}^2-2\delta\,(1+\delta)}{1-\delta^2}}\,,\qquad\qquad
c=\frac{(2-\delta)\,{\tt K}^2-\frac{1}{4}\,\delta\,(1+\delta)\,(3-\delta)}{
\sqrt{(1-\delta^2) (4{\tt K}^2-2\delta\,(1+\delta))}}\ .
\ee
It is straightforward to check that if $\psi$ solves the differential equation
\eqref{askjdjh21A},\,\eqref{askjdjh21B} with $\{(\varpi_i,c_i)\}_{i=1}^{2}$ as described above,  then
$\tilde{\psi}$ defined through the relation
\bea\label{dsjfu823huhbsdj}
\tilde{\psi}&=&\frac{1}{(\re^{2v}-w)\sqrt{(\epsilon^2-\varpi_1^2)(\epsilon^2-\varpi_2^2)}}\ \bigg(
\Big(2-\frac{1-\delta}{w}\,\epsilon\,\re^v\Big)\,
\bigg(\partial_v -\sum_{i=1}^2\frac{\delta\,\varpi_i^2}{%
\epsilon^2-\varpi_i^2}\bigg)
\nonumber\\[0.2cm]
&+&1-3\delta+\frac{1-\delta^2}{2w}\,\epsilon\,\re^v
+\frac{1-\delta}{2w}\,\big(\re^{2v}-w\big)\,\Big(\epsilon^2-\frac{4w}{(1-\delta)^2}\Big)\bigg)\,\psi\,,
\eea
where
$
w=-\frac{(1-\delta)^2}{4}\,\varpi_1\varpi_2
$,
satisfies
\be\label{askj932jsd}
\bigg(- \partial_v^2 +{\tt K}^2+\re^{2v}+\epsilon\,\re^v+
\frac{8\re^{4v}}{(\re^{2v}-w)^2}-
\frac{2\,(3+\delta)\,\re^{2v}}{\re^{2v}-w}\,\bigg)\,\tilde{\psi}=0\,.
\ee
Through the change of variables $x=\big(\frac{2}{1-\delta}\,\re^v\big)^{\frac{1-\delta}{2}}$,
$\Psi=\big(\frac{2}{1-\delta}\,\re^v\big)^{\frac{1-\delta}{4}}\tilde{\psi}$
and parameters  \eqref{ks8djhwebjjdAA}, the last equation
becomes  the Schr\"{o}dinger equation with monster potential 
\eqref{sa903hjbdf32}, where ${\tt L}=1$ and
\be
z_1=
-\varpi_1\varpi_2
\ee
 is the unique solution to \eqref{jssysys}.
\medskip

For the cases ${\tt J}=4,6$ the explicit form of the transformation taking the ODE \eqref{askjdjh21A}-\eqref{kasj892jhsa}
to the Schr\"{o}dinger equation
is rather cumbersome and will not be presented here. However, our analysis suggests that eq.\,\eqref{dsjfu823huhbsdj}
for ${\tt J}=4,6,\ldots$
 is generalized to 
\be
\tilde{\psi}=\bigg(\prod_{j=1}^{\frac{{\tt J}}{2}}\re^{2v}-\frac{(1-\delta)^2}{4}\, z_j\bigg)^{-1}\bigg(\,
\prod_{i=1}^{{\tt J}}\epsilon^2-\varpi_i^2\bigg)^{-{\frac{1}{2}}}\,\bigg(P_1(\epsilon,\re^{v})\,
\bigg(\partial_v-\sum_{i=1}^{\tt J}\frac{\delta\,\varpi_i^2}{\epsilon^2-\varpi_i^2}\bigg)+
P_2(\epsilon,\re^{v})\bigg)\,\psi\ .
\ee
Here 
 $\{z_j\}_{j=1}^{\frac{{\tt J}}{2}}$  is the  solution set of \eqref{jssysys} with ${\tt L}=\frac{{\tt J}}{2}$
corresponding to the particular solution 
$\{(\varpi_i,c_i)\}_{i=1}^{{\tt J}}$ of the system \eqref{kasj892jhsa},\eqref{as8923hjshghsd}. Also,
 $P_1$ and $P_2$   are polynomials in $\epsilon$ of the form
\be
P_1=\sum_{n=0}^{{\tt J}-1} P_{1,n}(\re^{v})\,\epsilon^{n}\,,\qquad\qquad
P_2=\sum_{n=0}^{{\tt J}} P_{2,n}(\re^{v})\,\epsilon^{n}\ .
\ee
In their turn, $P_{1,n}$ are polynomials in $\re^v$ of order
${\tt J}-1$ for $n$ odd and ${\tt J}-2$ for $n$ even, while
$P_{2,n}$ 
is a  polynomial in $\re^{v}$ of order ${\tt J}$ and ${\tt J}-1$ for $n$ even and odd, respectively.
The coefficients of $P_{1,n}$, $P_{2,n}$ depend  on $\delta$ and ${\tt K}$, but are independent of $\mu$.

\section{Discussion}
The subject matter of the paper is the 
 critical behavior of the inhomogeneous six-vertex model/\\ inhomogeneous
XXZ spin-$\frac{1}{2}$ chain.
We focus on the case when the lattice system possesses ${\cal Z}_r$ symmetry with
$r=1,3,5,\ldots$ in the critical regime, where 
the anisotropy parameter
$\gamma\in\big(\tfrac{\pi}{2}(1-\tfrac{1}{r}),\tfrac{\pi}{2}(1+\tfrac{1}{r})\big)$.
It was discussed the space of states occurring in the scaling limit
of the lattice system as well as  the  algebra of extended conformal symmetry. The central result
is the proposal for the class of differential equations, which describe the scaling limit of the eigenvalues of the 
Baxter $Q$-operators. While our conjectures were not proven, extensive numerical
work was performed in their support. 

\medskip
Below we would like to bring attention to some
problems, which turned out to be beyond the scope of our current study  and deserve to be investigated further.

\subsection{Integrable structure of the conformal field theory}
A full description of the ODE/IQFT correspondence requires one to construct the 
commuting family of operators acting in the field theory space of states
as well as the joint eigenbasis that diagonalizes them. In the case at hand,
the generating function of this family is played by the operators 
${\mathlarger{\mathlarger{\mathlarger {\boldsymbol  a}}}}_\pm(\mu)$
that occur in the scaling limit of the Baxter $Q$-operators \eqref{jkas8912jhasA} along with their barred counterparts.
One expects that
\be
{\mathlarger{\mathlarger{\mathlarger {\boldsymbol  a}}}}_\pm(\mu)={\bf 1}+\sum_{m=1}^\infty
{\mathlarger{\mathlarger{\mathlarger {\boldsymbol  a}}}}_{\pm,m}\,\mu^m\,,
\ee
so the  problem is to provide a well defined expression for the Taylor coefficients ---
the non-local IM.
\medskip

The joint eigenbasis can  be obtained, in principle, by means of the direct diagonalization of 
the non-local IM ${\mathlarger{\mathlarger{\mathlarger {\boldsymbol  a}}}}_{\pm,m}$.
However, this turns out to be a technically difficult task in practice and has not been realized even for
quantum KdV, the simplest non-trivial integrable field theory. 
Thus, one usually focuses on the operators  that occur in the large-$\mu$ asymptotic expansion
of ${\mathlarger{\mathlarger{\mathlarger {\boldsymbol  a}}}}_\pm(\mu)$. 
We have  described the general form of the
 expansion for the logarithm of the eigenvalues of this operator in 
eq.\,\eqref{aysayasy}. It involves the asymptotic coefficients  ${\mathfrak C}_+ $, ${Q}_{\frac{2m-1}{r}}$
and $\tilde{H}_m$. The coefficient $Q_1$,
up to an overall 
multiplicative normalization factor,
coincides with the eigenvalue of the local IM ${\bf I}_1$, whose density is the chiral component
of the energy-momentum tensor. The  coefficients ${Q}_{\frac{2m-1}{r}}$
are expected to be the eigenvalues of certain operators ${\bf Q}_{\frac{2m-1}{r}}$, which, in our terminology,
are referred to as the semi-local IM for $2m-1$ not divisible by $r$ and local IM otherwise.
In all likelihood ${\mathfrak C}_+ $ and $\tilde{H}_m$ are also the eigenvalues of  
operators from the commuting family (dual non-local IM).
 For $m=1,2,\ldots,\frac{r-1}{2}$ the
operators  ${\bf Q}_{\frac{2m-1}{r}}$ describe
 the scaling limit of the lattice quasi-shift operators, as was explained in sec.\,\ref{sec62}, while
the asymptotic coefficient ${\mathfrak C}_+ $  appears in the scaling limit
of various products over the Bethe roots (see sec.\ref{sec63}).
\medskip

The explicit formulae for 
${ Q}_{\frac{2m-1}{r}}$ with $m=1,2,\ldots,\frac{3r+1}{2}$ 
as well as $\mathfrak{C}_+$ in terms of the parameters of the ODE are presented 
in Appendix \ref{AppA}.
However, even with such knowledge, 
we were not able to construct any of  the IM for $r=3,5,\ldots$ apart from ${\bf Q}_1$
 as  operators acting in the space ${\cal H}_{\mathfrak{j},P}$, the irrep of 
the Kac-Moody algebra $\widehat{\mathfrak{su}}_1(r)\oplus\widehat{\mathfrak{u}}(1)$. 
In the case of ${\bf Q}_3$, whose eigenvalues $Q_3$ are given by a rather
cumbersome expression 
(see eqs.\eqref{hagf} and \eqref{hagf1} together with the  definition of $X_{3}$ \eqref{asyuysaysayas}),
we assumed that it takes the form of an integral over a  Lorentz spin 4 density, which is
a  local differential polynomial built out of the spin 1 currents ${\cal J}^A$ and ${\cal J}$ \eqref{asj9812ja}.
 Despite that this ansatz contains a finite number of parameters, we were not able to fix them such that
the direct diagonalization of the operator restricted to  ${\cal H}_{\mathfrak{j},P}^{({\tt L})}$  
matches the data coming from the formulae presented in Appendix \ref{AppA} away from the case $\delta=0$.
In our attempts to construct the simplest semi-local IM ${\bf Q}_{\frac{1}{r}}$
  an interesting, not  a priori expected  phenomenon, was observed.
  Namely, we found that it is always possible to
reproduce its whole spectrum in the level subspace ${\cal H}_{\mathfrak{j},P}^{({\tt L})}$
via a
${\cal D}_{\mathfrak{j}}({\tt L})\times{\cal D}_{\mathfrak{j}}({\tt L})$ matrix of the form
\be
\bm{A}+P\bm{B}\in{\rm End}\big({\cal H}_{\mathfrak{j},P}^{({\tt L})}\big)
\ee
with $P$-independent matrices $\bm{A}$ and $\bm{B}$. 
\medskip

We believe that the explicit
construction of the commuting family of operators is an important open problem, which may yield deeper insights
into the algebraic foundation of integrable quantum field theory.

\medskip
The spectrum of the IM for the quantum KdV theory admits an alternative description presented in the work 
of Litvinov \cite{Litvinov:2013zda},
which is different from that occurring in the context of the ODE/IQFT correspondence. 
The relation between the two has remained unclear for over 10 years now. It would be interesting to explore whether
the results of ref.\cite{Litvinov:2013zda} can be generalized to the integrable structure discussed in this paper 
for any odd $r$.

\subsection{ODEs for the system with softly broken ${\cal Z}_r$ symmetry }
 In ref.\cite{Gehrmann:2024tue}
 the scaling limit of  the inhomogeneous
XXZ spin-$\frac{1}{2}$ chain  with  softly broken ${\cal Z}_r$ symmetry was studied. By ``softly broken''
we mean that the  inhomogeneities 
$\eta_\ell\equiv \eta_{\ell+r}$ are not fixed to $(-1)^r\re^{\frac{\ri\pi}{r}(2\ell-1)}$ 
corresponding to the ${\cal Z}_r$ invariant case, but instead assigned a certain $N$-dependence
such that they achieve these values only in the $N\to\infty$ limit.  Among the results of the work \cite{Gehrmann:2024tue}
is  the differential equation that describes the scaling limit of the $Q$-functions for the ground state
in the regime, where the anisotropy parameter $\gamma\in\big(\frac{\pi}{2}(1-\frac{1}{r}),\frac{\pi}{2}(1+\frac{1}{r})\big)$
and $r$ is odd. In the notations used in this paper, it reads as
\be\label{i8923uqw}
\Big(-\partial_v^2+{\tt K}^2+\re^{2v}+\re^v\ F(\epsilon)\Big)\psi=0\,,
\ee
where
\bea
F(\epsilon)=\epsilon^r+\sum_{j=0}^{\frac{r-3}{2}}F_{2j+1}\ \epsilon^{2j+1}
\eea
and $\epsilon=\mu\ \re^{-\delta v}$.  The above ODE contains the $\frac{r-1}{2}$ constants $F_{2j+1}$. They 
are related to the $\frac{r-1}{2}$ RG invariants
\be\label{asn3hg27821}
{\mathfrak a}_{2j+1}=\frac{1}{2j+1}\  \bigg(\frac{2N}{\pi}\bigg)^{1-\frac{2j+1}{r}}\ \frac{1}{r}\ \sum_{\ell=1}^r(\eta_\ell)^{-2j-1}
\ \ \ \  \ \qquad \big(j=1,\ldots,\tfrac{r-1}{2}\big)\,,
\ee
which are assumed to be fixed as $N\to\infty$.\footnote{%
The differential equation \eqref{i8923uqw} becomes
eq.\,(2.18) from the work \cite{Gehrmann:2024tue} 
upon the  change of variables $v=\frac{n+r}{2}\,y+\log(\frac{2}{n+r})$.
 Also the parameters $\mu$, $\delta$, ${\tt K}$ and $F_{2j+1}$  
should be expressed in terms of $E$, $n$, $p$  and  $c_{2j+1}$  from that paper
as:
$$\mu=\re^{-\frac{\ri\pi}{2r}(r+1)}\ \Big(\frac{2}{n+r}\Big)^{\frac{2n}{r(n+r)}}\ E\ ,\ \ \ 
\qquad \ \ \ \ \delta=\frac{n-r}{r(n+r)}\ ,
\ \ \ \qquad \ \ \ \ \ {\tt K}=\frac{2p}{n+r}$$
 and
$$F_{2j+1}=\re^{-\frac{\ri\pi (r-1)}{2r}(2j+1)}\  \Big(\frac{2}{n+r}\Big)^{1-\frac{2j+1}{r}}\ c_{2j+1} \ .$$
Formula (2.83) in \cite{Gehrmann:2024tue} gives the relation between the RG invariants $\mathfrak{a}_{2j+1}$
and the coefficients $c_{2j+1}$.}
\medskip

In sec.\,\ref{sec3} it was presented the line of arguments  that takes one from the vacuum 
equation \eqref{as9021jjbnsd} to the class of ODEs corresponding 
to the excited states  \eqref{askjdjh21A}-\eqref{kasj892jhsa}. This can be repeated for the
 ODE \eqref{i8923uqw}, yielding
\be\label{askjdjh21AAAA}
\big(-\partial_v^2+U(v)\big)\psi=0\,,
\ee
where
\be\label{askjdjh21BsssssAAAAA}
U(v)={\tt K}^2+\re^{2v}+\re^v\ F(\epsilon)+\sum_{i=1}^{{\tt J}}\bigg(
\frac{3\delta^2\varpi_i^2\epsilon^2}{(\epsilon^2-\varpi_i^2)^2}+\frac{2\delta}{\epsilon^2-\varpi_i^2}\,\big(c_i\epsilon^2+
\varpi_i\epsilon\,\re^v\big)\bigg)\ .
\ee
Now the parameters $\{(\varpi_i,c_i)\}_{i=1}^{{\tt J}}$ specifying the position of the apparent singularities  and their residues
are determined through the algebraic system
\bea\label{kasj892jhsdddaaaa}
2c_j&=&F(\varpi_j)-1+2\delta\sum_{i\ne j}\frac{\varpi_i\varpi_j}{\varpi_j^2-\varpi_i^2}\\[0.2cm]
c_j^2&=& {\tt K}^2+\delta c_j-\frac{\delta^2}{4}+\sum_{i\ne j}\bigg(
\frac{3\delta^2\varpi_i^2\varpi_j^2}{(\varpi_i^2-\varpi_j^2)^2}+\frac{2\delta\,c_i\varpi_j^2}{\varpi_j^2-\varpi_i^2}\bigg)
\qquad\qquad (j=1,2,\ldots,{\tt J})\, ,\nonumber
\eea
which generalizes  \eqref{kasj892jhsa}.
A full analysis of the scaling limit of the $Q$-functions for the
 inhomogeneous
XXZ spin-$\frac{1}{2}$ chain with  softly broken ${\cal Z}_r$ symmetry
requires a significant amount of numerical work  and we leave it for future study.
Some
preliminary results concerning  the spectrum of the spin chain Hamiltonian  have already been  reported in sec.\,2.5.2 of 
ref.\cite{Gehrmann:2024tue}.

\bigskip

\section*{Acknowledgments}
\noindent
The authors acknowledge fruitful discussions with Sascha Gehrmann.
\smallskip

\noindent
GK would like to thank Holger Frahm for his interest in the work and  for his support.
Part of the research was carried out during GK's visits to the NHETC at Rutgers University. 
He is grateful for the  hospitality he received during those stays.
\smallskip

\noindent
The research of SL and DS is supported by the NSF under grant number NSF-PHY-2210187.
\bigskip
\bigskip

\appendix

\section{Some coefficients in the large-$\mu$ asymptotic expansion of $D_+(\mu)$ \label{AppA}}

\subsection{Asymptotic coefficient ${\mathfrak C}_+$\label{AppAa}}

The coefficient  ${\mathfrak C}_+$,  which appears in the large-$\mu$ asymptotic expansion of the spectral determinant \eqref{aysayasy}, can  be expressed in terms of the connection coefficient of the Fuchsian differential equation
\bea
\label{askjdjhAA}
\Bigg[-\partial_z^2+\frac{{\tt p}^2-1}{4z^2}+\sum_{i=1}^{{\tt J}}\bigg(
\frac{3\varpi_i^2 }{ 4 z ( z-\varpi_i^2)^2}+\frac{\gamma_i } { 2 z(z-\varpi_i^2)}\,\bigg)
\Bigg]{\tilde \psi}=0\, .
\eea
Here the parameters ${\tt p}$ and $\{\varpi_i\}_{i=1}^{\tt J}$  should be treated as independent, while $\{\gamma_i\}_{i=1}^{\tt J}$ are determined via  the algebraic  system
\bea\label{kasj89hqsa}
\gamma_j^2= {\tt p}^2+ \gamma_j-\frac{1}{4}+\sum_{i\ne j}\bigg(
\frac{3\varpi_i^2\varpi_j^2}{(\varpi_i^2-\varpi_j^2)^2}+\frac{2\gamma_i\varpi_j^2}{\varpi_j^2-\varpi_i^2}\bigg)
\qquad\qquad (j=1,2,\ldots,{\tt J})\, .
\eea
The latter ensures that all the  singularities of the ODE  are apparent except for the  regular singular points at  $z=0$ and $z=\infty$.

\medskip
The differential equation \eqref{askjdjhAA} admits the solution $\tilde{\psi}_{\tt p}$
distinguished by the condition that its analytic
continuation along a small counterclockwise contour  around $z=0$ yields
$-\re^{\ri\pi{\tt p}}\tilde{\psi}_{\tt p}$ together with the behavior
\bea
{\tilde \psi}_{\tt p}\to z^{\frac{1}{2}({\tt p}+1)}\ \ \ \  \ \ {\rm as} \ \ \  \ z\to 0\ .
\eea
Its asymptotic at large  $z$  takes the form
\bea\label{aksj982h981278sajh}
{\tilde \psi}_{\tt p}\to \re^{\frac{\ri\pi}{2}{\tt J}}\  C_{\tt p}\ z^{\frac{1}{2}({\tt p}-{\tt M}+1)}\ \ \ \  \ \ {\rm as} \ \ \  \ z\to\infty\,,
\eea
where  ${\tt M}$ is an integer such that
\bea
2\sum_{i=1}^{\tt J}\gamma_i=-2{\tt M}\, {\tt p}+{\tt M}^2\ .
\eea
The  connection  coefficient $C_{\tt p}$  is determined  by
 the independent parameters  ${\tt p}$ and $\{\varpi_i\}_{i=1}^{\tt J}$,  as well as  the particular solution  $\{\gamma_i\}_{i=1}^{\tt J}$ of the algebraic system \eqref{kasj89hqsa}. In other words,
\bea\label{ususasy}
C_{\tt p}=C_{\tt p}\big(\{\varpi_i\}_{i=1}^{\tt J}|\{\gamma_i\}_{i=1}^{\tt J}\big)\ .
\eea
The solution $\tilde{\psi}_{\tt p}$ can be constructed explicitly.
For example, in the simplest case ${\tt J}=1$,  eq.\eqref{kasj89hqsa}  gives 
 $\gamma_1=\frac{1}{2}\mp {\tt p}$ so that  ${\tt M}=\pm 1$. 
The  corresponding solution $ {\tilde \psi}_{\tt p}$ reads  as
\bea
\psi_{\tt p}=\frac{z^{\frac{1}{2}({\tt p} +1)}}{\sqrt{1-\varpi^{-2} z}}\times 
\begin{cases}
1\ \ \ \ &{\rm for}\ \ {\tt M}=+1\\[0.1in]
1-\frac{{\tt p}}{1+{\tt p}}\ \varpi^{-2}\ z\ \ \ \ &{\rm for}\ \ {\tt M}=-1
\end{cases}\ ,
\eea
where $\varpi^2=\varpi_1^2$. It follows that
\bea
C_{\tt p}=  \begin{cases}
\varpi\ \ \ \ &{\rm for}\ \ {\tt M}=+1\\[0.1in]
-\frac{{\tt p}}{1+{\tt p}}\, \varpi^{-1}\ \ \ \ &{\rm for}\ \ {\tt M}=-1\\[0.1in]
\end{cases}\ .
\eea
In the  general case,  $\tilde{\psi}_{\tt p}$ has the  form
\bea
\tilde{\psi}_{\tt p}=\frac{ z^{\frac{1}{2}({\tt p} +1)}}{\sqrt{\prod_{i=1}^{\tt J}\big(1-\varpi_i^{-2}z \big)}}\ \prod_{j=1}^{\frac{{\tt J}-{\tt M}}{2}}\Big(1-\frac{z}{x_j}\Big)\ .
\eea
Substituting this expression into the differential equation  \eqref{askjdjhAA}
yields the following algebraic system for the zeroes of  $\tilde{\psi}_{\tt p}$:
\bea
\sum_{m=1\atop m\not=j}^{\frac{{\tt J}-{\tt M}}{2}}
\frac{2}{x_m-x_j}-\sum_{i=1}^{{\tt J}}\frac{1}{\varpi_i^2-x_j}-\frac{{\tt p}+1}{x_j}=0\ \  \ \ \ \ \ \ \ \ \ \Big(j=1,\ldots,\frac{{\tt J}-{\tt M}}{2}\Big)\ .
\eea
This substitution also provides an explicit expression for the parameters $\gamma_i$ in terms of the sets $\{\varpi_i\}_{i=1}^{\tt J}$ and $\{ x_j\}_{i=1}^{\frac{{\tt J}-{\tt M}}{2}}$:
\bea
\gamma_i=\frac{1}{2}-{\tt p}+\sum_{m=1}^{\frac{{\tt J}-{\tt M}}{2}}
\frac{2\varpi_i^2}{x_m-\varpi^2_i}-\sum_{m=1\atop m\not=i}^{{\tt J}}\frac{\varpi_i^2}{\varpi^2_m-\varpi_i^2}\ \ \ \ \ \ \ \  (i=1,\ldots,{\tt J})\ .
\eea
The  connection coefficient  \eqref{aksj982h981278sajh},\,\eqref{ususasy}  is then given by
\bea
C_{\tt p}=(-1)^{\frac{{\tt J}-{\tt M}}{2}}\  \frac{\prod_{i=1}^{\tt J}{\varpi }_i}{ \prod_{j=1}^{\frac{{\tt J}-{\tt M}}{2}}x_j} \ .
\eea
Note that, using the  results of ref.\cite{Tarasov},  one can derive    
closed-form expressions for   $\psi_{\tt p}$ and  $C_{\tt p}$ in terms of the 
sets $\{\varpi_i\}_{i=1}^{\tt J}$  and  $\{\gamma_i\}_{i=1}^{\tt J}$.
The corresponding formulae, however,  will not be presented here.

\medskip

 Having obtained $C_{\tt p}$, the asymptotic coefficient ${\mathfrak C}_+$ is given by 
\bea\label{asoisisia}
\mathfrak{C}_+= 2^{-\frac{1}{2}-{\tt K}}\  \big(1-r\delta\big)^{\frac{1}{2}+\frac{2}{1-r\delta}\,({\tt K}+{\tt M}\,\delta)}\ \  \frac{\Gamma\big(1+\frac{2}{1-r\delta}\,({\tt K}+{\tt M}\,\delta)\big)}{\Gamma(1+{\tt K})}\ 
\  C_{-{\tt K}/\delta}\big(\{\varpi_i\}_{i=1}^{\tt J}|\{c_i/\delta\}_{i=1}^{\tt J}\big)\  .
\eea
This formula  can be obtained along the following line.

\medskip
Let us assume that $-\frac{1}{r}<\delta<0$, while $\mu$ is a real, positive number and
\bea -\delta\ \log(\mu)\gg 1\ .
\eea
Introduce a parameter $\Lambda$ such that
 \bea
 1\ll \Lambda\ll -\delta\ \log(\mu)\ .
 \eea
For  $v\in(-\infty,-\Lambda)$,  the ODE \eqref{askjdjh21A},\,\eqref{askjdjh21B} can be replaced  by
\bea\label{adjh21A}
\bigg[-\partial_v^2+
{\tt K}^2+\sum_{i=1}^{{\tt J}}\bigg(
\frac{3\delta^2\varpi_i^2\epsilon^2}{(\epsilon^2-\varpi_i^2)^2}+\frac{2\delta c_i\epsilon^2}{\epsilon^2-\varpi_i^2}\
\bigg)\Bigg]\psi=0\,,
\eea
while  in the complimentary domain   $v\in(-\Lambda,+\infty)$,
the equation  is approximated  as 
 \be\label{asjjbnsd}
\big(- \partial_v^2 +({\tt K}+{\tt M}\,\delta)^2+\re^{2v}+\epsilon^r\,\re^v\big)\psi=0
\ee
(recall that $\epsilon=\mu\re^{-\delta v}$).
To obtain the first three terms  in the large-$\mu$ asymptotic  expansion of the spectral determinant \eqref{aysayasy}, one should glue appropriate   solutions of the above differential
equations in the overlap region  $v\sim-\Lambda$. Now   we  note that eq.\eqref{adjh21A} can be brought to the form \eqref{askjdjhAA} via the  change of variables
$z=\mu^2\re^{-2\delta\, v}$,\ ${\tilde\psi}=z^{\frac{1}{2}}\psi$ and the parameters 
${\tt K}=-
{\tt p}\,\delta $,\ $c_i=\frac{\gamma_i}{\delta}$.
The analysis of the ODE \eqref{asjjbnsd} can be  carried out using the  standard WKB method. Performing the calculations then leads  to    \eqref{asoisisia}. For the case $0<\delta<\frac{1}{r}$,  a similar treatment results in the same formula for $\mathfrak{C}_+$.

\subsection{Asymptotic coefficients $Q_{\frac{2m-1}{r}}$\label{AppAb}}

Here we present   explicit  expressions  for  the coefficients $Q_{\frac{2m-1}{r}}$ for $m=1,2,\ldots ,\frac{3r+1}{2}$ appearing in
the asymptotic expansion \eqref{aysayasy}.
\medskip

Introduce the notation
\bea
S_{l}=
\begin{cases}
\sum_{i=1}^{{\tt J}}(\varpi_i)^{l}\ \ \ \ &{\rm for}\ \ \  l=1,3,5,\ldots\\[0.1in]
\sum_{i=1}^{\tt J}\big(\,\tfrac{3}{2}\, l\delta+2\, c_i\,\big)\,(\varpi_i)^{l}\ \ \ \ \ & {\rm for}\ \ \  l=2,4,6,\ldots\\[0.1in]
0\ \ \ \ & {\rm for}\ \ \  l\leq 0
\end{cases}
\eea
and define
\bea\label{asyuysaysayas}
&&X_{\frac{2m-1}{r}}= 
S_{2m-1} 
+
\frac{(2m-1)\delta }{1+r\delta -2(2m-1) \delta}\ 
S_{ 2m-r-1}\\[0.2in]
&&\,+\ \frac{(2 m-1)  \big(1+r \delta-(2 m-1) \delta\big)\delta }{6 (1+r \delta) \big(1+r \delta-2 (2 m-1) \delta\big)}\ 
\big(1- (2 m-1)\delta+(2m-1-2 r)(2 m-1-r)\, \delta^2\big)\, S_{2m-2 r-1}
\nonumber\\[0.2in]
&&\,+\ \frac{(2m-1)\delta^2}{1+ r\delta }\  \sum_{i=1}^{m-\frac{r+1}{2}}   S_{2i-1}S_{2m-r-2 i}-
\frac{2 (2 m-1) \big(1+r\delta - (2 m-1)\delta\big)\, \delta}
{(1+r\delta ) \big(1+r\delta -2( 2m-1) \delta\big)}\   ({\tt K}+ {\tt M}\delta)^2\, S_{2m-2r -1}
\nonumber \\[0.2in]
&&
\,-\ \frac{2 (2 m-1) \big(1+r\delta - (2 m-1)\delta\big)\, \delta^2}
{(1+r\delta ) \big(1+r\delta -2( 2m-1) \delta\big)}\ \sum_{i=1}^{m-r-1}S_{2i-1}S_{2(m-r-i)}\nonumber\\[0.2in]
&&\,-\  \frac{2 (2 m-1) \big(1+r\delta - (2 m-1)\delta\big)\, \delta^3}
{3 (1+r \delta  )^2} \ \sum_{i+j+k=m-r+1}
{ S}_{2 i-1} { S}_{2 j-1} { S}_{2 k-1}\ .\nonumber
\eea
Note that, as  follows from the above definitions,
\bea\label{askjru92jhsfdh3yyuyA}
X_{\frac{2m-1}{r}}=S_{2m-1}=\sum_{i=1}^{{\tt J}}(\varpi_i)^{2m-1}\ \ \ \ \ {\rm for}\ \ \ m=1,2,\ldots,\tfrac{r-1}{2}\ .
\eea
Furthermore, the first equation in \eqref{kasj892jhsa} together with  \eqref{askj902jhbdssa} imply that
\bea
X_1=S_r=2\,{\tt M}{\tt K}+ {\tt M}^2\,\delta+{\tt   J}\ .
\eea

For    $1\leq 2 m-1<3r$ and when  $2 m-1$ is not divisible by $r$, the asymptotic coefficients  $Q_{\frac{2m-1}{r}}$  are proportional to  $X_{\frac{2m-1}{r}}$. Specifically, one can show that
\be\label{askjru92jhsfdh3yyuyB}
Q_{\frac{2 m-1}{r}}= \frac{\Gamma(1-\frac{(2 m-1) \delta }{1+r\delta})\Gamma(\frac{1}{2}+\frac{(2 m-1)\delta  }{1+r\delta})}{(2m-1)\sqrt{\pi}}\ 
X_{\frac{2m-1}{r}}\ \ \ \  \   \big(1\leq 2 m-1<3r\,;\  2 m-1\not =0\ ({\rm mod}\, r)\big)
\ee
At the same time, for  $2m-1=(2k-1) r$, 
 the corresponding coefficients
are given by:
\bea\label{hagf}
Q_{2k-1}=(-1)^{k} \ 
  \frac{\Gamma\big(- (k-\frac{1}{2})\,\frac{1-r\delta}{1+r\delta}\big)\Gamma\big(\, (k-\frac{1}{2})\,\frac{2}{1+r\delta }\big)}
{  2 (1+r\delta) \sqrt{\pi}\, k!}\  \bigg(\frac{2(1-r\delta)}{r}\bigg)^{k }\ {\tilde I}_{2k-1}\ ,
\eea
where  the first two ${\tilde I}_{2 k-1}$ read as
\bea\label{hagf1}
{\tilde I}_1&=& \frac{ r\,({\tt K}+  {\tt M}\,\delta)^2}{2(1-r\delta)}-\frac{r}{24}+\frac{1}{2} \ ( 2\,{\tt M}{\tt K}+ {\tt M}^2\,\delta+{\tt   J}) \\[0.1in]
{\tilde I}_3&=&\frac{ r^2\,({\tt K}+ {\tt M}\,\delta)^4}{4 (1-r\delta)^2}
-\frac{r^2({\tt K}+ {\tt M}\,\delta)^2}{8(1-r\delta)}
+\frac{r^2(1-2 r\delta)(7+r\delta)}{960 (1-r\delta)}
+ \frac{r (1+r\delta)(1-5 r\delta)} {4(2-r\delta) (1-2 r\delta)(1-r\delta)}\ X_{3}\ .\nonumber
\eea
All other ${\tilde I}_{2k-1}$ with $k\geq 3$ are not  known  in  closed-form. 
However, they satisfy the following asymptotic behavior:
\bea
{\tilde I}_{2k-1}\to P^{2k}\ \ \ \ \ \ {\rm as}\ \ \ \ \ \ \ \  P\equiv \sqrt{\frac{r }{2(1-r\delta)}}\ \Big({\tt K}+\frac{{\tt M}}{r}\Big)\to \infty\ .
\eea
It is expected that ${\tilde{ I}}_{2k-1}$ are the eigenvalues of certain local IMs, normalized by the condition
\bea
{\tilde {\bf I}}_{2k-1}=\int_{0}^{2\pi}\frac{\rd x}{2\pi}\ \big({\cal J}^{2k}+\ldots\big)\ ,
\eea 
where the ellipses  denotes  terms involving  lower powers of the ${\hat {\mathfrak u}}(1)$ current ${\cal J}$. 
Note that ${\tilde {\bf I}}_{1}$ coincides with the local IM   defined in eqs.\eqref{kja09jh21jhjh}  and  \eqref{kaui32jnbas}, namely,
\bea
{\tilde {\bf I}}_{1}={ {\bf I}}_{1}=\int_{0}^{2\pi}\frac{\rd x}{2\pi}\ \Big({\cal J}^{2}+\frac{ q_{AB}}{2(r+1)}\ {\cal J}^A{\cal J}^B \Big)\ .
\eea
However, for $k>1$ the operators  ${\tilde {\bf I}}_{2k-1}$ at the free fermion point  differ 
from ${ {\bf I}}_{2k-1}$ \eqref{kjas9023hjjhd}
in their overall multiplicative normalization.
Taking into account eq.\eqref{9032jksdbnbsaaaaa}, one can verify that
\bea
{ {\bf I}}_{2k-1}=\frac{1}{k}\ \Big(\frac{2}{r}\Big)^{k-1}\ {\tilde  {\bf I}}_{2k-1}\big|_{\delta=0}\ .
\eea
\bigskip

\section{Symmetry transformations in the class of ODEs \eqref{askjdjh21A}-\eqref{kasj892jhsa} \label{AppB}}
The class of ODEs  \eqref{askjdjh21A}-\eqref{kasj892jhsa} possesses the symmetry transformations, 
which preserves
their monodromy properties. In particular, the spectral determinants of 
the differential equations  related via such transformations are the same. 
The formal symmetry group coincides with $\mathbb{Z}$ and the action
of its generator on the ODEs was briefly discussed in sec.\,\ref{sec5}. 
It takes the differential equation \eqref{askjdjh21A},\,\eqref{askjdjh21B} 
depending on the parameter ${\tt K}$ and with the given
set $\{(\varpi_i,c_i)\}_{i=1}^{{\tt J}}$ satisfying \eqref{kasj892jhsa} as well as the condition
\eqref{9023jdjnkjsd} with some integer ${\tt M}$ to the  equation
\eqref{askjui23jhdsn}-\eqref{asjhj3298u32hj} with $\tilde{{\tt M}}={\tt M}-r$.
 Here we describe the relation between the solution sets 
$\{(\varpi_i,c_i)\}_{i=1}^{{\tt J}}$ and $\{(\tilde{\varpi}_i,\tilde{c}_i)\}_{i=1}^{\tilde{\tt J}}$ entering into these two 
differential equations as well as the functions $F_1$ and $F_2$  appearing in the change of variables 
$\psi\mapsto\tilde{\psi}$ \eqref{sak90kj21jh3290A}.
\bigskip

\medskip

Let $\psi$ and $\tilde{\psi}$ be the functions related according to \eqref{sak90kj21jh3290A} with some $F_1=F_1(v)$ and $F_2=F_2(v)$. 
Then, if $\psi$
 is a solution of the ODE \eqref{askjdjh21A},\,\eqref{askjdjh21B} the condition
that $\tilde{\psi}$ obeys the differential equation \eqref{askjui23jhdsn},\,\eqref{askjui23jhdsnA}
leads to the coupled system:
\begin{subequations}\label{askd2hjb}
\bea
&&F_2\,\big(U-\tilde{U}\big)-2F_1'\,\tilde{U}-F_1\,\tilde{U}'-F_2'' =0\label{askd2hjbA}\\[0.2cm]
&&F_1\,\big(U-\tilde{U}\big)-2F_2'-F_1''=0\ ,
\eea
\end{subequations}
where the prime stands for the derivative w.r.t. to the argument, i.e., $v$.
Multiplying the first line by $F_1$ and the second line by $F_2$ and taking the difference of the resulting equations yields
\be
F_1^2\,\tilde{U}'+2F_1\,F_1'\tilde{U}-2F_2\,F_2'-F_2\,F_1''+F_1\,F_2''=0\,.
\ee
This may be  integrated to give
\be\label{askjdj12hbfd}
F_2^2-F_1^2\,\tilde{U}+F_2\,F_1'-F_1\,F_2'=C
\ee
with $C$ being the integration constant. The latter must be different from zero, since otherwise the transformation
\eqref{sak90kj21jh3290A} can not be inverted to express $\tilde{\psi}$ in terms of $\psi$ and its first derivative.
In view of the freedom in the choice of the overall multiplicative normalization of $\tilde{\psi}$, one may set
$C$ to any non-vanishing value.
\medskip

After working through some explicit examples, we were  motivated to search for 
a solution of the system \eqref{askd2hjb}  within the ansatz
\be\label{ashbdhgv32198}
F_1=\frac{\re^{-v}\,g(\epsilon)}{\sqrt{\Big(\prod_{i=1}^{\tt J} \epsilon^2-\varpi_i^2\Big)
\Big(\prod_{i=1}^{\tilde{\tt J}}\epsilon^2-\tilde{\varpi}_i^2\Big)}}\,,\qquad\qquad F_2=
\frac{\re^{-v}\,f_1(\epsilon)+f_2(\epsilon)}{\sqrt{\Big(\prod_{i=1}^{\tt J} \epsilon^2-\varpi_i^2\Big)
\Big(\prod_{i=1}^{\tilde{\tt J}}\epsilon^2-\tilde{\varpi}_i^2\Big)}}\ .
\ee
Here the parameters $\tilde{\varpi}_i$ are taken to be free for the moment, while
$g(\epsilon)$ and $f_i(\epsilon)$ depend on the parameter $\mu$ and variable $v$  through
the combination $\epsilon\equiv\mu\re^{-\delta v}$ only. 
Substituting \eqref{ashbdhgv32198} as well as  $\tilde{U}(v)$ in the form
\be
\tilde{U}(v)=\tilde{U}^{(0)}(\epsilon)+\tilde{U}^{(1)}(\epsilon)\,\re^v+\re^{2v}
\ee
  into \eqref{askjdj12hbfd},
yields three separate equations:
\begin{subequations}
\bea\label{askdyggh2v34}
&& f_1^2-g^2\,\tilde{U}^{(0)}+\delta\, \epsilon\,
 \big(g\,f_1'-f_1\,g'\big)=0\\[0.2cm]
&&f_2\,\big(2f_1-g\big)
-g^2\,\tilde{U}^{(1)}
+
\delta\, \epsilon  \,\big(g\, f_2'-f_2\,g'\big)=0\label{askjdbb2h3123aa}
\eea
and
\be\label{askjd892hhdb}
f_2^2-g^2=4\,\bigg(\prod_{i=1}^{\tt J} \epsilon^2-\varpi_i^2\bigg)
\bigg(\prod_{i=1}^{\tilde{\tt J}}\epsilon^2-\tilde{\varpi}_i^2\bigg)\ ,
\ee
\end{subequations}
where, in writing the last equation, we set the integration constant $C=4$.
The above is a system of first order differential equations that allows one, in principle,
to determine the functions $f_1$, $f_2$ and $g$. However, rather than solving them directly,
 simple expressions for $g$ and $f_2$ can be obtained by combining \eqref{askjd892hhdb}
with
\be\label{asj32hg98h1aa}
f_2\,\bigg(\sum_{i=1}^{{\tt J}}\frac{\varpi_i}{\epsilon^2-\varpi_i^2}-\sum_{i=1}^{\tilde{\tt J}}
\frac{\tilde{\varpi_i}}{\epsilon^2-\tilde{\varpi}_i^2}\bigg)-g\,
\bigg(\sum_{i=1}^{{\tt J}}\frac{\epsilon}{\epsilon^2-\varpi_i^2}+\sum_{i=1}^{\tilde{\tt J}}
\frac{\epsilon}{\epsilon^2-\tilde{\varpi}_i^2}\bigg)+g'=0 \ ,
\ee
which is a consequence of \eqref{askd2hjbA}.
This way, one finds
\bea\label{asldjh3298hi}
g(\epsilon)&=&\Big(\prod_{i=1}^{\tt J}\epsilon+\varpi_i\Big)\,\Big(
\prod_{i=1}^{\tilde{ \tt J}}\epsilon-\tilde{\varpi}_i\Big)-\Big(\prod_{i=1}^{\tt J}\epsilon-\varpi_i\Big)\,\Big(
\prod_{i=1}^{\tilde{ \tt J}}\epsilon+\tilde{\varpi}_i\Big) \,
 \\[0.2cm]
f_2(\epsilon)&=&
\Big(\prod_{i=1}^{\tt J}\epsilon+\varpi_i\Big)\,\Big(
\prod_{i=1}^{\tilde{ \tt J}}\epsilon-\tilde{\varpi}_i\Big)+\Big(\prod_{i=1}^{\tt J}\epsilon-\varpi_i\Big)\,\Big(
\prod_{i=1}^{\tilde{ \tt J}}\epsilon+\tilde{\varpi}_i\Big)\ .\nonumber
\eea
 Having at hand $g$ and $f_2$,  a further analysis of \eqref{askd2hjb} leads to the following expression for $f_1$:
\be\label{asldjh3298hiA}
f_1(\epsilon)=\frac{\delta}{2}\,\epsilon\,g'(\epsilon)+g(\epsilon)\,\bigg(\tilde{\tt K}-
\delta\sum_{i=1}^{{\tilde{\tt J}}}\frac{\epsilon^2}{\epsilon^2-\tilde{\varpi}_i^2}\bigg)\ .
\ee
\medskip

It should be pointed out, that though the unknown functions in the ansatz \eqref{ashbdhgv32198} have  been determined,
the equations \eqref{askd2hjb} are still not satisfied for an arbitrary set $\{(\tilde{\varpi}_i,\tilde{c}_i)\}_{i=1}^{{\tilde{\tt J}}}$. However,
it turns out that if this set is chosen properly then \eqref{askd2hjb}  holds true. 
To find the corresponding conditions   note that
the l.h.s. of eq.\,\eqref{askdyggh2v34} is a rational function of $\epsilon$. We consider its residues at $\epsilon=\tilde{\varpi}_j$. 
Setting them to zero  leads to
\be
-\tilde{\varpi}_j\,\frac{g'(\tilde{\varpi}_j)}{g(\tilde{\varpi}_j)}
+\sum_{i\ne j}\frac{2\tilde{\varpi}_j^2}{\tilde{\varpi}_j^2-\tilde{\varpi}_i^2}+
1-\frac{2}{\delta}\,\big(\tilde{{\tt K}}+\tilde{c}_j\big)= 0\qquad\qquad \qquad (j=1,2,\ldots,\tilde{{\tt J}})
\ee
or, explicitly, 
\be\label{kasjhd8ihfds}
\bigg(\prod_{i=1\atop i\ne j}^{\tilde{\tt{J}}}
\frac{\tilde{\varpi}_j-\tilde{\varpi}_i}{\tilde{\varpi}_j+\tilde{\varpi}_i}\bigg)\,
\bigg(\prod_{i=1}^{\tt{J}}\frac{\tilde{\varpi}_j+{\varpi}_i}{\tilde{\varpi}_j-{\varpi}_i}\bigg)
+
\sum_{i=1\atop i\ne j}^{\tilde{\tt{J}}}\frac{2\tilde{\varpi}_j}{\tilde{\varpi}_j-\tilde{\varpi}_i}-
\sum_{i=1}^{\tt{J}}\frac{2\tilde{\varpi}_j}{\tilde{\varpi}_j-{\varpi}_i}+1-\frac{4}{\delta}\,\big(\tilde{c}_j+\tilde{\tt K}\big)=0\ .
\ee
Strictly speaking, the obtained formulae  for $g$, $f_i$ and the  above relations are just necessary conditions that eqs.\,\eqref{askd2hjb}
are satisfied. Nevertheless the numerical analysis of various examples shows that there is a unique solution (up to permutations) of the system
\bea
2\tilde{c}_j&=&\tilde{\varpi}_j^r-1+2\delta\sum_{i\ne j}\frac{\tilde{\varpi}_i\tilde{\varpi}_j}{\tilde{\varpi}_j^2-\tilde{\varpi}_i^2}\\[0.2cm]
\tilde{c}_j^2&=& \tilde{{\tt K}}^2+\delta \tilde{c}_j-\frac{\delta^2}{4}+\sum_{i\ne j}\bigg(
\frac{3\delta^2\tilde{\varpi}_i^2\tilde{\varpi}_j^2}{(\tilde{\varpi}_i^2-\tilde{\varpi}_j^2)^2}+\frac{2\delta\,
\tilde{c}_i\tilde{\varpi}_j^2}{\tilde{\varpi}_j^2-\tilde{\varpi}_i^2}\bigg)
\qquad\qquad \qquad (j=1,2,\ldots,\tilde{{\tt J}})\,,\nonumber
\eea
supplemented by
\be\label{aksb328776155}
\sum_{i=1}^{\tilde{{\tt J}}}\tilde{c}_i=\tilde{{\tt M}}\tilde{{\tt K}}+\frac{\tilde{{\tt M}}^2}{2}\,\delta\,,
\ee
where
\be
\tilde{{\tt K}}={\tt K}+1\,,\qquad\qquad \tilde{\tt J}= {\tt J}+r-2{\tt M}\,,\qquad\qquad\tilde{\tt M}={\tt M}-r\,,
\ee
which in addition obeys \eqref{kasjhd8ihfds}. 
With such a choice of $\{(\tilde{\varpi}_i,\tilde{c}_i)\}$, 
eqs.\,\eqref{askd2hjb} hold true.
\medskip

Finally we note that, since the change of variables \eqref{sak90kj21jh3290A} is an invertible  transformation,
the sets  $\{(\varpi_i,c_i)\}$ and $\{(\tilde{\varpi}_i,\tilde{c}_i)\}$,
together with the relations \eqref{kasjhd8ihfds},
also obey
\be\label{ksa9012n43AA}
\bigg(\prod_{i=1\atop i\ne j}^{{\tt J}}\frac{\varpi_j-\varpi_i}{\varpi_j+\varpi_i}\bigg)\,
\bigg(\prod_{i=1}^{\tilde{{\tt J}}}\frac{\varpi_j+\tilde{\varpi}_i}{\varpi_j-\tilde{\varpi}_i}\bigg)+
\sum_{i=1\atop i\ne j}^{{\tt J}}\frac{2\varpi_j}{\varpi_j-\varpi_i}-
\sum_{i=1}^{\tilde{{\tt J}}}\frac{2\varpi_j}{\varpi_j-\tilde{\varpi}_i}
+1-\frac{4}{\delta}\,\big({c}_j-{\tt K}\big)=0\ .
\ee


\begin{thebibliography}{100}

\bibitem{Baxter}
  R.~J.~Baxter, \emph{Partition function of the eight-vertex lattice model},
  \href{https://www.sciencedirect.com/science/article/pii/0003491672903351?via\%3Dihub}{Ann. Phys.  \textbf{70} (1972) 193-228}.\hfil
  
  
  
\bibitem{Bazhanov:1996dr}
  V.~V.~Bazhanov, S.~L.~Lukyanov and A.~B.~Zamolodchikov,
\emph{Integrable structure of conformal field theory II. Q-operator and DDV equation},
\href{https://doi.org/10.1007/s002200050240}{Commun. Math. Phys.  {\bf 190} (1997) 247-278}
\href{http://arxiv.org/abs/hep-th/9604044}{{\ttfamily [arXiv:hep-th/9604044]}}.\hfil
%

\bibitem{Bazhanov:1998dq} 
  V.~V.~Bazhanov, S.~L.~Lukyanov and A.~B.~Zamolodchikov,
\emph{Integrable structure of conformal field theory III. The
Yang-Baxter relation},
\href{https://link.springer.com/article/10.1007/s002200050531}{Commun.\ Math.\ Phys.\  {\bf 200} (1999) 297-324}
\href{https://arxiv.org/abs/hep-th/9805008}{{\ttfamily [arXiv:hep-th/9805008]}}.\hfil


\bibitem{Gromov:2010km}
N.~Gromov, V.~Kazakov, S.~Leurent and Z.~Tsuboi,
\emph{Wronskian Solution for AdS/CFT Y-system},
\href{https://link.springer.com/article/10.1007/JHEP01(2011)155}{JHEP \textbf{01}, (2011) 155}
\href{https://arxiv.org/abs/1010.2720}{\ttfamily [arXiv:hep-th/1010.2720]}.\hfil

\bibitem{Jeong:2018qpc}
S.~Jeong and N.~Nekrasov,
\emph{Opers, surface defects, and Yang-Yang functional}, \href{https://link.intlpress.com/JDetail/1805560350156201986}{Adv. Theor. Math. Phys. \textbf{24}, no.7 (2020) 1789-1916}
\href{https://arxiv.org/abs/1806.08270}{\ttfamily [arXiv:hep-th/1806.08270]}.\hfil

  
  
\bibitem{Costello:2021zcl}
K.~Costello, D.~Gaiotto and J.~Yagi,
 \emph{Q-operators are {\textquoteright}t Hooft lines},
 \href{https://link.springer.com/article/10.1007/JHEP11(2024)003}{JHEP \textbf{11} 003 (2024) 003}\hfil\\
 \href{https://arxiv.org/abs/2103.01835}{\ttfamily [arXiv:2103.01835]}.

\bibitem{Voros:1994}
A.~Voros, \emph{Exact quantization condition for anharmonic oscillators (in one
dimension)},
\href{https://iopscience.iop.org/article/10.1088/0305-4470/27/13/038/meta}{J. Phys. {\bf A} {\textbf{27}} (1994) 4653-4661}.\hfil



  
  

\bibitem{Dorey:1998pt} 
  P.~Dorey and R.~Tateo,
  \emph{Anharmonic oscillators, the thermodynamic Bethe Ansatz, and nonlinear integral equations},
\href{https://doi.org/10.1088/0305-4470/32/38/102}{J. Phys. {\textbf A} {\textbf {32}} (1999) L419-L425}
\href{https://arxiv.org/abs/hep-th/9812211}{{\ttfamily [arXiv:hep-th/9812211]}}.\hfil






\bibitem{Bazhanov:1998wj} 
  V.~V.~Bazhanov, S.~L.~Lukyanov and A.~B.~Zamolodchikov, \emph{Spectral determinants for
  Schroedinger equation and Q operators of conformal field theory},
\href{https://doi.org/10.1023/A:1004838616921}{J.\ Stat.\ Phys.\  {\textbf 1}{\textbf 0}{\textbf 2} (2001) 567-576}
\href{https://arxiv.org/abs/hep-th/9812247}{{\ttfamily [arXiv:hep-th/9812247]}}.\hfil



                                                         
\bibitem{Bazhanov:2003ni} 
  V.~V.~Bazhanov, S.~L.~Lukyanov and A.~B.~Zamolodchikov,
 \emph{Higher level eigenvalues of Q operators and Schroedinger equation},
\href{https://doi.org/10.4310/ATMP.2003.v7.n4.a4}{Adv.\ Theor.\ Math.\ Phys.\  {\bf 7} (2003) 711-725}
\href{https://arxiv.org/abs/hep-th/0307108}{{\ttfamily[arXiv:hep-th/0307108]}}.\hfil


   \bibitem{Bazhanov:1994ft}
  V.~V.~Bazhanov, S.~L.~Lukyanov and A.~B.~Zamolodchikov,
 \emph{Integrable structure of conformal field theory, quantum KdV theory and
  thermodynamic Bethe ansatz},
\href{https://doi.org/10.1007/BF02101898}{Commun.\ Math.\ Phys.\  {\bf 177} (1996) 381-398}
\href{http://arxiv.org/abs/hep-th/9412229}{{\ttfamily [arXiv:hep-th/9412229]}}.\hfil











  

\bibitem{Masoero}
R.~Conti and D.~Masoero, \emph{Counting monster potentials}, \href{https://link.springer.com/article/10.1007/JHEP02(2021)059}
{JHEP\,{\bf 02}\,(2021)\,059}\hfil\\
\href{https://arxiv.org/abs/2009.14638}{\ttfamily [arXiv:hep-th/2009.14638]}.\hfil



\bibitem{Kotousov:2019ygw} 
  G.~A.~Kotousov and S.~L.~Lukyanov,
  \emph{Bethe state norms for the Heisenberg spin chain in the scaling limit},
  \href{https://doi.org/10.1016/j.nuclphysb.2019.114748}{Nucl.\ Phys.\ {\bf B} {\bf 947} (2019) 114748}
  \href{https://arxiv.org/abs/1906.07081}{{\ttfamily [arXiv:hep-th/1906.07081]}}.\hfil




\bibitem{Baxter:1971} 
R.~J.~Baxter,
\emph{Generalized ferroelectric model on a square lattice}, \href{https://doi.org/10.1002/sapm197150151}{Stud. Appl. Math. {\bf 50} (1971) 51-69}.\hfil


\bibitem{Bazhanov:2020new} 
  V.~V.~Bazhanov, G.~A.~Kotousov, S.~M.~Koval and S.~L.~Lukyanov,
  \emph{Some algebraic aspects of the inhomogeneous six-vertex model},
\href{https://doi.org/10.3842/SIGMA.2021.025}{SIGMA \textbf{17} (2021) 025}
\href{http://arxiv.org/abs/2010.10615}{{\ttfamily [arXiv:math-ph/2010.10615]}}.\hfil





\bibitem{Jacobsen:2005xz} 
  J.~L.~Jacobsen and H.~Saleur,
  \emph{The antiferromagnetic transition for the square-lattice Potts model},
 \href{https://doi.org/10.1016/j.nuclphysb.2006.02.041}{Nucl.\ Phys.\ {\bf B} {\bf 743} (2006) 207-248}
 \href{https://arxiv.org/abs/cond-mat/0512058}{\ttfamily [arXiv:cond-mat/0512058]}.\hfil

\bibitem{Ikhlef:2008zz} 
  Y.~Ikhlef, J.~Jacobsen and H.~Saleur,
  \emph{A staggered six-vertex model with non-compact continuum limit},
 \href{https://doi.org/10.1016/j.nuclphysb.2007.07.004}{Nucl.\ Phys.\ {\bf B} {\bf 789} (2008) 483-524}
 \href{https://arxiv.org/abs/cond-mat/0612037}{\ttfamily [arXiv:cond-mat/0612037]}.


\bibitem{IJS2}
Y.~Ikhlef, J.~L.~Jacobsen and H.~Saleur,
\emph{The $\mathbb{Z}_2$ staggered vertex model and its
applications}, \href{https://iopscience.iop.org/article/10.1088/1751-8113/43/22/225201}{J. Phys. {\bf A 43} (2010) 225201}
\href{https://arxiv.org/abs/0911.3003}{{\ttfamily [arXiv:math-ph/0911.3003]}}.\hfil



\bibitem{Ikhlef:2011ay} 
  Y.~Ikhlef, J.~L.~Jacobsen and H.~Saleur,
  \emph{An integrable spin chain for the SL(2,R)/U(1) black hole sigma model}, \href{https://doi.org/10.1103/PhysRevLett.108.081601}{Phys. Rev. Lett.  {\bf 108} (2012) 081601}
  \href{https://arxiv.org/abs/1109.1119}{{\ttfamily [arXiv:hep-th/1109.1119]}}.\hfil

  
  
  
  


\bibitem{Candu:2013fva} 
  C.~Candu and Y.~Ikhlef,
 \emph{Nonlinear integral equations for the SL(2,R)/U(1) black hole sigma model}, \href{https://doi.org/10.1088/1751-8113/46/41/415401}{J.\ Phys.\ {\bf A} {\bf 46} (2013) 415401}
 \href{https://arxiv.org/abs/1306.2646}{{\ttfamily [arXiv:hep-th/1306.2646]}}.\hfil
  
  
\bibitem{Frahm:2013cma} 
  H.~Frahm and A.~Seel,
  \emph{The staggered six-vertex model: conformal invariance and corrections to scaling}, \href{https://doi.org/10.1016/j.nuclphysb.2013.12.015}{Nucl.\ Phys.\ {\bf B} {\bf 879} (2014) 382-406}
  \href{https://arxiv.org/abs/1311.6911}{{\ttfamily [arXiv:comd-mat/1311.6911]}}.\hfil


\bibitem{Bazhanov:2019xvy} 
  V.~V.~Bazhanov, G.~A.~Kotousov, S.~M.~Koval and S.~L.~Lukyanov,
\emph{On the scaling behaviour of the alternating spin chain}, \href{https://doi.org/10.1007/JHEP08(2019)087}{JHEP {\bf 08} (2019) 087}
\href{https://arxiv.org/abs/1903.05033}{{\ttfamily[arXiv:hep-th/1903.05033]}}.\hfil

  
 

\bibitem{Bazhanov:2019xvyA} 
V.~V.~Bazhanov, G.~A.~Kotousov, S.~M.~Koval and S.~L.~Lukyanov,
\emph{Scaling limit   of the ${\cal Z}_2$ invariant inhomogeneous
  six-vertex model}, \href{https://www.sciencedirect.com/science/article/pii/S0550321321000341?via\%3Dihub}{Nucl.\ Phys.\ {\bf B} {\bf 965} (2021) 115337}\hfil\\
\href{https://arxiv.org/abs/2010.10613}{{\ttfamily [arXiv:math-ph/2010.10613]}}.\hfil
  


\bibitem{Robertson:2020imc}
N.~F.~Robertson, J.~L.~Jacobsen and H.~Saleur,
\emph{Lattice regularisation of a non-compact boundary conformal field theory}, \href{https://doi.org/10.1007/JHEP02(2021)180}{JHEP \textbf{02} (2021) 180}
\href{https://arxiv.org/abs/2012.07757}{\ttfamily [arXiv:hep-th/2012.07757]}.\hfil



%
\bibitem{Frahm:2021ohj}
H.~Frahm and S.~Gehrmann,
\emph{Finite size spectrum of the staggered six-vertex model with U$_{q}$($ \mathfrak{sl} $(2))-invariant boundary conditions},
\href{https://doi.org/10.1007/JHEP01(2022)070}{JHEP \textbf{01} (2022) 070}
\href{https://arxiv.org/abs/2111.00850}{\ttfamily [arXiv:cond-mat/2111.00850]}.\hfil

\bibitem{Frahm:2023ery}
H.~Frahm, S.~Gehrmann and G.~A. Kotousov, \emph{Scaling limit of the staggered six-vertex
model\\ with
 $U_q\big(\mathfrak{sl}(2)\big)$ invariant boundary conditions},
\href{https://doi.org/10.21468/SciPostPhys.16.6.149}{SciPost Phys. \textbf{16} (2024) 149}\hfil\\
\href{https://arxiv.org/abs/2312.11238}{\ttfamily [arXiv:hep-th/2312.11238]}.\hfil


\bibitem{Kluemper}
M.~Azhari and A.~Kl\"{u}mper, \emph{Managing singular kernels and logarithmic corrections 
in the staggered six-vertex model}, \href{https://doi.org/10.1007/JHEP12(2024)040}{JHEP \textbf{12} (2024) 040}
\href{https://arxiv.org/abs/2406.09889}{\ttfamily [arXiv:cond-mat/2406.09889]}.\hfil


\bibitem{Kotousov:2021vih}
G.~A.~Kotousov and S.~L.~Lukyanov,
\emph{ODE/IQFT correspondence for the generalized affine $ \mathfrak{sl} $(2) Gaudin model},
\href{https://doi.org/10.1007/JHEP09(2021)201}{JHEP \textbf{09} (2021) 201} \href{http://arxiv.org/abs/2106.01238}{\tt [arXiv:hep-th/2106.01238]}.\hfil

\bibitem{Kotousov:2023zps}
G.~A.~Kotousov and S.~L.~Lukyanov,
\emph{On the scaling behaviour of an integrable spin chain with ${\cal Z}_r$ symmetry}, \href{https://doi.org/10.1016/j.nuclphysb.2023.116269}{Nucl. Phys. {\bf B} \textbf{993} (2023) 116269}
\href{https://arxiv.org/abs/2305.03620}{\tt [arXiv:hep-th/2305.03620]}.\hfil






\bibitem{Gehrmann:2024tue}
S.~Gehrmann, G.~A.~Kotousov and S.~L.~Lukyanov,
\emph{Scaling limit of the ground state Bethe roots for the inhomogeneous XXZ spin-$\frac{1}{2}$ chain},
\href{https://doi.org/10.1016/j.nuclphysb.2024.11662}{Nucl. Phys. {\bf B} \textbf{1006} (2024) 116624}
\href{https://arxiv.org/abs/2406.12102}{\tt [arXiv:math-ph/2406.12102]}.\hfil



\bibitem{Gao}  
 G.~Cui, Y.~Gao, H.~H.~Rugh and L.~Tan,
 \emph{Rational maps as Schwarzian primitives},
 \href{https://link.springer.com/article/10.1007/s11425-016-5140-7}{Sci. China Math. {\textbf{59}}  (2016) 1267-1284}
 \href{https://arxiv.org/abs/1511.04246}{\ttfamily [arXiv:math.CV/1511.04246]}.\hfil

\bibitem{MasoeroRaimondo}
D.~Masoero, E.~Mukhin and A.~Raimondo,
\emph{$Q$-functions for lambda opers},
\href{https://doi.org/10.1007/s11005-025-01988-z}{Lett. Math. Phys. {\bf 115}  (2025) 103}
\href{https://arxiv.org/abs/2312.08842}{\tt [arXiv:math-ph/2312.08842]}.




  

\bibitem{Cardy:1986ie} 
  J.~L.~Cardy,
 \emph{Operator content of two-dimensional conformally invariant theories},
  \href{https://doi.org/10.1016/0550-3213(86)90552-3}{Nucl.\ Phys.\ {\bf B} {\bf 270} (1986) 186-204}.
  



\bibitem{BPZ}
A.~A.~Belavin, A.~M.~Polyakov and A.~B.~Zamolodchikov,
\emph{Infinite conformal symmetry in two-dimensional quantum field theory},
\href{https://doi.org/10.1016/0550-3213(84)90052-X}{Nucl. Phys. {\bf B} \textbf{241} (1984) 333-380}.\hfil


\bibitem{Knizhnik:1984nr}
V.~G.~Knizhnik and A.~B.~Zamolodchikov,
\emph{Current algebra and Wess-Zumino model in two-dimensions},
\href{https://doi.org/10.1016/0550-3213(84)90374-2}{Nucl. Phys. {\bf B} \textbf{247} (1984) 83-103}.\hfil


\bibitem{Zamolodchikov:1985wn} 
  A.~B.~Zamolodchikov,
  \emph{Infinite additional symmetries in two-dimensional conformal quantum field theory},
\href{https://doi.org/10.1007/BF01036128}{Theor.\ Math.\ Phys.\  {\bf 65} (1985) 1205-1213}
\href{http://www.mathnet.ru/php/archive.phtml?wshow=paper&jrnid=tmf&paperid=5141&option_lang=rus}{[Teor.\ Mat.\ Fiz.\  {\bf 65} (1985) 347-359]}.\hfil
 

\bibitem{Witten:1983ar}
E.~Witten,
\emph{Nonabelian bosonization in two-dimensions},
\href{https://doi.org/10.1007/BF01215276}{Commun. Math. Phys. \textbf{92} (1984) 455-472}. 



\bibitem{Itzykson}
C.~Itzykson, \emph{Level one Kac-Moody characters and modular invariance}, 
\href{https://www.sciencedirect.com/science/article/pii/0920563288903787}{Nucl. Phys. {\bf B 5} (Proc. Suppl.)  (1988) 150}.\hfil

\bibitem{rep}
G. A. Kotousov, S. L. Lukyanov and D. A. Shabetnik, \emph{Dataset: Numerical data for the
inhomogeneous ${\cal Z}_r$ invariant XXZ spin-$\frac{1}{2}$ chain}, 
\href{https://doi.org/10.25835/c00z8k7d}{Research data repository,
Leibniz Universit\"{a}t Hannover (2025)}.
  
\bibitem{Litvinov:2013zda}
A.~V.~Litvinov,
\emph{On spectrum of ILW hierarchy in conformal field theory},
\href{https://doi.org/10.1007/JHEP11(2013)155}{JHEP \textbf{11} (2013) 155}\hfil\\
\href{https://arxiv.org/abs/1307.8094}{\tt [arXiv:hep-th/1307.8094]}.\hfil





\bibitem{Tarasov}
A.~ Eremenko and V.~Tarasov,  \emph{Fuchsian equations with three non-apparent singularities},
\href{https://doi.org/10.3842/SIGMA.2018.058}{SIGMA \textbf{14} (2018) 058}
\href{https://arxiv.org/abs/1801.08529}{\ttfamily [arXiv:math.CA/1801.08529]}.\hfil
\end{thebibliography}
\end{document}